\documentclass[
    prd,
    amsmath,
    amssymb,
    longbibliography,
    floatfix,
    reprint,
    superscriptaddress,
]{revtex4-1}

\usepackage{bm}
\usepackage[toc, page]{appendix}
\usepackage[]{siunitx}
\usepackage{tikzscale}
\usepackage[normalem]{ulem}
\usepackage{pgfplotstable}
\usepackage[mode=buildnew]{standalone}
\usepackage{bigints}
\usepackage{booktabs}
\usepackage{ifthen}
\usepackage{tensor}
\usepackage[hidelinks]{hyperref}
\usepackage[nameinlink]{cleveref}
\usepackage{xfrac}
\usepackage[
	top    = 1.95cm,
	bottom = 1.9cm,
	left   = 1.5cm,
	right  = 1.5cm]
{geometry}
\usepackage{xparse}
\usepackage{xstring}
\usepackage{mathtools}
\usepackage{comment}

\definecolor{color1}{RGB}{202,0,32}
\definecolor{color2}{RGB}{244,165,130}
\definecolor{color3}{RGB}{146,197,222}
\definecolor{color4}{RGB}{5,113,176}

\crefname{appendix}{Appendix}{Appendices}
\crefname{subappendix}{Appendix}{Appendices}

\makeatletter

\g@addto@macro\bfseries{\boldmath}
\makeatother

\hypersetup{
	colorlinks,
	linkcolor={color4},
	citecolor={color4},
	urlcolor={color4}
}

\AtBeginDocument{
	\heavyrulewidth=.08em
	\lightrulewidth=.05em
	\cmidrulewidth=.03em
	\belowrulesep=.65ex
	\belowbottomsep=0pt
	\aboverulesep=.4ex
	\abovetopsep=0pt
	\cmidrulesep=\doublerulesep
	\cmidrulekern=.5em
	\defaultaddspace=.5em
}

\pgfplotsset{compat=1.15}

\newcommand{\dd}{\textrm{d}}
\newcommand{\dx}{\dd x}

\newcommand{\I}{\mathrm{i} \mkern1mu} 

\newcommand{\Hnum}[1]{\mathcal{H}_{#1}}

\newcommand{\Cin}{\mathrm{Cin}}
\newcommand{\Si}{\mathrm{Si}}

\newcommand{\cosnhalf}[2][]{\cos^{#1}\!\left(#2/{2}\right)}
\newcommand{\sinnhalf}[2][]{\sin^{#1}\!\left(#2/{2}\right)}
\newcommand{\secnhalf}[2][]{\sec^{#1}\!\left(#2/{2}\right)}
\newcommand{\cscnhalf}[2][]{\csc^{#1}\!\left(#2/{2}\right)}

\newcommand{\cosn}[2][]{\cos^{#1}{#2}}
\newcommand{\sinn}[2][]{\sin^{#1}{#2}}

\newcommand{\vect}[1]{\bm{\mathrm{#1}}}

\newcommand{\vq}{\vect{\hat{q}}}
\newcommand{\vqp}{\vect{\hat{q}}^{\prime}}
\newcommand{\vk}{\vect{k}}

\newcommand{\vn}{\vect{\hat{n}}}
\newcommand{\vm}{\vect{\hat{m}}}

\newcommand{\vdeltan}{\vect{\delta n}}

\newcommand{\vux}{\vect{\hat{u}}^{x}}
\newcommand{\vuy}{\vect{\hat{u}}^{y}}
\newcommand{\vuth}{\vect{\hat{u}}^{\theta}}
\newcommand{\vuph}{\vect{\hat{u}}^{\phi}}

\newcommand{\dn}{d_{\vn}}
\newcommand{\dm}{d_{\vm}}

\newcommand{\s}[1][]{%
	\ifthenelse{\equal{#1}{}}{S(\Theta)}{S^{#1}(\Theta)}%
}

\NewDocumentCommand{\Gorf}{ m m O{} O{} O{} O{} }{%
    \Gamma^{#3}_{#1#2\ifstrempty{#5#6}{}{,#5\IfBeginWith{#6}{-}{,}{}#6}}\ifstrempty{#4}{}{\!\left(#4\right)}%
}

\newcommand{\Gij}{\Gorf{i}{j}}
\newcommand{\GijP}{\Gorf{i}{j}[P]}

\newcommand{\GxthpTh}[2]{\Gorf{x}{\theta}[+][\Theta][#1][#2]}
\newcommand{\GyphpTh}[2]{\Gorf{y}{\phi}[+][\Theta][#1][#2]}

\newcommand{\GxphpTh}[2]{\Gorf{x}{\phi}[+][\Theta][#1][#2]}
\newcommand{\GythpTh}[2]{\Gorf{y}{\theta}[+][\Theta][#1][#2]}

\newcommand{\GxthpcTh}[2]{\Gorf{x}{\theta}[+, \times][\Theta][#1][#2]}
\newcommand{\GyphpcTh}[2]{\Gorf{y}{\phi}[+, \times][\Theta][#1][#2]}

\newcommand{\GxphpcTh}[2]{\Gorf{x}{\phi}[+, \times][\Theta][#1][#2]}
\newcommand{\GythpcTh}[2]{\Gorf{y}{\theta}[+, \times][\Theta][#1][#2]}

\newcommand{\GxthsTh}[2]{\Gorf{x}{\theta}[S][\Theta][#1][#2]}
\newcommand{\GyphsTh}[2]{\Gorf{y}{\phi}[S][\Theta][#1][#2]}

\newcommand{\GxphsTh}[2]{\Gorf{x}{\phi}[S][\Theta][#1][#2]}
\newcommand{\GythsTh}[2]{\Gorf{y}{\theta}[S][\Theta][#1][#2]}

\newcommand{\GxthxyTh}[2]{\Gorf{x}{\theta}[X,Y][\Theta][#1][#2]}
\newcommand{\GyphxyTh}[2]{\Gorf{y}{\phi}[X,Y][\Theta][#1][#2]}

\newcommand{\GxphxyTh}[2]{\Gorf{x}{\phi}[X,Y][\Theta][#1][#2]}
\newcommand{\GythxyTh}[2]{\Gorf{y}{\theta}[X,Y][\Theta][#1][#2]}

\newcommand{\GxthlTh}[2]{\Gorf{x}{\theta}[L][\Theta][#1][#2]}
\newcommand{\GyphlTh}[2]{\Gorf{y}{\phi}[L][\Theta][#1][#2]}

\newcommand{\GxphlTh}[2]{\Gorf{x}{\phi}[L][\Theta][#1][#2]}
\newcommand{\GythlTh}[2]{\Gorf{y}{\theta}[L][\Theta][#1][#2]}

\catcode`,\active

\catcode`\,12

\ExplSyntaxOn
\DeclareExpandableDocumentCommand{\RepQuad}{m}
{\int_compare:nT { #1 > 0 }
	{\quad \prg_replicate:nn { #1 - 1 } {\quad}}}
\ExplSyntaxOff

\renewcommand{\thetable}{\arabic{table}}

\makeatletter 
	\renewcommand{\fnum@figure}{Fig.~\thefigure}
\makeatother

\makeatletter 
	\renewcommand{\fnum@table}{Table~\thetable}
\makeatother

\makeatletter
\DeclareRobustCommand{\rcite}[1]{%
  \rcite@aux#1,\@nil{#1}%
}
\def\rcite@aux#1,#2\@nil#3{%
  \if\relax#2\relax
    Ref.~\cite{#3}%
  \else
    Refs.~\cite{#3}%
  \fi
}
\makeatother

\let\orgautoref\autoref
\renewcommand{\autoref}{%
    \def\equationautorefname{Eq.}%
    \def\figureautorefname{Fig.}%
    \def\sectionautorefname{Section}%
    \def\subsectionautorefname{Section}%
    \def\subsubsectionautorefname{Section}%
    \orgautoref
}

\allowdisplaybreaks[4]

\newcommand{\gw}{\textsc{gw}}
\newcommand{\sgwb}{\textsc{sgwb}}
\newcommand{\ligo}{\textsc{ligo}}
\newcommand{\kagra}{\textsc{kagra}}
\newcommand{\lvk}{\textsc{lvk}}
\newcommand{\gr}{\textsc{gr}}
\newcommand{\lisa}{\textsc{lisa}}
\newcommand{\pta}{\textsc{pta}}

\begin{document}
\raggedbottom
\setlength{\parskip}{0pt}
\setlength{\parindent}{1em}

\title[Astrometric Effects of anisotropic Gravitational Wave Backgrounds]{Astrometric effects of anisotropic gravitational wave backgrounds}

\author{Petar D. Dimitrov}
\email{pdimitrov@phys.uni-sofia.bg}
\affiliation{Department~of~Astronomy,~Faculty~of~Physics,~Sofia~University~``St~Kliment~Ohridski'',\\ 5 James Bourchier Blvd, 1164 Sofia, Bulgaria}

\author{Deyan P. Mihaylov}
\email{deyan@phys.uni-sofia.bg}
\email{me@deyanmihaylov.net}
\affiliation{Department~of~Astronomy,~Faculty~of~Physics,~Sofia~University~``St~Kliment~Ohridski'',\\ 5 James Bourchier Blvd, 1164 Sofia, Bulgaria}
\affiliation{CERCA/ISO, Department of Physics, Case Western Reserve University, 10900 Euclid Avenue, Cleveland, OH 44106, USA}

\begin{abstract}
\noindent 
Precise astrometric observations of distant objects such as stars and quasars can be used to probe gravitational wave signals at nanohertz frequencies. 
A stochastic gravitational wave background (SGWB) produces correlated apparent motions of pairs of objects on the sky. 
That correlation as a function of their angular separation is the astrometric analog of the Hellings--Downs curve in the context of pulsar timing measurements. 
Upper limits on the SGWB have been obtained from very-long baseline interferometry and Gaia astrometry, and future Gaia data releases are expected to improve them. 
An astrophysical background from inspiraling supermassive black hole binaries is expected to be significantly anisotropic. 
In this work, we compute the astrometric correlation functions for an anisotropic SGWB, known as the generalized overlap reduction functions, for all six gravitational wave polarizations. 
We present closed-form expressions in all cases where the distant-light source limit can be applied, following the approach used for pulsar timing arrays. 
We find that the correlations differ qualitatively between the tensor, vector, and scalar polarizations. 
Finally, we discuss the applicability of these results in future astrometry searches for arbitrarily polarized anisotropic gravitational wave backgrounds.
\end{abstract}

\maketitle

\section{Introduction}
\label{sec:intro}
\noindent
The first directly observed gravitational wave (\gw) signal, emitted from the merger of two stellar-mass black holes, was detected by the Laser Interferometer Gravitational-Wave Observatory (\ligo) detectors on 14 September 2015, thus marking the beginning of the era of \gw\ astronomy~\cite{LIGOScientific:2016aoc}. 
Since then, the \ligo-Virgo-\kagra\ (\lvk) Collaboration has confirmed over 300 \gw\ events, among them a handful of binary neutron star and neutron star--black hole mergers \cite{LIGOScientific:2026wfs, LIGOScientific:2017vwq, LIGOScientific:2021qlt}. 
Data from these observations has allowed us to learn about compact objects and the evolution of compact binaries -- the inspiral, merger, and subsequent ringdown of the binary, via the \gw\ signal produced by them.
In particular, this has enabled us to test general relativity (\gr) in the strong-field regime and has been complemented by electromagnetic observations from binary neutron stars \cite{LIGOScientific:2016lio, LIGOScientific:2017zic}. 

However, current ground-based \gw\ observatories are limited to detections in a narrow high-frequency range, with the four \lvk\ detectors operating in a nominal frequency band of \qtyrange[range-units = single]{10}{5000}{\hertz} (although the peak sensitivity is around \qty{100}{Hz}) during their fourth observing run \cite{Capote:2024rmo}. Future ground-based \gw\ observatories are under development: the Cosmic Explorer and the Einstein Telescope are designed to be able to detect frequencies as low as \qty{5}{\hertz} and \qty{1}{\hertz}, respectively, and are slated to have significantly greater sensitivity than the \lvk\ detectors \cite{Reitze:2019iox, Punturo:2010zz}. The future space-based mission Laser Interferometer Space Antenna (\lisa) will have a frequency range of \qtyrange[range-units=single]{e-4}{e-1}{\hertz} (with peak sensitivity at around millihertz) and will allow us to observe massive black hole binaries (\qtyrange[range-units=single]{e4}{e7}{M_\odot}), extreme-mass-ratio inspirals (inspirals of compact objects into supermassive black holes), and white dwarf binaries, among other objects \cite{LISA:2017pwj}.

However, these planned \gw\ observatories are not sensitive to signals with frequencies lower than about a microhertz \cite{Capote:2024rmo, Moore:2014lga, Reitze:2019iox, Punturo:2010zz, LISA:2017pwj}. 
At frequencies lower than that (corresponding to waves with periods of months or even years), we expect to find signals from a variety of astrophysical and cosmological sources. 
On the one hand, supermassive black hole binaries in our local region of the Universe are expected to emit approximately monochromatic individual \gw s, although there is no strong evidence for such signals yet \cite{Sesana:2008xk, NANOGrav:2023pdq}. 
Other sources in this band include S-stars orbiting Sagittarius A*, although their signals are well below the sensitivity of current observatories \cite{Cai:2018fkc}. 
On the other hand, we expect inspiraling supermassive black hole binaries to form a stochastic gravitational wave background (\sgwb) in the nanohertz frequency range \cite{Rajagopal:1994zj, Phinney:2001di, Jaffe:2002rt, Wyithe:2002ep, Enoki:2004ew, Sesana:2008mz, Sesana:2012ak, Kelley:2017lek, Burke-Spolaor:2018bvk, Sykes:2022rzh}. 
Additionally, various sources from the early Universe, such as cosmic strings and first-order phase transitions, can also lead to a nanohertz \sgwb\ \cite{Caprini:2018mtu}. 

One way to probe the nanohertz \sgwb\ is by using pulsar timing arrays (\pta s) \cite{Sazhin:1978myk, Detweiler:1979wn, Romani:1989, Foster:1990}. 
\pta s consist of a collection of millisecond pulsars that are observed over the course of many years. 
Passing \gw s affect the time of arrival of light pulses from the pulsars in a predictable manner \citep{Hellings_Downs:1983}. 
The \gw\ signal can then be detected by measuring the two-point correlation function of the (integrated) differences between the measured and expected times of arrival of pulses (timing residuals) from pairs of pulsars on the sky. 
The dependence of the two-point correlation function on the angular separation of the two pulsars is called the overlap reduction function. 
This function was first established for a Gaussian, zero-mean, stationary, isotropic, and unpolarized \sgwb\ with \gw s consistent with \gr\ by Ron Hellings and George Downs \citep{Hellings_Downs:1983}, and is known as the Hellings--Downs curve. 

It is generally expected that a population of inspiraling supermassive black hole binaries can lead to an \sgwb\ with significant anisotropy \cite{Mingarelli:2017fbe, Sato-Polito:2023spo, Yang:2024mqz, Lin:2026owz}. 
Correlation functions analogous to the Hellings--Downs curve can also be considered in the case of an anisotropic \sgwb. 
In such cases, the overlap reduction function can be expressed as a sum of \textit{generalized} overlap reduction functions, by first decomposing the supposed anisotropy of the background in an appropriate basis. 
The generalized overlap reduction functions for the two \gr\ polarizations were first calculated up to quadrupole order in \rcite{Mingarelli:2013dsa} and then generalized to non-Einsteinian polarizations and arbitrary anisotropy in \rcite{Gair:2014rwa, Gair:2015hra}. 

General relativity predicts only two \gw\ polarizations: \(+\) and \(\times\), while modified theories of gravity allow for up to six polarization states \cite{Will:2014kxa}. 
Since these overlap reduction functions have been shown to be qualitatively different for separate polarizations, measurements of the two-point correlations could provide a useful test for \gr\ \cite{Gair:2015hra}. 
The generalized overlap reduction functions were later also calculated using a bipolar spherical harmonic description and generalized for linear and circular polarizations \cite{Kato:2015bye, Hotinli:2019tpc, Liu:2022skj, AnilKumar:2023yfw, AnilKumar:2023hza, Sato-Polito:2021efu}. 
In 2023, the North American Nanohertz Observatory for Gravitational Waves (\textsc{nano}Grav) announced evidence for an \sgwb\ in the nanohertz frequency band \cite{NANOGrav:2023gor}, but to date no statistically significant evidence for its origin or possible non-isotropy has been found \cite{NANOGrav:2023hvm, NANOGrav:2023tcn}. 
Other \pta\ projects, such as the European Pulsar Timing Array (\textsc{epta}), the Indian Pulsar Timing Array (In\pta), the Parkes Pulsar Timing Array (\textsc{ppta}), the Chinese Pulsar Timing Array (\textsc{cpta}) and the Meer\textsc{kat} Pulsar Timing Array (\textsc{mpta}) have also reported evidence for an \sgwb\ \cite{EPTA:2023fyk, Reardon:2023gzh, Xu:2023wog, Miles:2024seg}.

An alternative method for detecting the \sgwb\ is to consider the astrometric deflections of distant objects such as stars or quasars due to passing gravitational waves \cite{Braginsky:1989pv, Pyne:1995iy}. 
Astrometric measurements can be used to provide an independent cross-check of \pta\ results, while joint \pta--astrometry analyses may improve constraints \cite{Cruz:2024diu, Vaglio:2025tex}. 
Although these astrometric deflections have not yet been directly observed, several upper limits have already been obtained from very-long-baseline interferometry (\textsc{vlbi}) and Gaia observations \cite{Gwinn:1996gv, Darling:2018hmc, Aoyama:2021xhj, Jaraba:2023djs, Darling:2024myz}. 
The upcoming Gaia Data Release 4 (DR4), expected in December 2026, will be an opportunity to improve the constraints on the \sgwb\ \cite{Geyer:2024lvk}. 
Although part of the \gw\ signal is likely lost when fitting the standard Gaia astrometric solution, a significant part goes into the residuals and can in principle be detected using careful analysis \cite{Geyer:2024lvk, Klioner:2017asb}. 

The astrometric analog of the Hellings--Downs curve for an isotropic \sgwb\ consisting of \gw s with \gr\ polarizations was derived in \rcite{Book:2010pf}, while the correlations for the cases of non-Einsteinian polarizations \cite{Mihaylov:2018uqm, OBeirne:2018slh, Qin:2018yhy} and non-luminal \gw s \cite{Mihaylov:2019lft} have also been calculated. 
\rcite{Inomata:2024kzr} used a bipolar spherical harmonics description to consider the case of an anisotropic \sgwb\ with linear and circular polarizations and calculated the astrometric generalized overlap reduction functions for the tensor and vector polarizations. 
The authors express their results in terms of an infinite linear combination of bipolar spherical harmonics with the coefficients in this expansion shown in Eq.~(65), Eq.~(77) and Tables II--IV of \rcite{Inomata:2024kzr}. 
In this work, we present closed-form expressions for the astrometric generalized overlap reduction functions for all of the six possible \gw\ polarizations for an anisotropic \sgwb\ using the methods of \rcite{Gair:2014rwa, Gair:2015hra}, including results for the transverse-scalar and scalar longitudinal polarizations. 
We assume that the observed objects are sufficiently far away that we can use the distant-light source approximation, which is the assumption used, for example, in \rcite{Mihaylov:2018uqm}. 
All our analytic results are convergent, except for the case of \(m = 0\) for two of the four components of the generalized overlap reduction function for the scalar longitudinal polarization. 
These two components of the generalized overlap reduction function diverge as the angular separation of the pairs of stars approaches zero and hence the distant-light source approximation cannot be used. 
However, when we compare these divergent generalized overlap reduction functions with the exact numerical results for pairs of stars sufficiently distant from Earth, the obtained analytic results are a faithful approximation for all but the smallest angular separations.

Our work is structured as follows. In \autoref{sec:theory}, we describe the theory related to the astrometric deflections caused by an \sgwb. In \autoref{sec:results}, we show our results for the generalized overlap reduction functions, and in \autoref{sec:discussion} we present our discussion and conclusions.
 
\section{Theoretical background}
\label{sec:theory}
\noindent
In this section, we describe the theory relating to the correlated astrometric deflection fields. 
Our presentation will follow that of \rcite{Mihaylov:2018uqm}, with the difference that we will instead consider an anisotropic \sgwb. 3-vectors are denoted by bold symbols and with a hat if they are unit vectors (e.g. $\vq$, $\vn$, $\vm$). The spatial components of a vector or tensor are denoted using Latin letters (e.g. $i$, $j$, $k$), while the components of a 4-vector are denoted using Greek letters (e.g. $\mu$, $\nu$, $\rho$). We work in units where $c=1$.

\begin{figure}[b]
    \centering
    \includegraphics[scale=1]{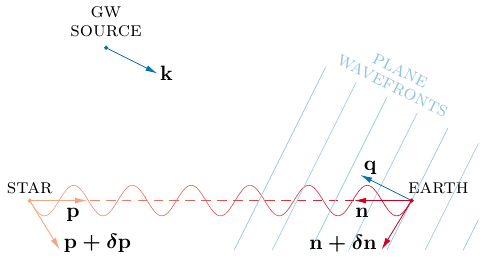}
    \caption{
        Diagram of geometric setup taken from \rcite{Mihaylov:2018uqm}. 
        A passing \gw\ with wavevector $\vk$ emitted from a \gw\ source in the direction $\vq = -\frac{1}{\omega}\vk$ changes the angular position of a star in the direction $\vn$ as viewed from the Earth by $\vdeltan$.}
    \label{fig:setup}
\end{figure}

A passing gravitational wave changes the apparent positions of distant objects. 
The geometry is shown in \autoref{fig:setup} taken from \rcite{Mihaylov:2018uqm}. 
We will call the idealized stationary observer ``Earth", to be consistent with the terminology in the literature, but the observer can be elsewhere, say, at the L2 point like Gaia. 
In any case, we have to correct for the motion of the detector. On Earth in the absence of \gw s we observe light with frequency $\Omega$ in the direction $\vn$ emitted from an object which we will call the ``star" but could also be, say, a quasar. 
We consider the weak-field regime with $g_{\mu\nu}=\eta_{\mu\nu}+h_{\mu\nu}$ where $\eta_{\mu\nu}=\mathrm{diag}\left(-1,1,1,1\right)$ is the Minkowski metric. 
We assume monochromatic, plane-fronted \gw s with frequency $\omega$ due to a distant \gw\ source in the direction $\vq$:
\begin{equation}
    h_{\mu\nu}\left(t, x^i\right)=\Re\left\{H_{\mu\nu}\exp\left(\I \mathrm{k}_{\rho}\mathrm{x}^{\rho}\right)\right\},
\end{equation}
where $\mathrm{k}^{\rho}\!=\!\left(\omega,-\omega q^i\right)$ is the wavevector of the \gw s and $H_{\mu\nu}$ are small complex constants that describe the strength and polarization of the \gw s. It can be shown that the angular deflection of a star due to passing monochromatic, plane-fronted \gw s is (see e.g. \rcite{Book:2010pf, Mihaylov:2018uqm})
\begin{widetext}
\begin{align}\label{eq:exact}
\begin{aligned}
    \delta n_{i} &= \Bigg[\Bigg(\Bigg\{1+\frac{\I \left(2 - q^rn_r\right)}{\omega \lambda_S \Omega\left(1 - q^wn_w\right)}\left(1 - \exp\left(-\I \omega\Omega\lambda_S \left(1 - q^vn_v\right)\right)\right)\Bigg\} n_{i} \\
    & \qquad\qquad
    -\Bigg\{1+\frac{\I}{\omega \lambda_S \Omega\left(1 - q^wn_w\right)}\left(1 - \exp\left(-\I \omega\Omega\lambda_S \left(1 - q^vn_v\right)\right)\right)\Bigg\} q_{i}\Bigg) \frac{H_{jk} n^j n^k}{2\left(1 - q^sn_s\right)} \\
    & \qquad\qquad\qquad\qquad
    -\Bigg\{\frac{1}{2} + \frac{\I}{\omega \lambda_S \Omega\left(1 - q^wn_w\right)}\left(1 - \exp \left(-\I \omega\Omega\lambda_S \left(1 - q^vn_v\right)\right)\right) \Bigg\} H_{ij} n^j\Bigg] \exp\left(-\I \omega t_{0}\right),
\end{aligned}
\end{align}
\end{widetext}
where we define $\lambda_S\equiv-|\bm{\mathrm{x}}_S|/\Omega$ ($|\bm{\mathrm{x}}_S|$ is the distance between the star and Earth) and $t_0$ is the time of arrival of the light signal at Earth in the absence of \gw s. This result is calculated in a time-independent orthonormal tetrad basis. In the distant-light source limit, $\omega|\lambda_S|\Omega\gg1$, we can approximate the angular displacement as 
\begin{equation} \label{eq:approximation}
    \delta n_{i} = \frac{1}{2}\left(\frac{n_{i}-q_{i}}{1-q^sn_s}h_{jk}\left(\mathrm{E}\right)n^{j}n^{k}-h_{ij}\left(\mathrm{E}\right)n^j\right),
\end{equation}
so that the angular displacement now only depends on the metric perturbation at Earth, $h_{jk}\left(\mathrm{E}\right)=H_{jk}\exp\left(-\I \omega t_{0}\right)$. 
This is similar to neglecting the pulsar term in \pta\ analysis. Note that by the distant-light source limit (sometimes just called the distant-source limit in the literature), we mean that we take the distance to the star to be infinity, not the distance to the \gw\ source.

Henceforth, we shall use Eq.~\eqref{eq:approximation} whenever possible. However, as we will later show, in some cases when the gravitational wave source is in the same direction as the star ($q^sn_s = 1$), Eq.~\eqref{eq:approximation} diverges, and we cannot use the distant-light source approximation. We will comment on these cases when we present our results for the astrometric correlation functions.

According to \gr, gravitational waves can have two possible polarizations --- the transverse-traceless $+$ and $\times$ modes. Alternative theories of gravity permit up to four additional polarizations --- a transverse-scalar mode ($S$), two vectorial polarizations with mixed transverse and longitudinal components ($X$ and $Y$), and a purely longitudinal scalar polarization ($L$) \cite{Will:2014kxa}. If we define the standard orthonormal basis associated with the spherical polar coordinates as
\begin{subequations}\label{eq:spherical_basis}
\begin{align}
        \hat{e}^r_i & = \left(\sin{\theta}\cos{\phi}, \sin{\theta}\sin{\phi}, \cos{\theta}\right), \\
        \hat{e}^{\theta}_i & = \left(\cos{\theta}\cos{\phi}, \cos{\theta}\sin{\phi}, -\sin{\theta}\right), \\
        \hat{e}^{\phi}_i & = \left(-\sin{\phi}, \cos{\phi}, 0\right),
\end{align}
\end{subequations}
then we can express the constants $H_{ij}$ for a \gw\ traveling from a source in the direction $q_i=\hat{e}^r_i$ as
\begin{equation}
    H_{ij}=A_+\epsilon^+_{ij}+A_{\times}\epsilon^{\times}_{ij} + A_{S}\epsilon^{S}_{ij} + A_{X}\epsilon^{X}_{ij} + A_{Y}\epsilon^{Y}_{ij} + A_{L}\epsilon^{L}_{ij},
\end{equation}
where we define the polarization basis tensors as

\vspace{0pt}
\begin{subequations}\label{eq:pol_basis_vectors}
\noindent
\begin{minipage}[t]{0.48\linewidth}
    \setlength{\abovedisplayskip}{0pt}
    \setlength{\belowdisplayskip}{0pt}
    \begin{align}
        \epsilon^+_{ij}\left(\vq\right) &= \hat{e}^{\theta}_i\hat{e}^{\theta}_j-\hat{e}^{\phi}_i\hat{e}^{\phi}_j,\label{subeq:plus_pol} \\  
        \epsilon^{\times}_{ij}\left(\vq\right) &= \hat{e}^{\theta}_i\hat{e}^{\phi}_j+\hat{e}^{\phi}_i\hat{e}^{\theta}_j,\label{subeq:cross_pol} \\
        \epsilon^S_{ij}\left(\vq\right) &= \hat{e}^{\theta}_i\hat{e}^{\theta}_j+\hat{e}^{\phi}_i\hat{e}^{\phi}_j,\label{subeq:s_pol}
    \end{align}
\end{minipage}%
\hfill
\begin{minipage}[t]{0.48\linewidth}
    \setlength{\abovedisplayskip}{0pt}
    \setlength{\belowdisplayskip}{0pt}
    \begin{align}
        \epsilon^X_{ij}\left(\vq\right) &= \hat{e}^{\theta}_i\hat{e}^{r}_j+\hat{e}^{r}_i\hat{e}^{\theta}_j,\label{subeq:x_pol} \\
        \epsilon^Y_{ij}\left(\vq\right) &= \hat{e}^{\phi}_i\hat{e}^{r}_j+\hat{e}^{r}_i\hat{e}^{\phi}_j,\label{subeq:y_pol} \\
        \epsilon^L_{ij}\left(\vq\right) &= \sqrt{2} \, \hat{e}^{r}_i\hat{e}^{r}_j. \label{subeq:l_pol}
    \end{align}
\end{minipage}
\end{subequations}
\par
\vspace{\belowdisplayskip}
\noindent

These expressions are the same as the expressions for the polarization basis tensors in Cartesian coordinates for a \gw\ traveling along the z-axis, with the replacement $\left(x,y,z\right)\leftrightarrow\left(\theta,\phi,r\right)$. 
The factor of $\sqrt{2}$ in Eq.~\eqref{subeq:l_pol} is added for normalization purposes.

We will now consider the astrometric response due to an anisotropic \sgwb. 
This generates stochastic patterns of astrometric deflections over the sky that in general are different compared with the case of an isotropic background. The metric at Earth can be Fourier decomposed as
\begin{equation}
    h_{ij}\!\left(t\right) = \Re\!\left[\sum_{P} \! \int_0^{\infty} \!\!\!\! \dd f\int_{S^{2}} \!\!\! \dd \Omega_{\vq}\,A_{P}\!\left(\vq, f\right) e^{-2\pi \I f t} \epsilon^P_{ij}\!\left(\vq\right)\right],
\end{equation}
where \(f = \omega / (2 \pi)\) is the linear \gw\ frequency, $A_P\left(\vq,f\right)$ are the Fourier coefficients, $\mathrm{d}\Omega_{\vq}$ is the area element on the sphere, and $\vq$ is the direction towards the \gw\ source. We integrate over all directions in the sky and over all frequencies, and sum over all possible \gw\ polarizations, $P \in \left\{+, \times, S, X, Y, L\right\}$.

We assume that the \gw\ background is Gaussian, zero-mean, stationary and unpolarized, but unlike \rcite{Mihaylov:2018uqm}, we do not assume it to be isotropic. Therefore, the expectation values of the Fourier coefficients are
\begin{widetext}
\begin{subequations}\label{eq:fc_expectations}
\begin{align}
    \left\langle A_{P}\left(\vq, f\right) A^{*}_{P^{\prime}}\left(\vqp,f^{\prime}\right) \right\rangle&=C\left(f\right)P\left(\vq\right) \, \delta_{PP^{\prime}} \, \delta_{S^2}\left(\vq,\vqp\right) \, \delta\left(f-f^{\prime}\right), \label{subeq:correlations_rc} \\
    \left\langle A_P\left(\vq,f\right) A_{P^{\prime}}\left(\vqp,f^{\prime}\right) \right\rangle&=0, \label{subeq:correlations_rr}
\end{align}
\end{subequations}
\end{widetext}
where $C(f)$ and $P\left(\vq\right)$ describe the frequency and the angular dependence of the power spectrum, respectively. 
For an isotropic background, $P\left(\vq\right)=\mathrm{const}$. 
We assume that this angular distribution of the background is independent of the frequency \(f\) -- this does not change the generalized overlap reduction functions. We mention the differences in the calculations below.

Using the distant-light source approximation described by Eq.~\eqref{eq:approximation}, we can express the astrometric response as
\begin{align}\label{eq:astrometric_response_single_star}
\begin{aligned}
    & \delta n_i \! \left(\vn,t\right) = \Re\Bigg[\sum_P\int_0^{\infty} \!\!\!\! \dd f \, e^{-2\pi \I f t} \\
    & \qquad\qquad\qquad \times \int_{S^2} \!\!\! \dd\Omega_{\vq}\,A_P\left(\vq,f\right) \Delta_i^{jk}\left(\vn,\vq\right)\epsilon^P_{jk}\left(\vq\right)\Bigg] ,
\end{aligned}
\end{align}
where
\begin{equation} \label{eq:triangle}
    \Delta_i^{jk}\left(\vn,\vq\right)\equiv\frac{1}{2}\left(\frac{n_i-q_i}{1 - q^sn_s} \, n^j n^k - \delta_i^j n^k\right).
\end{equation}

\begin{figure}[b]
    \centering
    \includegraphics[scale=1]{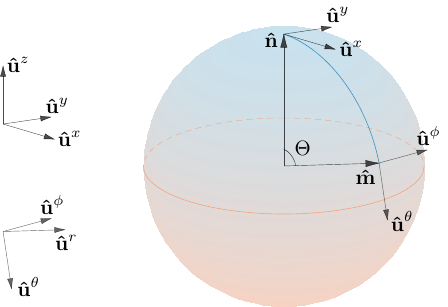}
    \caption{Coordinate system and basis chosen for calculating the generalized overlap reduction functions, taken from \rcite{Mihaylov:2018uqm}. The star in the direction $\vn$ is placed on the North pole, while the star in the direction $\vm$ is placed in the xz plane. We consider the geodesic connecting the two points. At the point $\vn\,(\vm)$ we choose as a basis a vector parallel to the geodesic, $\vux\,(\vuth)$, and a vector perpendicular to the geodesic, $\vuy\,(\vuph)$.}
    \label{fig:unit_vectors}
\end{figure}

The astrometric response depends linearly on the metric perturbation, which ensures that it is a Gaussian random variable and that its statistical properties depend only on the two-point correlations. Using Eqs.~\eqref{eq:fc_expectations} and \eqref{eq:astrometric_response_single_star}, the expectation value of the product of the astrometric responses of two stars, $\delta n_i$ and $\delta m_j$, in directions $\vn$ and $\vm$, can be expressed as 
\begin{equation}\label{eq:two_point_corr_approx}
    \left\langle\delta n_i\left(\vn,t\right)\delta m_j\left(\vm,t'\right)\right\rangle=T\left(t,t^{\prime}\right)\Gamma_{ij}\left(\vn,\vm\right) . 
\end{equation}
The temporal dependence of the correlation is
\begin{equation}
    T\!\left(t,t^{\prime}\right) = \frac{1}{4}\int_0^{\infty}\!\!\!\dd f\, C\!\left(f\right)\left(e^{2\pi \I f \left(t - t^{\prime}\right)} + e^{-2\pi \I f \left(t - t^{\prime}\right)}\right) .
\end{equation}
The factorization in Eq.~\eqref{eq:two_point_corr_approx} follows from the assumption in Eq.~\eqref{subeq:correlations_rc} that the power spectrum can be written as \(C(f) P(\vq)\). 
This is a simplifying assumption which would not necessarily hold for some foreground signals. 
For a frequency-dependent power spectrum, \(\mathcal{P} (f, \vq)\), Eq.~\eqref{eq:two_point_corr_approx} would no longer be valid, and would need to be replaced by a non-separable integral over \(f\). 

The spatial correlation factor in Eq.~\eqref{eq:two_point_corr_approx} is equal to
\begin{equation}
    \Gamma_{ij}\left(\vn,\vm\right) = \sum_{P} \GijP \! \left(\vn, \vm\right) ,
\end{equation}
where for each polarization state we define the overlap reduction functions
\begin{subequations}
\begin{align} \label{eq:correlation_function_general}
\Gamma^P_{ij}\left(\vn,\vm\right)&=\int_{S^2}\dd \Omega_{\vq}\,P\left(\vq\right)\delta n_i^P\left(\vn,\vq\right)\delta m_j^P\left(\vm,\vq\right), \\
\mathrm{where} \;\:&\delta n_i^P\left(\vn,\vq\right)\equiv\Delta_i^{jk}\left(\vn,\vq\right)\epsilon^P_{jk}\left(\vq\right) . \label{subeq:delta_ni}
\end{align}
\end{subequations}

We can expand the power spectrum \(P\left(\vq\right)\) in terms of spherical harmonics, as was done in the case of \pta s in e.g. \rcite{Mingarelli:2013dsa}:
\begin{equation}\label{eq:power_expansion}
    P\!\left(\vq\right) = \sum_{\ell=0}^{\infty}\sum_{m=-\ell}^{\ell}c_{\ell}^mY_{\ell}^m\left(\vq\right) = \sum_{\ell m}c_{\ell}^mY_{\ell}^m\left(\vq\right),
\end{equation}
where $\ell$ and $m$ are the harmonic numbers which can take values $\ell=0, 1, \ldots, \infty$ and $m = -\ell, -\ell + 1, \ldots, \ell - 1, \ell$. 
The spherical harmonic functions are defined in \cref{app:math}. 
Note that in the case of a frequency-dependent background distribution \(\mathcal{P}(f, \vq)\) (instead of the separable form \(C(f) P(\vq)\) in Eq.~\eqref{subeq:correlations_rc}), the coefficients \(c_{\ell}^{m}\) would need to be replaced by the  frequency dependent \(c_{\ell}^{m} (f)\). 
If we plug in Eq.~\eqref{eq:power_expansion} into Eq.~\eqref{eq:correlation_function_general}, we obtain the result
\begin{equation}
    \GijP \left(\vn, \vm\right) = \sum_{\ell m}c_{\ell}^{m} \, \Gamma^P_{ij, \ell m}\left(\vn,\vm\right), 
\end{equation}
where we define 
\begin{equation}\label{eq:gorf}
\begin{aligned} 
    \Gamma^{P}_{ij, \ell m}\left(\vn,\vm\right) = \! \int_{S^2} \!\!\! \dd\Omega_{\vq} \, Y_{\ell}^{m}\!\left(\vq\right)\delta n_i^{P}\!\left(\vn, \vq\right) \delta m_j^{P}\!\left(\vm, \vq\right).
\end{aligned}
\end{equation}
We will henceforth call $\Gamma^P_{ij, \ell m}\left(\vn, \vm\right)$ the generalized overlap reduction functions (as was done in \rcite{Mingarelli:2013dsa} in the context of \pta s), and in the rest of the paper we will calculate Eq.~\eqref{eq:gorf} for the various polarization states of the gravitational wave background. 
Eq.~\eqref{eq:gorf} differs from the isotropic case due to the presence of the spherical harmonic functions, $Y_{\ell}^m\left(\vq\right)$, in the integral over the sphere. 
The spatial correlation functions for an isotropic background calculated in \rcite{Mihaylov:2018uqm} are the same as $\Gamma^P_{ij, 00}\left(\vn,\vm\right)$ up to a multiplication factor and are used as a test for our calculations.

\begin{figure*}[p]
\centering
\includegraphics{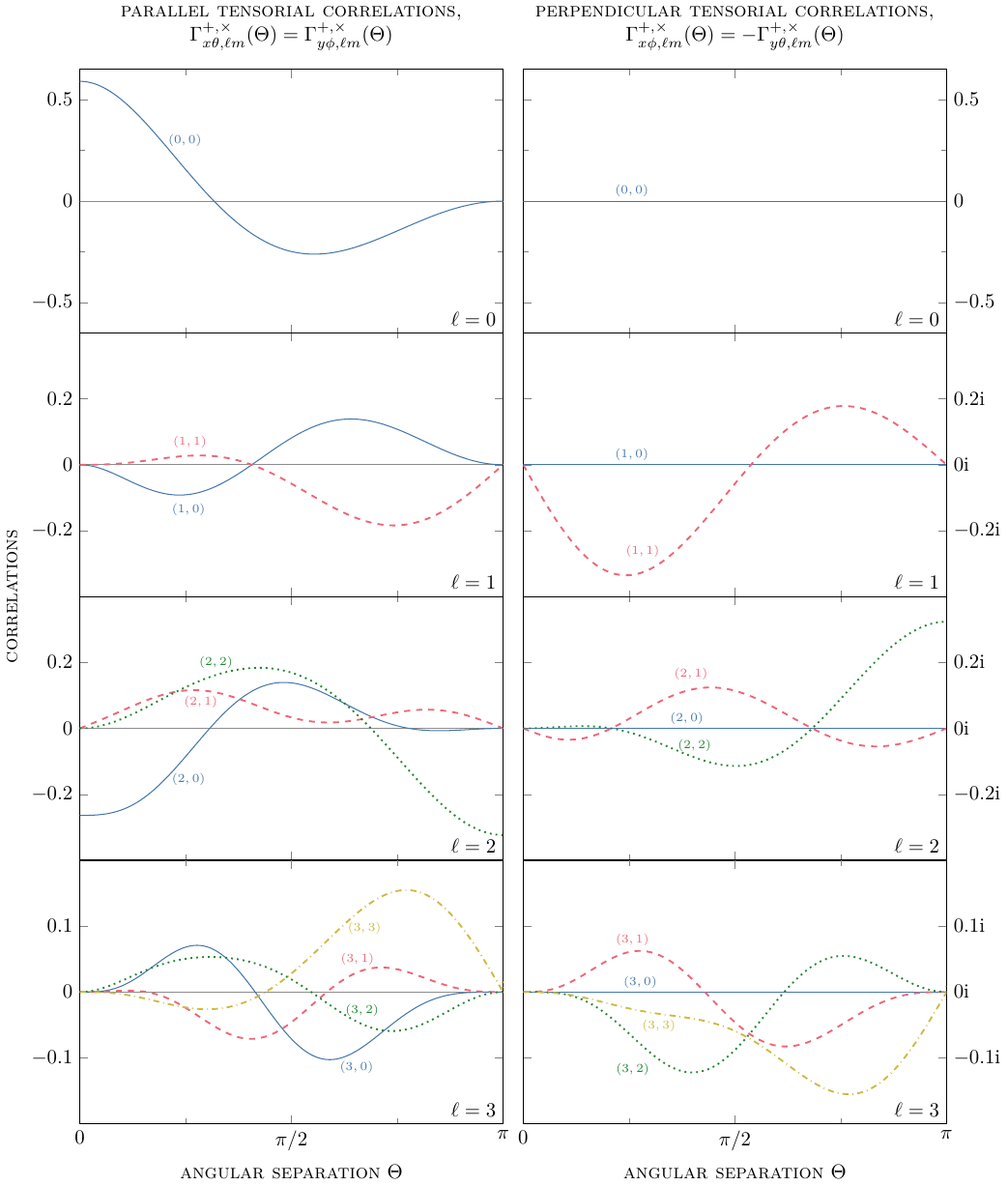}
\caption{Generalized overlap reduction functions due to an anisotropic \sgwb\ containing \gw s with the \gr\ polarizations, $+$ and $\times$, for all anisotropies $m\geq0$ up to $\ell=3$ and $m=3$. The correlations with negative $m$ are related to those with positive $m$ via Eqs. \eqref{eq:gr_pm}. The parallel astrometric correlations, $\GxthpcTh{\ell}{m}=\GyphpcTh{\ell}{m}$, are purely real. The perpendicular astrometric correlations, $\GxphpcTh{\ell}{m}=-\GythpcTh{\ell}{m}$, are purely imaginary and, except the case $m=0$, are in general non-zero, unlike the case of an isotropic \sgwb.}
\label{fig:gr_results}
\end{figure*}

We choose a coordinate system in which one star is at the North pole, while the other is in the x-z plane:
\vspace{0pt}
\begin{subequations}\label{coordinates_nm}
\noindent
\begin{minipage}[t]{0.4\linewidth}
    \setlength{\abovedisplayskip}{0pt}
    \setlength{\belowdisplayskip}{0pt}
    \begin{equation}
        \vn = \left(0, 0, 1\right) ,
        \label{subeq:n}
    \end{equation}
\end{minipage}%
\hfill
\begin{minipage}[t]{0.56\linewidth}
    \setlength{\abovedisplayskip}{0pt}
    \setlength{\belowdisplayskip}{0pt}
    \begin{equation}
        \vm = \left(\sin{\Theta}, 0, \cos{\Theta}\right) ,
        \label{subeq:m}
    \end{equation}
\end{minipage}
\end{subequations}
\par
\vspace{\belowdisplayskip}
\noindent
where $\Theta$ is the angle between the two stars. The coefficients $c_{\ell}^m$ must first be rotated into this frame, which can be done using Wigner D-matrices, see e.g. \cite{Gair:2014rwa}. At each of the two points on the sphere, the angular deflections, respectively $\delta n$ and $\delta m$, lie in the tangent spaces at the two points. These spaces are two-dimensional and are spanned by two basis vectors each. We define the following basis vectors using the geodesic connecting the two points (see \autoref{fig:unit_vectors} taken from \rcite{Mihaylov:2018uqm}):

\vspace{0pt}
\begin{subequations}\label{basis_vectors_def}
\noindent
\begin{minipage}[t]{0.4\linewidth}
    \setlength{\abovedisplayskip}{0pt}
    \setlength{\belowdisplayskip}{0pt}
    \begin{align}
        \vux &= \left(1,0,0\right), \label{subeq:ux} \\ 
        \vuy &= \left(0,1,0\right), \label{subeq:uy}
    \end{align}
\end{minipage}%
\hfill
\begin{minipage}[t]{0.56\linewidth}
    \setlength{\abovedisplayskip}{0pt}
    \setlength{\belowdisplayskip}{0pt}
    \begin{align}
        \vuth &= \left(\cos{\Theta},0,-\sin{\Theta}\right), \label{subeq:utheta} \\
        \vuph &=\left(0,1,0\right).\label{subeq:uphi}
    \end{align}
\end{minipage}
\end{subequations}
\par
\vspace{\belowdisplayskip}
\noindent

In this basis the astrometric deflections become
\begin{subequations}\label{eq:basisdefls}
    \begin{align}
        \delta n_i^P\left(\vq\right)&=\delta n_x^P\left(\vq\right)\hat{u}^x_i+\delta n_y^P\left(\vq\right)\hat{u}^y_i \, , \label{eq:basis1} \\
        \delta m_i^P\left(\vq\right)&=\delta m_{\theta}^P\left(\vq\right)\hat{u}^{\theta}_i+\delta m_{\phi}^P\left(\vq\right)\hat{u}^{\phi}_i \, , \label{eq:basis2}
    \end{align}
\end{subequations}
where the coefficients in the expansion are equal to

\vspace{0pt}
\begin{subequations}\label{eq:basis_vectors_dotting}
\noindent
\begin{minipage}[t]{0.4875\linewidth}
    \setlength{\abovedisplayskip}{0pt}
    \setlength{\belowdisplayskip}{0pt}
    \begin{align}
        \delta n_x^P \!\left(\vq\right)&=\delta n_i^P \!\left(\vq\right)\hat{u}^i_x, \label{subeq:coef_x} \\
        \delta n_y^P \!\left(\vq\right)&=\delta n_i^P \!\left(\vq\right)\hat{u}^i_y, \label{subeq:coef_y}
    \end{align}
\end{minipage}%
\hfill
\begin{minipage}[t]{0.5125\linewidth}
    \setlength{\abovedisplayskip}{0pt}
    \setlength{\belowdisplayskip}{0pt}
    \begin{align}
        \delta m_{\theta}^P \!\left(\vq\right) &= \delta m_i^P \!\left(\vq\right) \hat{u}^i_{\theta}, \label{subeq:coef_theta} \\
        \delta m_{\phi}^P \!\left(\vq\right) &= \delta m_i^P \!\left(\vq\right) \hat{u}^i_{\phi}. \label{subeq:coef_phi}
    \end{align}
\end{minipage}
\end{subequations}
\par
\vspace{\belowdisplayskip}
\noindent
We substitute Eqs.~\eqref{eq:basisdefls} into Eq.~\eqref{eq:gorf} and obtain
\begin{subequations}\label{eq:gorf_components}
    \begin{align}
           \Gij[P][\Theta][\ell][m] &= \hat{u}^x_i\hat{u}^{\theta}_j\int_{S^2}\mathrm{d}\Omega_{\vq}\,Y_{\ell}^m\left(\vq\right)\delta n_{x}^P\left(\vq\right)\delta m_{\theta}^P\left(\vq\right) \label{subeq:gorf_xtheta} \\
         &+\hat{u}^x_i\hat{u}^{\phi}_j\int_{S^2}\mathrm{d}\Omega_{\vq}\,Y_{\ell}^m\left(\vq\right)\delta n_x^P\left(\vq\right)\delta m_{\phi}^P\left(\vq\right) \label{subeq:gorf_xphi} \\
         &+\hat{u}^y_i\hat{u}^{\theta}_j\int_{S^2}\mathrm{d}\Omega_{\vq}\,Y_{\ell}^m\left(\vq\right)\delta n_y^P\left(\vq\right)\delta m_{\theta}^P\left(\vq\right) \label{subeq:gorf_ytheta} \\
         &+\hat{u}^y_i\hat{u}^{\phi}_j\int_{S^2}\mathrm{d}\Omega_{\vq}\,Y_{\ell}^m\left(\vq\right)\delta n_y^P\left(\vq\right)\delta m_{\phi}^P\left(\vq\right) \label{subeq:gorf_yphi}.
    \end{align}                 
\end{subequations}
Hence there are four components of the generalized overlap reduction functions, where the indices can take values $i\in\{x,y\}$ and $j\in\{\theta,\phi\}$. We will hence call the \eqref{subeq:gorf_xtheta} and \eqref{subeq:gorf_yphi} terms the parallel correlations, as $\vux$  and $\vuy$ are parallel-transported into $\vuth$ and $\vuph$, respectively, and we will call the \eqref{subeq:gorf_xphi} and \eqref{subeq:gorf_ytheta} terms the perpendicular correlations, since $\vux\perp\vuph$ and $\vuy\perp\vuth$. 
In the isotropic case where $\ell=0$ and $m=0$, due to symmetry, the perpendicular terms are equal to zero \cite{Mihaylov:2018uqm}. 
Unlike the isotropic case, in a general non-isotropic \gw\ background the perpendicular correlations can be non-zero.

In the next section, we will calculate the components of the generalized overlap reduction functions, or terms \eqref{subeq:gorf_xtheta}--\eqref{subeq:gorf_yphi}, for all six possible \gw\ polarizations.
\vspace*{10pt}

\section{Correlation functions}
\label{sec:results}

\subsection{Tensorial transverse-traceless polarizations}
\noindent
In this subsection, we derive the generalized overlap reduction functions due to an \sgwb\ consisting of the two \gr\ polarizations, $P \in \left\{+, \times\right\}$. 
We assume equal power in each of the modes and define
\begin{align}\label{gr_correlations_def}
    \Gij[+,\times][\Theta][\ell][m] &= \Gij[+][\Theta][\ell][m] + \Gij[\times][\Theta][\ell][m].
\end{align}
In \cref{app:gr_calculations} we evaluate the generalized overlap reduction function components in Eq.~\eqref{eq:gorf_components} for the $+$ and $\times$ polarizations and combine them according to Eq.~\eqref{gr_correlations_def} to obtain $\Gorf{i}{j}[+,\times][\Theta][\ell m]$. For the parallel components it turns out that $\GxthpcTh{\ell}{m}=\GyphpcTh{\ell}{m}$, while for the perpendicular components we find that $\GxphpcTh{\ell}{m}=-\GythpcTh{\ell}{m}$. As shown in \cref{app:gr_calculations}, we need different expressions for the different values of the azimuthal index $m$. We define the functions: 
\begin{widetext}
    \begin{subequations}
        \begin{align}
            & F_{q,r,\ell,m}^+ \equiv F_{q,r,\ell,m}^+(\cos{\Theta}) = \int_{\cos{\Theta}}^1 \dx \, \frac{(1-x)^q}{(1+x)^r}\frac{\dd^m}{\dx^m}P_\ell(x), \\
            & F_{q,r,\ell,m}^- \equiv F_{q,r,\ell,m}^-(\cos{\Theta}) = \int_{-1}^{\cos{\Theta}} \dx \, \frac{(1+x)^q}{(1-x)^r}\frac{\dd^m}{\dx^m}P_\ell(x), 
        \end{align}
    \end{subequations}
    where $P_\ell(x)$ are the ordinary Legendre polynomials. Then, starting with the parallel correlations and $m=0$, the generalized overlap reduction functions are:
    \begin{subequations}\label{eq:gr_results}
        \begin{align}
        & \begin{aligned}
            & \GxthpcTh{\ell}{0} = \GyphpcTh{\ell}{0} = \frac{1}{2}\sqrt{(2\ell+1)\pi}\Bigg(\frac{1}{3}\left(\cos{\Theta}-2\right)\delta_{\ell0} + \frac{1}{3}\cos{\Theta}\,\delta_{\ell1}-\frac{1}{15}\left(1+\cos{\Theta}\right)\delta_{\ell2}-\frac{1}{2}\cosnhalf[2]{\Theta}F_{1,0,\ell,0}^{+} \\
            & \qquad + \frac{1}{2}\sinnhalf[2]{\Theta}F_{0,-1,\ell,0}^+ + \sinnhalf[4]{\Theta}\secnhalf[2]{\Theta}F_{2,1,\ell,0}^-
            - \frac{3}{2}\sinnhalf[2]{\Theta}F_{1,0,\ell,0}^- + \frac{1}{2}\cosnhalf[2]{\Theta}F_{0,-1,\ell,0}^-\Bigg) , 
            \end{aligned} \\
            \intertext{where $\delta_{\ell m}$ is the Kronecker delta function. Similarly for $m=1$:}
            & \begin{aligned}
            & \GxthpcTh{\ell}{1} = \GyphpcTh{\ell}{1}  = \frac{1}{2}\sqrt{(2\ell+1)\pi\frac{(\ell-1)!}{(\ell+1)!}}\Bigg(\frac{1}{5}\sin{\Theta}\,\delta_{\ell2} + \frac{1}{2}\cosnhalf[3]{\Theta}\cscnhalf{\Theta}F_{2,0,\ell,1}^+ \\
            & \qquad - \frac{1}{2}\sinnhalf{\Theta}\cosnhalf{\Theta}F_{1,-1,\ell,1}^+  - \frac{1}{2}\sinnhalf[5]{\Theta}\secnhalf[3]{\Theta}F_{3,1,\ell,1}^- + \frac{1}{2}\sinnhalf[3]{\Theta}\secnhalf{\Theta}F_{2,0,\ell,1}^-\Bigg).
            \end{aligned} \\
            \intertext{For $m=2$:}
            & \GxthpcTh{\ell}{2} = \GyphpcTh{\ell}{2} = \frac{1}{2}\sqrt{(2\ell+1)\pi\frac{(\ell-2)!}{(\ell+2)!}}\Bigg(\frac{2}{5}\left(\cos{\Theta}-1\right)\delta_{\ell2} + \frac{1}{2}\left(1 + \left(-1\right)^{\ell}\right) - \frac{1}{2}\cosn{\Theta}\left(1 + \left(-1\right)^{\ell+1}\right) \nonumber\\  
            &\qquad - \frac{1}{2}\cosnhalf[4]{\Theta}\cscnhalf[2]{\Theta} F_{3,0,\ell,2}^+ + \frac{3}{4}\cosnhalf[2]{\Theta}F_{2,-1,\ell,2}^+ - \frac{1}{4}\sinnhalf[2]{\Theta}F_{1,-2,\ell,2}^+ \\ 
            &\qquad + \frac{1}{2}\sinnhalf[6]{\Theta}\secnhalf[4]{\Theta}F_{4,1,\ell,2}^- -\sinnhalf[4]{\Theta} \secnhalf[2]{\Theta}F_{3,0,\ell,2}^- + \frac{3}{4}\sinnhalf[2]{\Theta}F_{2,-1,\ell,2}^- -\frac{1}{4}\cosnhalf[2]{\Theta}F_{1,-2,\ell,2}^-\Bigg).
            \nonumber\\
            \intertext{And finally, for $m=3,4,\cdots$:}
            & \begin{aligned}
            & \GxthpcTh{\ell}{m} = \GyphpcTh{\ell}{m}  = (-1)^m\frac{1}{2}\sqrt{(2\ell+1)\pi\frac{(\ell-m)!}{(\ell+m)!}}\Bigg(\frac{1}{4}\cosnhalf[m]{\Theta}\cscnhalf[m]{\Theta}\Big[-2\cosnhalf[2]{\Theta}F_{m+1,0,\ell,m}^+ \\ 
            &\qquad + 4\sinnhalf[2]{\Theta}F_{m,-1,\ell,m}^+ - 2\sinnhalf[4]{\Theta}\secnhalf[2]{\Theta}F_{m-1,-2,\ell,m}^+\Big] + \frac{1}{4}\sinnhalf[m]{\Theta}\secnhalf[m]{\Theta} \\ 
            &\qquad \Big[2\sinnhalf[4]{\Theta}\secnhalf[2]{\Theta}F_{m+2,1,\ell,m}^- - 4\sinnhalf[2]{\Theta}F_{m+1,0,\ell,m}^- + 2\cosnhalf[2]{\Theta}F_{m,-1,\ell,m}^-\Big]\Bigg). 
            \end{aligned} \\
            \intertext{We now present the perpendicular correlations. For $m=0$ the correlations vanish:}
            & \GxphpcTh{\ell}{0} = \GythpcTh{\ell}{0} = 0. \\
            \intertext{For $m=1$ we obtain:}
            & \GxphpcTh{\ell}{1} = -\GythpcTh{\ell}{1} = \frac{\I}{2}\sqrt{(2\ell+1)\pi\frac{(\ell-1)!}{(\ell+1)!}}\Bigg(-\frac{1}{5}\sin{\Theta} \, \delta_{\ell2} - \frac{1}{2}\cosnhalf[3]{\Theta}\cscnhalf{\Theta}F_{2,0,\ell,1}^+  \nonumber\\
            &\qquad + \frac{1}{2}\sinnhalf{\Theta}\cosnhalf{\Theta}F_{1,-1,\ell,1}^+ - \frac{1}{2}\sinnhalf[5]{\Theta}\secnhalf[3]{\Theta}F_{3,1,\ell,1}^- + \frac{3}{2}\sinnhalf[3]{\Theta}\secnhalf{\Theta}F_{2,0,\ell,1}^- \\
            &\qquad - \sinnhalf{\Theta} \cosnhalf{\Theta}F_{1,-1,\ell,1}^-\Bigg). 
            \nonumber\\
            \intertext{Similarly for $m=2$:}
            & \GxphpcTh{\ell}{2} = -\GythpcTh{\ell}{2} =  \frac{\I}{2}\sqrt{(2\ell+1)\pi\frac{(\ell-2)!}{(\ell+2)!}}\Bigg(\frac{2}{5}\left(1 - \cos{\Theta}\right)\delta_{\ell2} - \frac{1}{2}\left(1 + \left(-1\right)^{\ell}\right) + \frac{1}{2}\cosn{\Theta}\left(1 + \left(-1\right)^{\ell+1}\right) \nonumber\\ 
            &\qquad + \frac{1}{2}\cosnhalf[4]{\Theta}\cscnhalf[2]{\Theta}F_{3,0,\ell,2}^+ - \frac{3}{4}\cosnhalf[2]{\Theta}F_{2,-1,\ell,2}^+ + \frac{1}{4}\sinnhalf[2]{\Theta}F_{1,-2,\ell,2}^+ \\ 
            &\qquad + \frac{1}{2}\sinnhalf[6]{\Theta}\secnhalf[4]{\Theta}F_{4,1,\ell,2}^- - \sinnhalf[4]{\Theta}\secnhalf[2]{\Theta}F_{3,0,\ell,2}^- + \frac{1}{4}\sinnhalf[2]{\Theta}F_{2,-1,\ell,2}^-+\frac{1}{4}\cosnhalf[2]{\Theta}F_{1,-2,\ell,2}^-\Bigg). 
            \nonumber\\
            \intertext{And finally, for $m=3,4,\cdots$:}
            & \begin{aligned}
            & \GxphpcTh{\ell}{m} = -\GythpcTh{\ell}{m} = (-1)^m\frac{\I}{2}\sqrt{(2\ell+1)\pi\frac{(\ell-m)!}{(\ell+m)!}}\Bigg(\frac{1}{4}\cosnhalf[m]{\Theta}\cscnhalf[m]{\Theta}\Big[2\cosnhalf[2]{\Theta}F_{m+1,0,\ell,m}^+ \\ 
            &\qquad - 4\sinnhalf[2]{\Theta}F_{m,-1,\ell,m}^+ + 2\sinnhalf[4]{\Theta}\secnhalf[2]{\Theta}F_{m-1,-2,\ell,m}^+\Big] + \frac{1}{4}\sinnhalf[m]{\Theta}\secnhalf[m]{\Theta} \\
            &\quad\Big[2\sinnhalf[4]{\Theta}\secnhalf[2]{\Theta}F_{m+2,1,\ell,m}^- - 4\sinnhalf[2]{\Theta}F_{m+1,0,\ell,m}^- + 2\cosnhalf[2]{\Theta}F_{m,-1,\ell,m}^-\Big]\Bigg).
            \end{aligned}
        \end{align}
    \end{subequations}
\end{widetext}
The parallel correlations, $\GxthpcTh{\ell}{m}=\GyphpcTh{\ell}{m}$, are purely real, while the perpendicular correlations, $\GxphpcTh{\ell}{m}=-\GythpcTh{\ell}{m}$, are purely imaginary. For negative values of $m$ the generalized overlap reduction functions are related to those with positive $m$ via
\begin{subequations}\label{eq:gr_pm}
    \begin{align}
        \GxthpcTh{\ell}{-m} &= \left(-1\right)^m\GxthpcTh{\ell}{m},\\
        \GyphpcTh{\ell}{-m} &= \left(-1\right)^m\GyphpcTh{\ell}{m},\\
        \GxphpcTh{\ell}{-m} &= \left(-1\right)^{m+1}\GxphpcTh{\ell}{m},\\
        \GythpcTh{\ell}{-m} &= \left(-1\right)^{m+1}\GythpcTh{\ell}{m}.
    \end{align}
\end{subequations}
In calculating the above correlation functions we have used the methods in Appendix E of \rcite{Gair:2014rwa}, where the authors derive the generalized overlap reduction functions in the context of \pta s in an anisotropic \sgwb. 
In particular, $F_{q,r,\ell,m}^{\pm}(\cos{\Theta})=\mathcal{F}_{q,r,\ell,m}^{\pm}(-\cos{\Theta})$. 
In \cref{app:Fpm_functions} we have provided closed-form expressions for $F_{q,r,\ell,m}^{\pm}(\cos{\Theta})$, confirming the results of Appendix~E.1 of \rcite{Gair:2014rwa}.

In \autoref{fig:gr_results} we have plotted the generalized overlap reduction functions for all anisotropies of the \sgwb\ for which $m\geq0$ up to $\ell=3$ and $m=3$ (cf. Fig. 2 in \rcite{Mingarelli:2013dsa} that shows the correlations for \pta s in an \sgwb\ with \gr\ polarizations). 
The result for $\ell=0$ and $m=0$ is the same up to a multiplicative constant as the overlap reduction function in an isotropic background in Eq.~(45) of \rcite{Mihaylov:2018uqm}. 
The perpendicular correlations, $\GxphpcTh{\ell}{m} = -\GythpcTh{\ell}{m}$, except the case $m=0$, are in general non-zero, in contrast to the case of an isotropic \sgwb.

\subsection{Scalar ``breathing" polarization}
\begin{figure*}[t]
\centering
\includegraphics{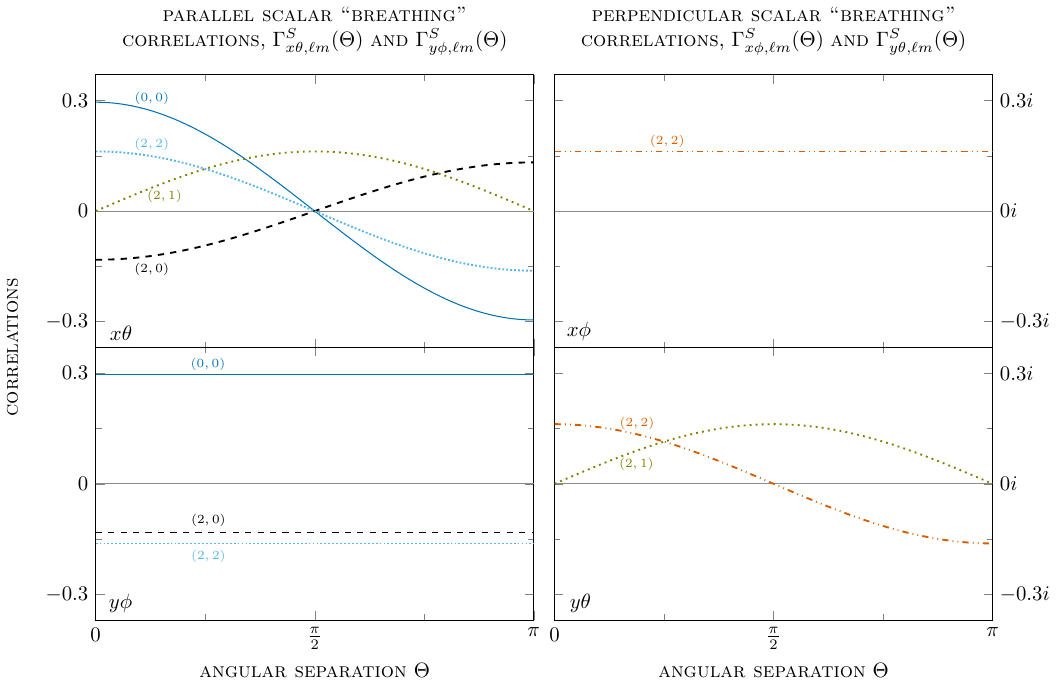}
\caption{All non-zero generalized overlap reduction functions with $m\geq0$ for an \sgwb\ containing transverse-scalar \gw s. The correlations with negative $m$ are related to those with positive $m$ via Eqs. \eqref{eq:s_pm}. The parallel correlations, $\GxthsTh{\ell}{m}$ and $\GyphsTh{\ell}{m}$, are purely real, while the perpendicular correlations, $\GxphsTh{\ell}{m}$ and $\GythsTh{\ell}{m}$, are purely imaginary. $\GyphsTh{\ell}{m}$ and $\GxphsTh{\ell}{m}$ do not depend on the angular separation $\Theta$.}
\label{fig:s_results}
\end{figure*}

\noindent
The generalized overlap reduction functions in an \sgwb\ with transverse-scalar \gw s were calculated to be (details of the calculations are shown in \cref{app:s_calculations})
\begin{widetext}
    \begin{subequations}\label{s_results}
        \begin{align}
            \GxthsTh{\ell}{m} &= \frac{1}{3}\pi N_{\ell}^0\cosn{\Theta}\,\delta_{m0}\left(\delta_{\ell0}-\frac{1}{5}\delta_{\ell2}\right) + \pi N_{\ell}^{|m|}\left(-1\right)^{m\mathbf{1}_{m<0}}\delta_{\ell2}\left(\frac{1}{5}\sinn{\Theta}\,\delta_{|m|,1}+\frac{2}{5}\cosn{\Theta}\,\delta_{|m|,2}\right), \\
            \GyphsTh{\ell}{m} &= \frac{1}{3}\pi N_{\ell}^0\,\delta_{m0}\left(\delta_{\ell0}-\frac{1}{5}\delta_{\ell2}\right) - \frac{2}{5}\pi N_{\ell}^{|m|}\left(-1\right)^{m\mathbf{1}_{m<0}}\,\delta_{|m|,2}\,\delta_{\ell2}, \\
            \GxphsTh{\ell}{m} &= \frac{2}{5} \, \I \pi N_{\ell}^{|m|}\left(-1\right)^{\left(m+1\right)\mathbf{1}_{m<0}} \, \delta_{|m|,2}\,\delta_{\ell2}, \\
            \GythsTh{\ell}{m} &= \I \pi N_{\ell}^{|m|}\left(-1\right)^{\left(m+1\right)\mathbf{1}_{m<0}}\delta_{\ell2}\left(\frac{1}{5}\sinn{\Theta}\,\delta_{|m|,1}+\frac{2}{5}\cosn{\Theta}\,\delta_{|m|,2}\right),
        \end{align}
    \end{subequations}
\end{widetext}
where $\mathbf{1}_{m<0}$ is the indicator function that equals $1$ when $m<0$ and is otherwise zero. Similarly to the correlations in \pta\ analysis (Eq.~(48) in \rcite{Gair:2015hra}), the astrometric correlations are non-zero only for a finite number of different values of $\ell$ and $m$. This means that using astrometry, we are sensitive to only a limited number of anisotropies of an \sgwb\ containing transverse-scalar \gw s. 
All non-zero astrometric correlations with $m\geq0$ are shown in \autoref{fig:s_results}. Like the $+$ and $\times$ polarizations, the correlations with negative $m$ are related to those with positive $m$ via:
\begin{subequations}\label{eq:s_pm}
    \begin{align}
        \GxthsTh{\ell}{-m} &= \left(-1\right)^m\GxthsTh{\ell}{m},\\
        \GyphsTh{\ell}{-m} &= \left(-1\right)^m\GyphsTh{\ell}{m},\\
        \GxphsTh{\ell}{-m} &= \left(-1\right)^{m+1}\GxphsTh{\ell}{m},\\
        \GythsTh{\ell}{-m} &= \left(-1\right)^{m+1}\GythsTh{\ell}{m}.
    \end{align}
\end{subequations}
Similarly as in the case of the \gr\ polarizations, the parallel correlations, $\GxthsTh{\ell}{m}$ and $\GyphsTh{\ell}{m}$, are purely real, while the perpendicular correlations, $\GxphsTh{\ell}{m}$ and $\GythsTh{\ell}{m}$, are purely imaginary. $\GyphsTh{\ell}{m}$ and $\GxphsTh{\ell}{m}$ turn out to be independent of the angular separation $\Theta$. 

\subsection{Vectorial polarizations}
\begin{figure*}[p]
\centering
\includegraphics{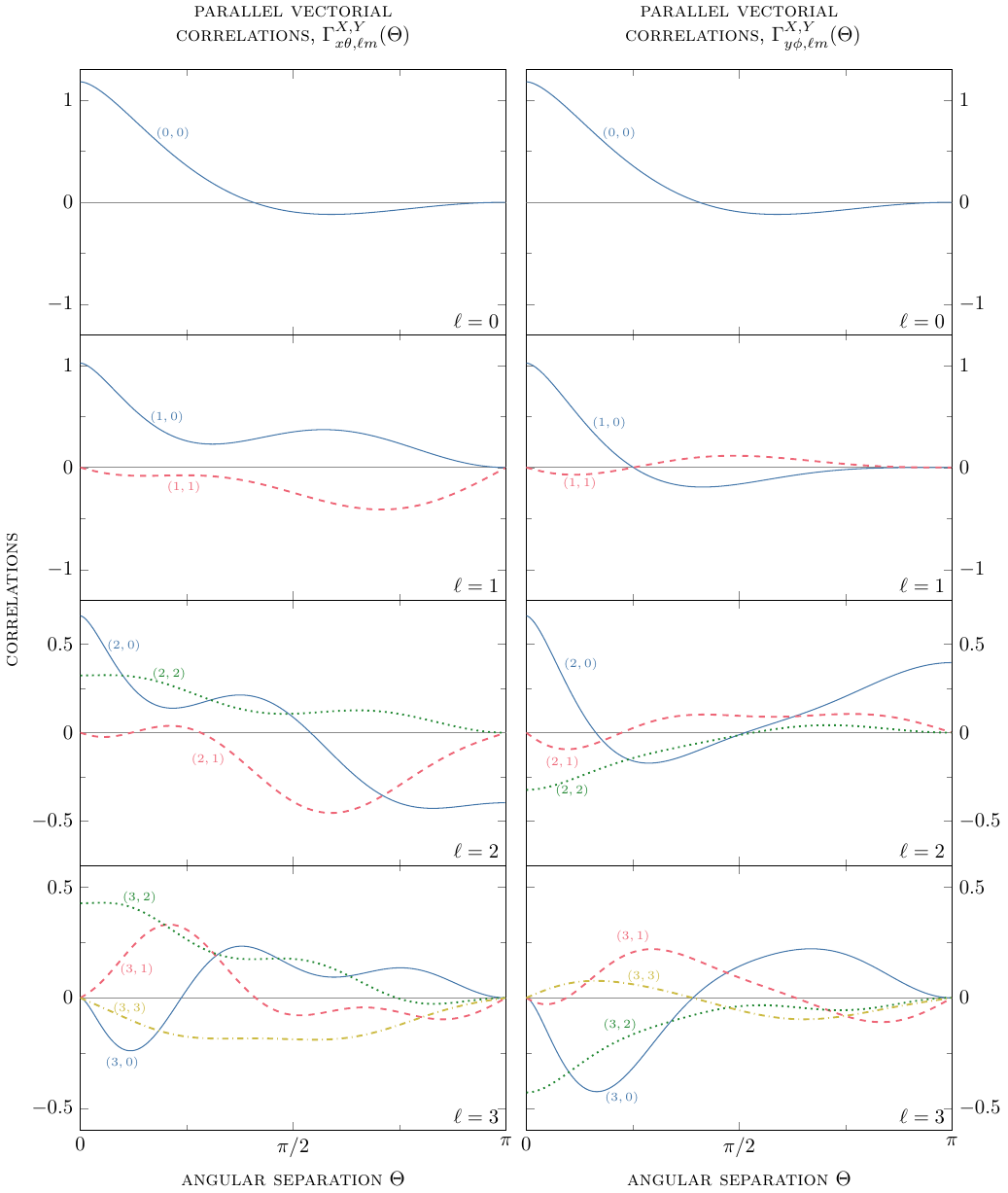}
\caption{The parallel generalized overlap reduction functions for the combined vectorial polarizations, $X$ and $Y$, in an anisotropic \sgwb\ for all anisotropies $m \geq 0$ up to $\ell=3$ and $m=3$. The correlations with negative $m$ are related to those with positive $m$ via Eqs. \eqref{eq:v_xtheta_pm} and \eqref{eq:v_yphi_pm}. The correlations are purely real. Excluding the isotropic case $\ell=0$ and $m=0$ shown in the top two panels, in general $\GxthxyTh{\ell}{m}\neq\GyphxyTh{\ell}{m}$.}
\label{fig:XY_parallel_results}
\end{figure*}

\begin{figure*}[p]
\centering
\includegraphics{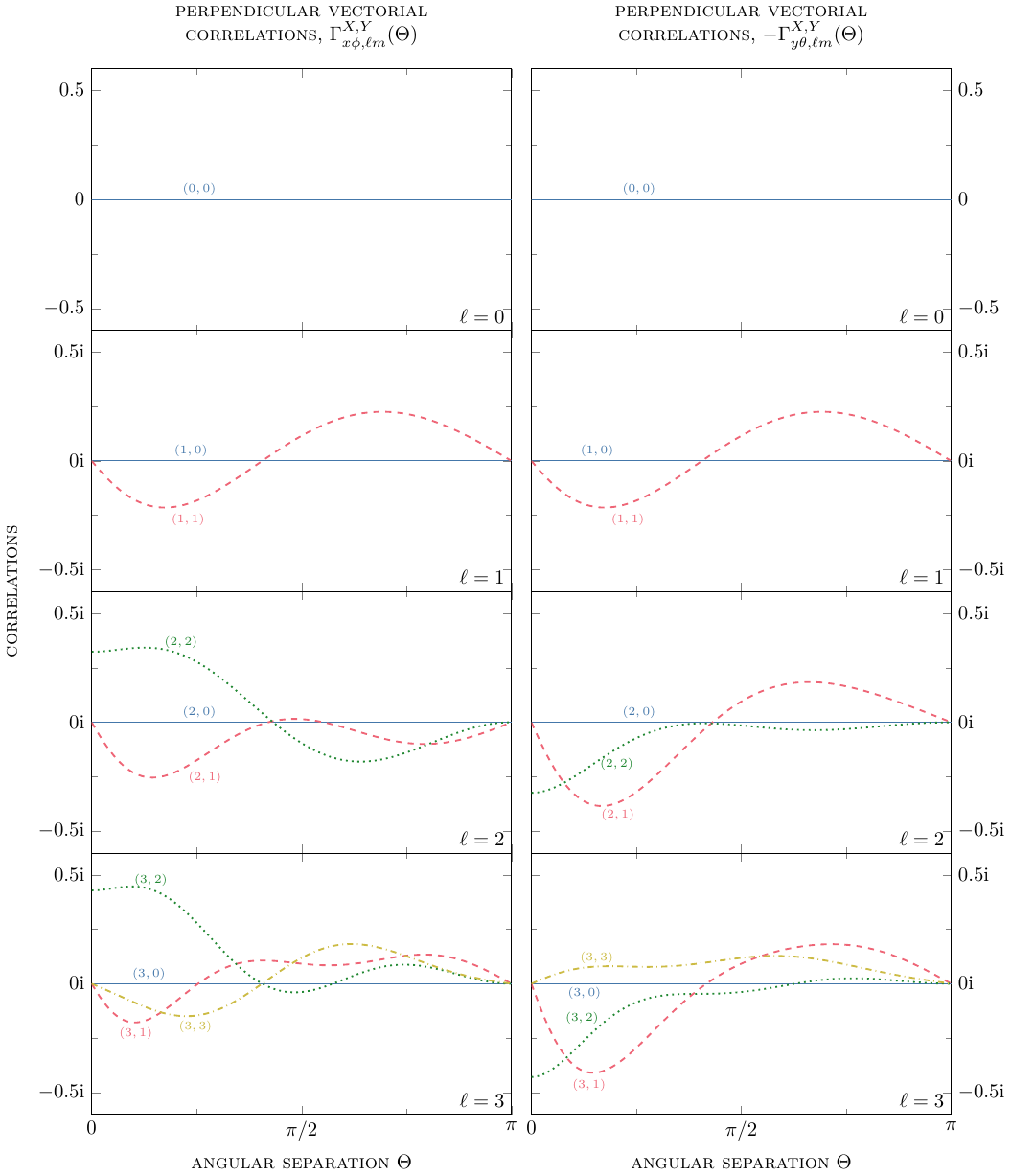}
\caption{The perpendicular generalized overlap reduction functions for the combined vectorial polarizations, $X$ and $Y$, in an anisotropic \sgwb\ for all anisotropies $m\geq0$ up to $\ell=3$ and $m=3$. 
The correlations with negative $m$ are related to those with positive $m$ via Eqs. \eqref{eq:v_xphi_pm} and \eqref{eq:v_ytheta_pm}. 
The correlations are purely imaginary and, except the case $m=0$, are non-zero. 
In general, $\GxphxyTh{\ell}{m}\neq-\GythxyTh{\ell}{m}$, with some exceptions like the case $\ell=1$ and $m=1$.}
\label{fig:XY_perpendicular_results}
\end{figure*}

\noindent
We consider the vectorial polarizations with mixed transverse and longitudinal components, $P\in\left\{X,Y\right\}$. We assume equal power in each polarization and define the correlation functions
\begin{align}\label{xy_correlations_def}
    \Gij[X,Y][\Theta][\ell][m] &= \Gij[X][\Theta][\ell][m] + \Gij[Y][\Theta][\ell][m].
\end{align}
In \cref{app:xy_calculations} we evaluate the components of the generalized overlap reduction functions defined in Eq.~\eqref{eq:gorf_components} for the $X$ and $Y$ polarizations and combine them according to Eq.~\eqref{xy_correlations_def} to obtain $\Gij[X,Y][\Theta][\ell][m]$. The analytical results for $\Gij[X,Y][\Theta][\ell][m]$ are presented in \cref{app:xy_calculations}. The parallel and perpendicular astrometric correlation functions for all anisotropies $m\geq0$ up to $\ell=3$ and $m=3$ are shown in \autoref{fig:XY_parallel_results} and \autoref{fig:XY_perpendicular_results}, respectively. Similarly as the \gr\ and transverse-scalar polarizations, the correlations with negative $m$ are related to those with positive $m$ by
\begin{subequations}
    \begin{align}
        \GxthxyTh{\ell}{-m} &= \left(-1\right)^m\GxthxyTh{\ell}{m},\label{eq:v_xtheta_pm}\\
        \GyphxyTh{\ell}{-m} &= \left(-1\right)^m\GyphxyTh{\ell}{m},\label{eq:v_yphi_pm}\\
        \GxphxyTh{\ell}{-m} &= \left(-1\right)^{m+1}\GxphxyTh{\ell}{m},\label{eq:v_xphi_pm}\\
        \GythxyTh{\ell}{-m} &= \left(-1\right)^{m+1}\GythxyTh{\ell}{m}.\label{eq:v_ytheta_pm}
    \end{align}
\end{subequations}
The isotropic results for $\ell=0$ and $m=0$ shown in the upper two panels of \autoref{fig:XY_parallel_results} are the same up to a multiplicative constant as Eqs.~(50) in \rcite{Mihaylov:2018uqm}. 
Except for the isotropic case, in general $\GxthxyTh{\ell}{m}\neq\GyphxyTh{\ell}{m}$ and $\GxphxyTh{\ell}{m}\neq-\GythxyTh{\ell}{m}$ (with some exceptions, like the case of $\ell=1$ and $m=1$ for the perpendicular correlations as seen in \autoref{fig:XY_perpendicular_results}). 
Similarly to the \gr\ results, the parallel correlations are purely real. 
The perpendicular correlations are purely imaginary and, except the case $m=0$, are non-zero, in contrast to the case of an isotropic \gw\ background.

\subsection{Scalar ``longitudinal" polarization}
\begin{figure*}[p]
\centering
    \includegraphics[scale=1]{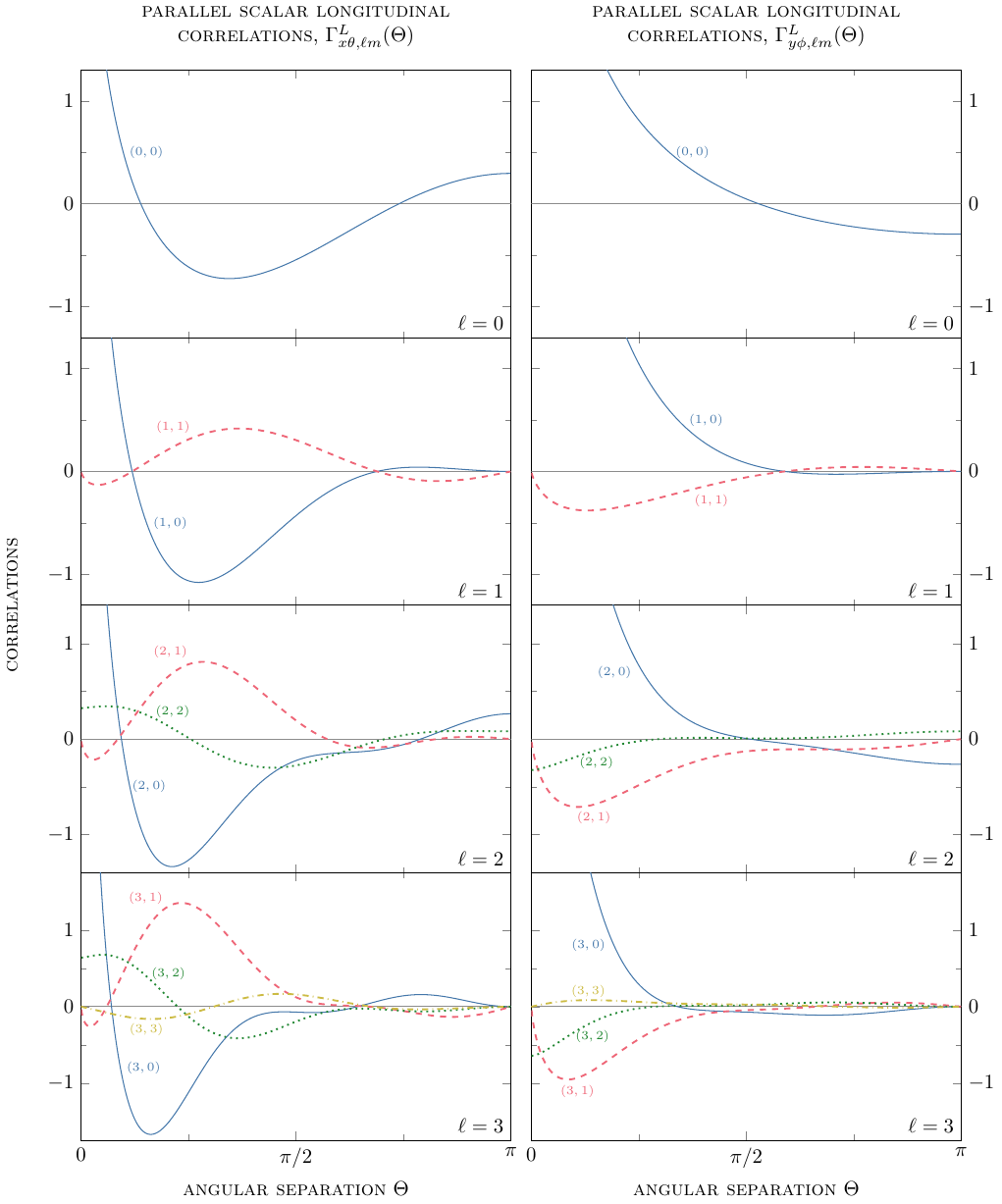}
    \caption{The parallel generalized overlap reduction functions for the scalar longitudinal polarization in an anisotropic \sgwb\ for all anisotropies $m\geq0$ up to $\ell=3$ and $m=3$. The correlations with negative $m$ are related to those with positive $m$ via Eqs. \eqref{eq:l_xtheta_pm} and \eqref{eq:l_yphi_pm}. For the case $m=0$, the overlap reduction functions diverge as $\Theta\rightarrow0$. The parallel correlations are purely real.}
    \label{fig:gLxtheta_results}
\end{figure*}

\begin{figure*}[p]
\centering
    \includegraphics[scale=1]{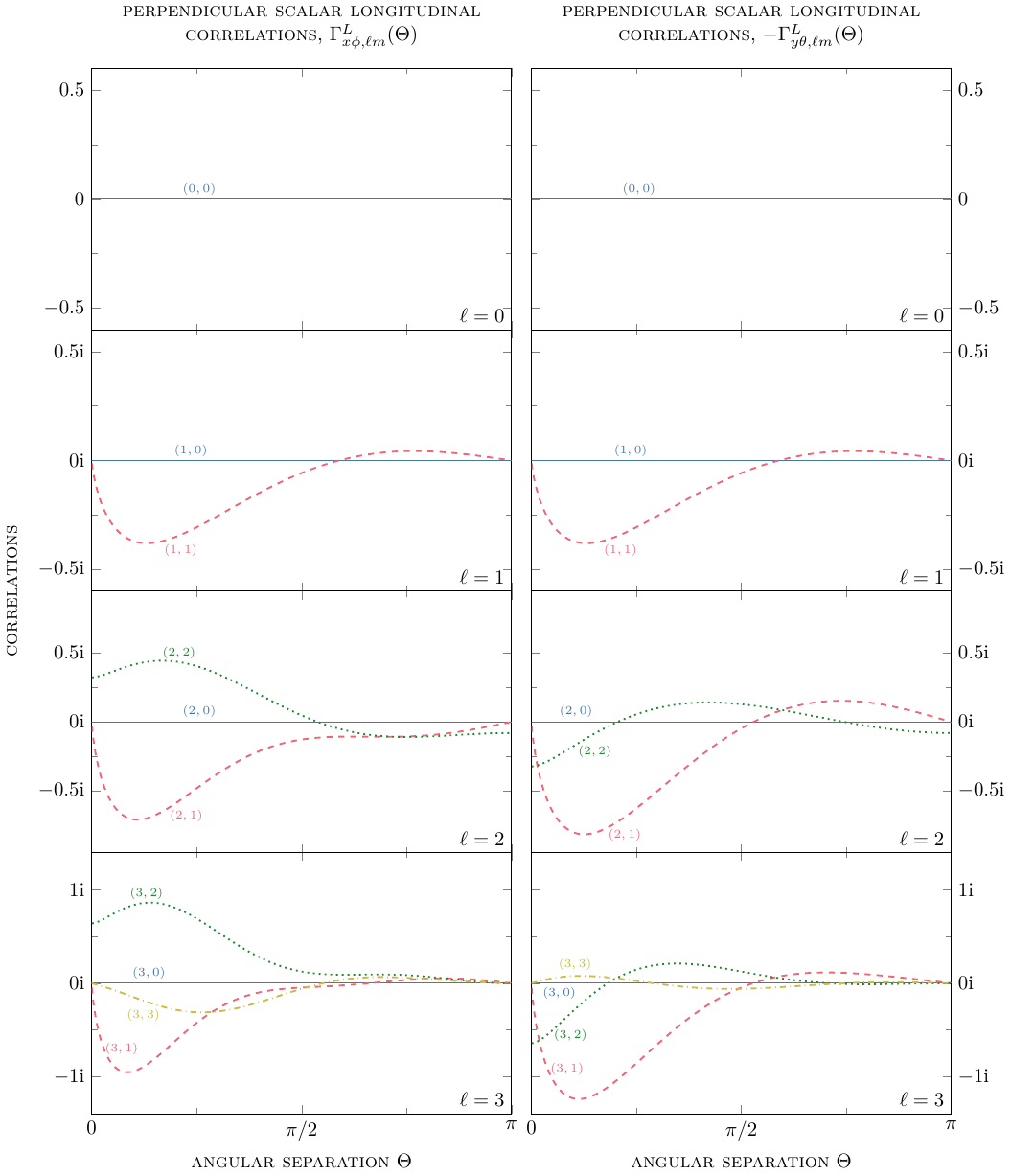}
    \caption{The perpendicular generalized overlap reduction functions for the scalar longitudinal polarization in an anisotropic \sgwb\ for all anisotropies $m\geq0$ up to $\ell=3$ and $m=3$. The correlations with negative $m$ are related to those with positive $m$ via Eqs. \eqref{eq:l_xphi_pm} and \eqref{eq:l_ytheta_pm}. All analytical results are finite for all values of $\Theta$. Except the case $m=0$, the correlation functions are in general non-zero. The perpendicular correlations are purely imaginary.}
    \label{fig:gLxphi_results}
\end{figure*}

\noindent
We finally consider the scalar ``longitudinal" polarization, $P=L$. In \cref{app:l_calculations}, we derive and present the results for the generalized overlap reduction functions for an \sgwb\ consisting of scalar longitudinal \gw s. 
The parallel and perpendicular astrometric correlation functions for all anisotropies $m\geq0$ up to $\ell = 3$ and $m=3$ are shown in \autoref{fig:gLxtheta_results} and \autoref{fig:gLxphi_results}, respectively. 
Analogous to the other polarizations discussed above, the astrometric correlations for negative \(m\) have the following symmetries
\begin{subequations}
    \begin{align}
        \GxthlTh{\ell}{-m} &= \left(-1\right)^m\GxthlTh{\ell}{m},\label{eq:l_xtheta_pm}\\
        \GyphlTh{\ell}{-m} &= \left(-1\right)^m\GyphlTh{\ell}{m},\label{eq:l_yphi_pm}\\
        \GxphlTh{\ell}{-m} &= \left(-1\right)^{m+1}\GxphlTh{\ell}{m},\label{eq:l_xphi_pm}\\
        \GythlTh{\ell}{-m} &= \left(-1\right)^{m+1}\GythlTh{\ell}{m}.\label{eq:l_ytheta_pm}
    \end{align}
\end{subequations}
The parallel correlations are purely real, while the perpendicular correlations are purely imaginary. 
Like the other polarizations, except the case $m=0$, the perpendicular correlations are generally non-zero.

Although the individual response functions $\delta n^L$ and $\delta m^L$ diverge as $\theta \rightarrow 0$ when $\Theta=0$, when integrated over the sphere and multiplied by the associated Legendre polynomials, $P_{\ell}^m$, the correlation integrals become finite. 
The only exceptions are the astrometric correlations $\GxthlTh{\ell}{0}$ and $\GyphlTh{\ell}{0}$, since the associated Legendre polynomials $P_{\ell}^0$ do not contain any $\sin{\theta}$ term to remove the divergence. 
Hence  $\GxthlTh{\ell}{0}$ and $\GyphlTh{\ell}{0}$ diverge as $\Theta\rightarrow0$ and we cannot apply the distant-light source approximation in Eq.~\eqref{eq:approximation}. 
In this case, it is useful to compare with numerical results. 
We need to use the exact response function given in Eq.~\eqref{eq:exact}, which leads to the modified expression for $\Delta_i^{jk}$:
\begin{widetext}
\begin{align}\label{eq:triangle_exact}
\begin{aligned}
    \Delta_i^{jk} \left(\vn, \vq, \dn\right) &= \Bigg(\Bigg\{1-\frac{\I \left(2 - q^rn_r\right)}{\dn\left(1 - q^wn_w\right)}\left(1 - \exp\left(\I \dn \left(1 - q^vn_v\right)\right)\right)\Bigg\} n_{i} \\
    & \qquad\qquad 
    -\Bigg\{1 - \frac{\I}{\dn \left(1 - q^wn_w\right)}\left(1 - \exp\left(\I \dn \left(1 - q^vn_v\right)\right)\right)\Bigg\} q_{i}\Bigg) \frac{n^{j} n^{k}}{2 \left(1 - q^sn_s\right)} \\
    & \qquad\qquad\qquad\qquad
    -\Bigg\{\frac{1}{2} - \frac{\I}{\dn \left(1 - q^wn_w\right)} \left(1 - \exp\left(\I \dn \left(1 - q^vn_v\right)\right)\right)\Bigg\} \delta_i^j n^k.
\end{aligned}
\end{align}
\end{widetext}
Now $\Delta_i^{jk}$ is also a function of the distance to the star in units of gravitational wavelengths times $2\pi$, or $\dn\equiv-\omega\lambda_S\Omega=2\pi |\bm{\mathrm{x}}_S| / \lambda_{\mathrm{GW}}$ (note that now the response implicitly depends on the frequency). 
$\Delta_i^{jk}$ is in general complex, which leads to the modified expressions for the two-point correlation functions
\begin{widetext}
    \begin{align}\label{eq:astrometric_correlation_exact}
        &\left\langle\delta n_i\left(\vn,t\right)\delta m_j\left(\vm,t'\right)\right\rangle=\frac{1}{4}\int_0^{\infty}\dd f\,C\left(f\right)\Big(e^{-2\pi \I f\left(t-t'\right)}\mathcal{L}_{ij}\left(\vn,\vm,\dn,\dm\right) + e^{2\pi \I f\left(t-t'\right)}\mathcal{L}^{*}_{ij}\left(\vn,\vm,\dn,\dm\right)\Big),
    \end{align}
\end{widetext}
where we define
\begin{equation}
    \mathcal{L}_{ij}\left(\vn,\vm,\dn,\dm\right)=\sum_P\mathcal{L}^P_{ij}\left(\vn,\vm,\dn,\dm\right) ,
\end{equation}
where for each polarization state we define
\begin{subequations}
\begin{align}
&\mathcal{L}^P_{ij}\left(\vn,\vm,\dn,\dm\right)=\nonumber\\
&=\int_{S^2}\dd\Omega_{\vq}\,P\left(\vq\right)\delta n_i^P\left(\vn,\vq,\dn\right)\delta m_j^{P*}\left(\vm,\vq,\dm\right), \\
&\mathrm{where} \;\:\delta n_i^P\left(\vn,\vq, \dn\right)=\Delta_i^{jk}\left(\vn,\vq,\dn\right)\epsilon^P_{jk}\left(\vq\right).\label{subeq:exact_response_definition}
\end{align}
\end{subequations}
Again expanding the power spectrum $P\left(\vq\right)$ in terms of spherical harmonics, we obtain
\begin{equation}
    \mathcal{L}^P_{ij}\left(\vn,\vm,\dn,\dm\right)=\sum_{\ell m}c_{\ell}^m \mathcal{L}^P_{ij, \ell m}\left(\vn,\vm,\dn,\dm\right), 
\end{equation}
where we define the modified expressions for the generalized overlap reduction functions to be:
\begin{align} \label{eq:gorf_complex}
   & \mathcal{L}^P_{ij, \ell m}\left(\vn, \vm, \dn, \dm\right) \nonumber\\
   & \, = \int_{S^2}\dd\Omega_{\vq}\,Y_{\ell}^m\left(\vq\right)\delta n_i^P\left(\vn,\vq,\dn\right)\delta m_j^{P*}\left(\vm,\vq,\dm\right).
\end{align}
Note that the authors of \rcite{Mihaylov:2018uqm} omit to take the complex conjugate of $\delta m_j^{P}\left(\vm,\vq,\dm\right)$ when defining the overlap reduction function for the scalar longitudinal mode in their Eq.~(F4). Their definition is not directly related to the astrometric correlation, while our definition is more naturally related to it via Eq.~\eqref{eq:astrometric_correlation_exact}.

\begin{figure*}[p]
\centering
    \includegraphics[scale=1]{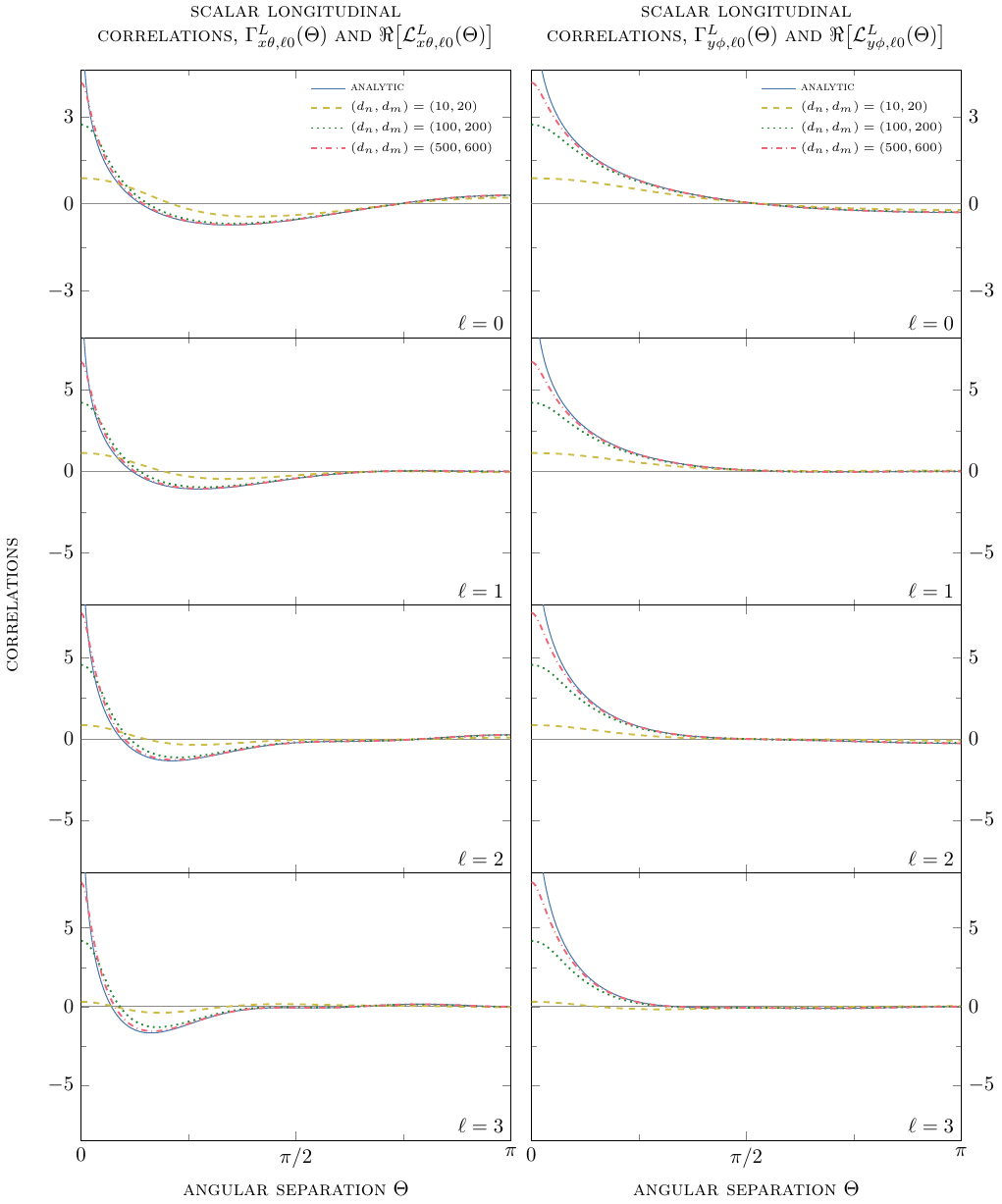}
    \caption{The generalized overlap reduction functions in the distant-light source approximation, $\GxthlTh{\ell}{0}$ and $\GyphlTh{\ell}{0}$, compared with numerical results of the real parts of the exact astrometric correlations, $\Re\big[\mathcal{L}_{x\theta,\ell 0}^L\big]$ and $\Re\big[\mathcal{L}_{y\phi,\ell 0}^L\big]$, for different distances to the pairs of stars, $(\dn,\dm)$ where we define $d\equiv2\pi|\bm{\mathrm{x}}_S| / \lambda_{\mathrm{GW}}$. As the distances to the stars from Earth increase, the analytic results approach the numerical results, with the agreement in general better for larger angular separations $\Theta$ (though note that the two results do not converge as $\Theta\rightarrow\pi$ for fixed distances).}
    \label{fig:gLxtheta_results_exact}
\end{figure*}

In \autoref{fig:gLxtheta_results_exact}, for $m=0$, we have plotted the divergent generalized overlap reduction functions obtained using the distant-light source approximation in Eq.~\eqref{eq:triangle}, $\GxthlTh{\ell}{0}$ and $\GyphlTh{\ell}{0}$, and the real parts of the numerical results for the astrometric correlations using the exact astrometric response in Eq.~\eqref{eq:triangle_exact}, $\Re\big[\mathcal{L}_{x\theta,\ell 0}^L\big]$ and $\Re\big[\mathcal{L}_{y\phi,\ell 0}^L\big]$. The second pair of functions corresponds to correlations measured at the same times, $t=t'$ (see Eq.~\eqref{eq:astrometric_correlation_exact}). We consider values of $(\dn,\dm)$ equal to $(10,20)$, $(100,200)$ and $(500,600)$. Although the analytic and numerical results diverge from each other as $\Theta\rightarrow0$, as the distances to the pairs of stars from Earth and the angle between them increase, the numerical results better approximate the closed-form expressions (though note that the two results do not converge as $\Theta\rightarrow\pi$ for fixed distances to the stars). 

We finally consider the exact solution for the special case of codirectional stars, or $\Theta=0$, for which the calculations are significantly simplified. The generalized overlap reduction functions in this case are shown in Eq.~\eqref{eq:gorf_y0} of \cref{app:scalar_long_y0} and plotted as a color map up to octopole anisotropy in \autoref{fig:l_y0_exact}.

\section{Discussion and conclusions}
\label{sec:discussion}
\noindent
In this paper, we have considered the astrometric effects of an anisotropic stochastic gravitational wave background. 
We have derived closed-form expressions for the generalized overlap reduction functions in the distant-light source limit for the six possible gravitational wave polarizations using the formalism of Refs. \citep{Gair:2014rwa}, \citep{Gair:2015hra} and \citep{Mihaylov:2018uqm}. Our closed-form expressions for the \gr, transverse-scalar, vectorial and scalar longitudinal polarizations are given in Eqs.~\eqref{eq:gr_results}, \eqref{s_results}, \eqref{eq:vectorial_results} and \eqref{eq:scalar_longitudinal_results}, respectively. 
We have provided graphs showing all correlation functions from monopole (\(\ell = 0\)) to octopole (\(\ell = 3\)) anisotropy in \autoref{fig:gr_results}, \autoref{fig:s_results}, \autoref{fig:XY_parallel_results}, \autoref{fig:XY_perpendicular_results}, \autoref{fig:gLxtheta_results} and \autoref{fig:gLxphi_results}. 
The results are expressed in terms of the functions \(F_{q,r,\ell,m}^{\pm}(\cos{\Theta})\) which themselves can be written as finite sums as shown in Eqs.~\eqref{eq:F_results}--\eqref{eq:F_recursive}. 
With the exception of the parallel scalar longitudinal correlations with \(m = 0\) ($\GxthlTh{\ell}{0}$ and $\GyphlTh{\ell}{0}$), all overlap reduction functions are computed in the distant-light source limit of Eq.~\eqref{eq:approximation}. 
The $\GxthlTh{\ell}{0}$ and $\GyphlTh{\ell}{0}$ correlations are divergent when computed by naively using the distant-light source limit, which can be rectified by using the exact formula for the astrometric displacement in Eq.~\eqref{eq:triangle_exact}, with the caveat that in that case the correlation functions cannot be computed analytically, except for select values of $\Theta$, e.g. $\Theta=0$. 
As the distances from the pairs of stars to Earth increase, the divergent analytic expressions are a good approximation to the real part of the exact result for all but the smallest angular separations, as shown in \autoref{fig:gLxtheta_results_exact}.

Future work will focus on deriving analogous closed-form expressions for the cross-correlations of redshift and astrometric displacements, following the formalism of \rcite{Mihaylov:2018uqm}. 
Redshift measurements carry independent information about the same background, and these cross-correlations are necessary to combine the two types of data consistently. 
Further work is also needed in order to connect these generalized overlap reduction functions to the data. 
Understanding these correlations in the context of angular scales, for example, in the formalism of \rcite{Mihaylov:2019lft}, would be a necessary first step. 
Furthermore, Gaia fits the positions, proper motions, and parallaxes of each source. 
This fit naturally absorbs part of the \gw\ signal, especially at periods which are comparable to, or longer than the observing baseline. 
Following \rcite{Klioner:2017asb, Geyer:2024lvk}, one needs to compute how much of the signal remains after this fit is performed, for each polarization and at each angular multipole. 
Finally, building a realistic noise model, compressing the data, and developing a forecasting pipeline are all necessary in order for us to be able to test how well astrometric data fits these correlation curves.

\section*{Acknowledgments}
\noindent
The authors thank Jonathan R. Gair, Christopher J. Moore, and Glenn D. Starkman for useful discussions and comments on the manuscript. 
P.D.D.\ and D.P.M.\ acknowledge support by the Bulgarian National Science Fund program ``VIHREN--2024'' project No. KP--06--DV/9/17.12.2024. 
D.P.M.\ acknowledges support from\ NASA\ ATP grant RES240737. 
We acknowledge the use of generative AI tools, namely the large language models (LLMs) ChatGPT 5.4 and 5.6 (OpenAI) and Claude Sonnet 5.5 (Anthropic). 
These were used to assist with the implementation of parts of the software used in this work and with proofreading and editing the final draft of the paper. 
All AI-assisted code was thoroughly reviewed, tested, and validated independently. 
AI-assisted text and references were also carefully checked, validated, and proofread.  
The authors take full responsibility for the correctness of the content and results of this paper.

\bibliographystyle{apsrev4-1}
\bibliography{main}

\onecolumngrid
\clearpage

\begin{appendices}
\numberwithin{equation}{section}
\renewcommand{\theequation}{\Alph{section}\arabic{equation}}
\renewcommand{\thesubsection}{\Roman{subsection}}
\section{Properties of spherical harmonic functions, associated Legendre polynomials and Legendre polynomials}
\label[appendix]{app:math}
\noindent
In this work, we define the spherical harmonic functions as
\begin{equation} \label{eq:spherical_harmonics_definition}
    Y_{\ell}^m\left(\vq\right)=N^m_{\ell}P^m_{\ell}\left(\cos{\theta}\right)e^{\I m\phi},
\end{equation}
where the normalization constants, $N^m_{\ell}$, are defined as 
\begin{equation} \label{eq:nlm}
    N^m_{\ell}=\sqrt{\frac{2\ell+1}{4\pi}\frac{\left(\ell-m\right)!}{\left(\ell+m\right)!}},
\end{equation} 
and $P^m_{\ell}\left(x\right)$ are the associated Legendre polynomials, which can be expressed in terms of the ordinary Legendre polynomials as
\begin{subequations}
\begin{align} \label{eq:associated_legendre}
    &P^m_{\ell}\left(x\right)=\left(-1\right)^m\left(1-x^2\right)^{m/2}\frac{\dd^m}{\dx^m}P_{\ell}\left(x\right) ,m\geq0, \\
    &P^{-m}_{\ell}\left(x\right)=\left(-1\right)^m\frac{\left(\ell-m\right)!}{\left(\ell+m\right)!}P^m_{\ell}\left(x\right).
\end{align}
\end{subequations}
From the above equations it can easily be seen that $P_{\ell}=P_{\ell}^0$. The associated Legendre polynomials have the orthogonality relations
\begin{equation}\label{eq:as_leg_norm}
    \int_{-1}^1\dx\, P^m_{\ell}\left(x\right)P^m_{\ell'}\left(x\right)=\frac{2}{2\ell+1}\frac{\left(\ell+m\right)!}{\left(\ell-m\right)!}\delta_{\ell\ell'}.
\end{equation}
The ordinary Legendre polynomials have the orthogonality relations (obtained from Eq.~\eqref{eq:as_leg_norm} by setting $m=0$)
\begin{equation}\label{eq:leg_norm}
    \int_{-1}^1\dx\, P_{\ell}\left(x\right)P_{\ell'}\left(x\right)=\frac{2}{2\ell+1}\delta_{\ell\ell'}.
\end{equation}
The Legendre polynomials satisfy the recursive relation:
\begin{equation}\label{eq:legendre_recursive}
    P_{\ell}\left(x\right)=\frac{\left(2\ell-1\right)}{\ell}P_{\ell-1}\left(x\right)-\frac{\left(\ell-1\right)}{\ell}P_{\ell-2}\left(x\right)-\frac{\left(2\ell-1\right)}{\ell}\left(1-x\right)P_{\ell-1}\left(x\right), \,\, \mathrm{for} \, \ell\geq2.
\end{equation}

\section{Calculation of astrometric integrals for GR polarizations}
\label[appendix]{app:gr_calculations}
\noindent
In this appendix, we show how to obtain Eqs.~\eqref{eq:gr_results}, the generalized overlap reduction functions for an \sgwb\ that contains \gw s with only \gr\ polarizations, $+$ and $\times$. We first focus on the $+$ polarization. As in the main text, we consider a coordinate system in which one star is at the North pole, while the other is in the xz plane, the position vectors described by Eqs.~\eqref{coordinates_nm}, so that the astrometric correlations are just a function of the angle between the two stars, $\Theta$. The integral we want to calculate is
\begin{align}
    \Gij[+][\Theta][\ell][m] &= \int_{S^2}\dd\Omega_{\vq}\,Y_{\ell}^m\left(\vq\right)\delta n_i^+\left(\vq\right)\delta m_j^+\left(\vq\right)=N^m_{\ell}\int_{0}^{\pi}\dd\theta\,\sin{\theta}\,P_{\ell}^m\left(\cos{\theta}\right)\int_0^{2\pi}\dd\phi\,e^{\I m\phi}\delta n_i^+\left(\vq\right)\delta m_j^+\left(\vq\right),
\end{align}
where $\vq=\left(\sin{\theta}\cos{\phi},\sin{\theta}\sin{\phi},\cos{\theta}\right)$ is the direction from which the \gw s originate, while $\delta n_i\left(\vq\right)$ and $\delta m_j\left(\vq\right)$ are the angular displacements defined in Eq.~\eqref{subeq:delta_ni}, where we henceforth drop the dependence on $\vn$ and $\vm$. 
$Y_{\ell}^m\left(\vq\right)=N^m_{\ell}P^m_{\ell}\left(\cos{\theta}\right)e^{\I m\phi}$ are the spherical harmonic functions, where $N^m_{\ell}$ are normalization constants and $P^m_{\ell}\left(\cos{\theta}\right)$ are the associated Legendre polynomials defined in \cref{app:math}. We will first focus on the case $m\geq0$. 
If we define $x\equiv\cos{\theta}$, we can rewrite the above integral as follows

\begin{align}\label{eq:integral_GR}
    \Gij[+][\Theta][\ell][m] &= N^m_{\ell}\int_{-1}^{1}\dx\,P_{\ell}^m\left(x\right)\int_0^{2\pi}\dd\phi\,e^{\I m\phi}\delta n_i^+\left(\vq\right)\delta m_j^+\left(\vq\right)=N^m_{\ell}\int_{-1}^{1}\dx\,P_{\ell}^m\left(x\right)\mathrm{I}^+_{ij,m}\left(\Theta,\theta\right),
\end{align}
where we define the azimuthal integral $\mathrm{I}^+_{ij,m}\left(\Theta,\theta\right)\equiv\int_0^{2\pi}\dd\phi\,e^{\I m\phi}\delta n_i^+\left(\vq\right)\delta m_j^+\left(\vq\right)$.

We calculate the angular displacements of the two stars, $\delta n_i^+\left(\vq\right)$ and $\delta m_i^+\left(\vq\right)$, using Eq.~\eqref{subeq:delta_ni}, plugging in the definition of $\Delta_i^{jk}\left(\vn,\vq\right)$ in Eq.~\eqref{eq:triangle} and the definition of the polarization tensor for $+$ polarized \gw s in Eq.~\eqref{subeq:plus_pol} defined in terms of the spherical coordinate basis vectors in Eqs.~\eqref{eq:spherical_basis}. We consider the basis vectors on the sphere defined in Eqs.~\eqref{basis_vectors_def}. Dotting the angular displacement vectors with the basis vectors on the sphere using Eqs.~\eqref{eq:basis_vectors_dotting}, we obtain the components of the astrometric deflections to be
\begin{subequations}
    \begin{align}
        & \delta n_x^+\left(\theta,\phi\right) = -\frac{1}{2}\sin{\theta}\cos{\phi},  
        \\[3ex]  
        & \delta n_y^+\left(\theta,\phi\right) = -\frac{1}{2}\sin{\theta}\sin{\phi}, 
        \\[3ex] 
        & \begin{aligned}
        & \delta m_{\theta}^+\left(\theta,\phi\right) = \frac{1}{1 - \cos{\Theta}\cos{\theta} - \sin{\Theta}\sin{\theta}\cos{\phi}}\Bigg(\frac{3}{16}\sin{\left(2\Theta\right)}\left(1 - \cos{\left(2\theta\right)}\right) + \frac{1}{4}\left(\cos{\left(2\Theta\right)}\sin{\left(2\theta\right)} - 2\cos{\Theta}\sin{\theta}\right)\cos{\phi} \\
        & \qquad + \frac{1}{16}\left(8\sin{\Theta}\cos{\theta} - 3\sin{\left(2\Theta\right)} -\sin{\left(2\Theta\right)}\cos{\left(2\theta\right)}\right)\cos{\left(2\phi\right)}\Bigg) , 
        \end{aligned}\raisetag{2.5\baselineskip} \\[3ex]
        & \begin{aligned}
        \delta m_{\phi}^+\left(\theta,\phi\right)&=\frac{1}{1-\cos\Theta\cos\theta-\sin\Theta\sin\theta\cos\phi}\Bigg(\frac{1}{2}\left(\cos\Theta\sin\theta\cos\theta-\cos\left(2\Theta\right)\sin\theta\right)\sin\phi\\
        & \qquad + \frac{1}{8}\left(2\sin\left(2\Theta\right)\cos\theta - 3\sin\Theta - \sin\Theta\cos\left(2\theta\right)\right)\sin\left(2\phi\right)\Bigg). 
        \end{aligned}
    \end{align}
\end{subequations}
Note that Eqs.~(B2b) and (B6b) in \rcite{Mihaylov:2018uqm} have typos. 
We can then express the products of the angular displacements as
\begin{subequations}
    \begin{align}
        \delta n_x^+\delta m_{\theta}^+&=\frac{1}{8}\cos{\Theta}\sin^2\theta\left(1+\cos(2\phi)\right)-\frac{1}{4}\sin{\Theta}\sin{\theta}\cos{\theta}\cos{\phi}+\frac{(\cos{\theta}-\cos{\Theta})\sin{\theta}\sin{\Theta}}{8(1-\cos{\Theta}\cos{\theta}-\sin{\Theta}\sin{\theta}\cos{\phi})}(\cos{\phi}-\cos{(3\phi)}), 
        \\[3ex]
        \delta n_y^+\delta m_{\phi}^+&=\frac{1}{8}(-1-\cos^2{\theta}+2\cos{\Theta}\cos{\theta})(1-\cos{(2\phi)})+\frac{(\cos{\theta}-\cos{\Theta})^2}{4(1-\cos{\Theta}\cos{\theta}-\sin{\Theta}\sin{\theta}\cos{\phi})}(1-\cos{(2\phi)}), 
        \\[3ex]
        \delta n_x^+\delta m_{\phi}^+&=\frac{1}{8}(-1-\cos^2{\theta}+2\cos{\Theta}\cos{\theta})\sin{(2\phi)}+\frac{(\cos{\theta}-\cos{\Theta})^2}{4(1-\cos{\Theta}\cos{\theta}-\sin{\Theta}\sin{\theta}\cos{\phi})}\sin{(2\phi)}, 
        \\[3ex]
        \delta n_y^+\delta m_{\theta}^+&=-\frac{1}{4}\sin{\Theta}\sin{\theta}\cos{\theta}\sin{\phi}+\frac{1}{8}\cos{\Theta}\sin^2{\theta}\sin(2\phi)+\frac{(\cos{\theta}-\cos{\Theta})\sin{\theta}\sin{\Theta}}{8(1-\cos{\Theta}\cos{\theta}-\sin{\Theta}\sin{\theta}\cos{\phi})}(3\sin{\phi}-\sin{(3\phi)}).
\end{align}
\end{subequations}
The azimuthal integrals become:
\begin{subequations}\label{azimuthal_int_GR}
    \begin{align}
        & \begin{aligned}
        &\mathrm{I}^+_{x\theta,m}\left(\Theta,\theta\right) = \int_0^{2\pi}\dd\phi\,e^{\I m\phi}\delta n_x^+\left(\vq\right)\delta m_{\theta}^+\left(\vq\right)=\frac{\pi}{8}\cos{\Theta}\sin^2\theta\left(2\delta_{m0} + \delta_{m2}\right) - \frac{\pi}{4}\sin{\Theta}\sin{\theta}\cos{\theta}\,\delta_{m1}\label{subeq:az_xtheta} \\
        & \qquad + \int_0^{2\pi}\dd\phi\,e^{\I m\phi}\frac{(\cos{\theta}-\cos{\Theta})\sin{\theta}\sin{\Theta}}{8(1-\cos{\Theta}\cos{\theta}-\sin{\Theta}\sin{\theta}\cos{\phi})}(\cos{\phi}-\cos{(3\phi)}), 
        \end{aligned}\\[3ex]
        & \begin{aligned}
        & \mathrm{I}^+_{y\phi,m}\left(\Theta,\theta\right) = \int_0^{2\pi}\dd\phi\,e^{\I m\phi}\delta n_y^+\left(\vq\right)\delta m_{\phi}^+\left(\vq\right)=\frac{\pi}{8}(-1 - \cos^2{\theta} + 2\cos{\Theta}\cos{\theta})(2\delta_{m0} - \delta_{m2}) \\
        &\qquad + \int_0^{2\pi}\dd\phi\,e^{\I m\phi}\frac{(\cos{\theta} - \cos{\Theta})^2}{4(1-\cos{\Theta}\cos{\theta} - \sin{\Theta}\sin{\theta}\cos{\phi})}(1 - \cos{(2\phi)}), 
        \end{aligned}\\[3ex]
        & \begin{aligned}
        & \mathrm{I}^+_{x\phi,m}\left(\Theta,\theta\right) = \int_0^{2\pi}\dd\phi\,e^{\I m\phi}\delta n_x^+\left(\vq\right)\delta m_{\phi}^+\left(\vq\right) = \frac{\I \pi}{8}(-1 - \cos^2{\theta} + 2\cos{\Theta}\cos{\theta})\delta_{m2} \\
        &\qquad + \int_0^{2\pi}\dd\phi\,e^{\I m\phi}\frac{(\cos{\theta} - \cos{\Theta})^2}{4(1 - \cos{\Theta}\cos{\theta} -\sin{\Theta}\sin{\theta}\cos{\phi})}\sin{(2\phi)}, 
        \end{aligned} \\[3ex]
        & \begin{aligned}
        & \mathrm{I}^+_{y\theta,m} \left(\Theta,\theta\right) = \int_0^{2\pi} \dd\phi\,e^{\I m\phi}\delta n_y^+\left(\vq\right)\delta m_{\theta}^+\left(\vq\right) = -\frac{\I \pi}{4}\sin{\Theta}\sin{\theta}\cos{\theta}\,\delta_{m1} + \frac{\I \pi}{8}\cos{\Theta}\sin^2{\theta}\,\delta_{m2} \\
        & \qquad + \int_0^{2\pi}\dd\phi\,e^{\I m\phi}\frac{(\cos{\theta} - \cos{\Theta})\sin{\theta}\sin{\Theta}}{8(1 - \cos{\Theta}\cos{\theta} - \sin{\Theta}\sin{\theta}\cos{\phi})}(3\sin{\phi} - \sin{(3\phi)}).
        \end{aligned}
\end{align}
\end{subequations}
The last term in each of the above equations can be integrated by writing $\cos{\left(n\phi\right)}$ and $\sin{\left(n\phi\right)}$ in terms of complex exponentials and using the result for the contour integral in Appendix B.1 of \rcite{Mihaylov:2018uqm} (there is a wrong sign in equation (B17a)):
\begin{equation} \label{eq:contour_integral}
    \int_0^{2\pi}\mathrm{d}\phi \, \frac{e^{\I n\phi}}{1-\cos{\Theta}\cos{\theta}-\sin{\Theta}\sin{\theta}\cos{\phi}}=\frac{2\pi}{|\cos{\theta}-\cos{\Theta}|}\left(\frac{1-\cos{\Theta}\cos{\theta}-|\cos{\theta}-\cos{\Theta}|}{\sin{\Theta}\sin{\theta}}\right)^{|n|}.
\end{equation}
Because of the $|n|$ in the exponent in Eq.~\eqref{eq:contour_integral}, we have different expressions for the correlation functions for different values of $m$.

The azimuthal integrals in Eqs.~\eqref{azimuthal_int_GR} are then multiplied by the associated Legendre polynomials $P_{\ell}^m\left(x\right)$ and integrated with respect to $x$ between \(-1\) and 1, as described by Eq.~\eqref{eq:integral_GR}. 
The $\delta_{m i}$ terms on the right hand sides of Eqs.~\eqref{azimuthal_int_GR} in most cases could be rewritten in terms of linear combinations of associated Legendre polynomials and then integrated using the orthogonality relations Eq.~\eqref{eq:as_leg_norm}. If not possible, Eq.~\eqref{eq:associated_legendre} was used to express the $P_{\ell}^m$ term in the integrand in terms of derivatives of Legendre polynomials, and then integration by parts was used to integrate the derivative and obtain ordinary Legendre polynomials for which we could then use the orthogonality relations in Eq.~\eqref{eq:leg_norm}. The results of the contour integration for the last terms of Eqs.~\eqref{azimuthal_int_GR} were expressed as linear combinations of $\frac{\left(1-x\right)^p}{\left(1+x\right)^q}$ and $\frac{\left(1+x\right)^p}{\left(1-x\right)^q}$. Because of the $|\cos{\theta}-\cos{\Theta}|$ term in the contour integral, we had to split the integration with respect to $x$ in two parts, one for $x\in\left(-1,\cos{\Theta}\right)$ and another for $x\in\left(\cos{\Theta},1\right)$. 

As a specific example, we show the calculation for the $x\theta$ component for the case $m=2$. We want to calculate the integral 
\begin{align}\label{eq:integral_GR_xtheta_l1}
    \GxthpTh{\ell}{2} &= N^2_{\ell}\int_{-1}^{1}\mathrm{d}x\,P_{\ell}^2\left(x\right)\int_0^{2\pi}\mathrm{d}\phi\,e^{2 \I \phi}\delta n_x^+\left(\vq\right)\delta m_{\theta}^+\left(\vq\right)=\sqrt{\frac{(2\ell+1)}{4\pi}\frac{(\ell-2)!}{(\ell+2)!}}\int_{-1}^{1}\mathrm{d}x\,P_{\ell}^2\left(x\right)\mathrm{I}^+_{x\theta,2}\left(\Theta,\theta\right),
\end{align}
where $\mathrm{I}^+_{x\theta,m}\left(\Theta,\theta\right)$ is given by Eq.~\eqref{subeq:az_xtheta}. Focusing first on the $\delta_{mi}$ terms in Eq.~\eqref{subeq:az_xtheta}, the only non-zero term is
\begin{equation}
    \frac{\pi}{8}\cos{\Theta}\sin^2{\theta}=\frac{\pi}{8}\cos{\Theta}\left(1-x^2\right)=\frac{\pi}{24}\cos{\Theta}P^2_2\left(x\right),
\end{equation}
where we have expressed this in terms of linear combinations of associated Legendre polynomials for which $m=2$ (just one associated Legendre polynomial was needed here). Using the orthogonality relations in Eq.~\eqref{eq:as_leg_norm}, the integral over $x$ then becomes
\begin{equation}\label{eq:free_term_ex}
    \int_{-1}^{1}\mathrm{d}xP_{\ell}^2\left(x\right)\left( \frac{\pi}{8}\cos{\Theta}\sin^2{\theta}\right)=\frac{\pi}{24}\cos{\Theta}\int_{-1}^{1}\mathrm{d}xP_2^2\left(x\right)P_{\ell}^2\left(x\right)=\frac{2\pi}{5}\cos{\Theta}\,\delta_{\ell2}.
\end{equation}
For $m=2$, the contour integral in Eq.~\eqref{subeq:az_xtheta} becomes
\begin{align}
\begin{aligned}
    &\int_0^{2\pi}\mathrm{d}\phi\,e^{2 \I \phi}\frac{(\cos{\theta}-\cos{\Theta})\sin{\theta}\sin{\Theta}}{8(1-\cos{\Theta}\cos{\theta}-\sin{\Theta}\sin{\theta}\cos{\phi})}(\cos{\phi}-\cos{(3\phi)})=
    \begin{dcases}\label{eq:count_int_res}
        \frac{\pi}{4}\left(x-\cos{\Theta}\right)\cot^4{\left(\Theta/2\right)}\frac{\left(1-x\right)^2}{\left(1+x\right)^2} \\
        \qquad \text{if } \cos{\Theta}\leq x \leq 1, \\[5pt] 
        \frac{\pi}{4}\left(x-\cos{\Theta}\right)\tan^4{\left(\Theta/2\right)}\frac{\left(1+x\right)^2}{\left(1-x\right)^2} \\
        \qquad \text{if } -1\leq x\leq\cos{\Theta}.
    \end{dcases}
\end{aligned}
\end{align}
Using Eq.~\eqref{eq:associated_legendre} to express $P_{\ell}^2\left(x\right)$ as
\begin{equation}
    P_{\ell}^2\left(x\right)=\left(1-x\right)\left(1+x\right)\frac{\mathrm{d^2}}{\mathrm{d}x^2}P_{\ell}\left(x\right),
\end{equation}
and rewriting 
\begin{align}
    x-\cos{\Theta}=\frac{1}{2}\left(1-\cos{\Theta}\right)\left(1+x\right)-\frac{1}{2}\left(1+\cos{\Theta}\right)\left(1-x\right)=\sinnhalf[2]{\Theta}\left(1+x\right)-\cosnhalf[2]{\Theta}\left(1-x\right),
\end{align}
when we multiply Eq.~\eqref{eq:count_int_res} by $P_{\ell}^2$ and integrate over $x$ we obtain
\begin{align}
    & \begin{aligned}
    &\int_{-1}^{1}\mathrm{d}xP_{\ell}^2\int_0^{2\pi}\mathrm{d}\phi\,e^{2 \I \phi}\frac{(\cos{\theta}-\cos{\Theta})\sin{\theta}\sin{\Theta}}{8(1-\cos{\Theta}\cos{\theta}-\sin{\Theta}\sin{\theta}\cos{\phi})}(\cos{\phi}-\cos{(3\phi)})=\\
    &\qquad - \frac{\pi}{4}\cosnhalf[6]{\Theta}\cscnhalf[4]{\Theta}\int_{\cos{\Theta}}^1\mathrm{d}x\frac{\left(1 - x\right)^4}{\left(1 + x\right)}\frac{\mathrm{d^2}}{\mathrm{d}x^2}P_\ell(x) + \frac{\pi}{4}\cosnhalf[4]{\Theta}\cscnhalf[2]{\Theta}\int_{\cos{\Theta}}^1\mathrm{d}x\left(1 - x\right)^3\frac{\mathrm{d^2}}{\mathrm{d}x^2}P_\ell(x)\\ 
    &\qquad + \frac{\pi}{4}\sinnhalf[6]{\Theta}\secnhalf[4]{\Theta}\int_{-1}^{\cos{\Theta}}\mathrm{d}x\frac{\left(1 + x\right)^4}{\left(1 - x\right)}\frac{\mathrm{d^2}}{\mathrm{d}x^2}P_\ell(x) - \frac{\pi}{4}\sinnhalf[4]{\Theta}\secnhalf[2]{\Theta}\int_{-1}^{\cos{\Theta}}\mathrm{d}x\left(1 + x\right)^3\frac{\mathrm{d^2}}{\mathrm{d}x^2}P_\ell(x). \label{eq:count_int_ex}
   \end{aligned}
\end{align}
Combining the results of Eqs.~\eqref{eq:free_term_ex} and \eqref{eq:count_int_ex}, we obtain the astrometric correlations for the $x\theta$ component for the case $m=2$ to be:
\begin{align}
        &\begin{aligned}
        &\GxthpTh{\ell}{2}=\frac{1}{2}\sqrt{(2\ell+1)\pi\frac{(\ell-2)!}{(\ell+2)!}}\Bigg(\frac{2}{5}\cos{\Theta}\,\delta_{\ell2} - \frac{1}{4}\cosnhalf[6]{\Theta}\cscnhalf[4]{\Theta}F_{4,1,\ell,2}^+ + \frac{1}{4}\cosnhalf[4]{\Theta}\cscnhalf[2]{\Theta}F_{3,0,\ell,2}^+\\ 
        &\qquad + \frac{1}{4}\sinnhalf[6]{\Theta}\secnhalf[4]{\Theta}F_{4,1,\ell,2}^- - \frac{1}{4}\sinnhalf[4]{\Theta}\secnhalf[2]{\Theta}F_{3,0,\ell,2}^-\Bigg),
        \end{aligned}
\end{align}
where, as in the main text, we define the functions
\begin{subequations}
    \begin{align}
        F_{q,r,\ell,m}^+ \equiv F_{q,r,\ell,m}^+(\cos{\Theta})&\equiv\int_{\cos{\Theta}}^1 \mathrm{d}x\frac{(1-x)^q}{(1+x)^r}\frac{\mathrm{d}^m}{\mathrm{d}x^m}P_\ell(x), \\
        F_{q,r,\ell,m}^- \equiv F_{q,r,\ell,m}^-(\cos{\Theta})&\equiv\int_{-1}^{\cos{\Theta}} \mathrm{d}x\frac{(1+x)^q}{(1-x)^r}\frac{\mathrm{d}^m}{\mathrm{d}x^m}P_\ell(x).
    \end{align}
\end{subequations}
We repeat the calculations for the other components of the generalized overlap reduction functions and the other values of $m$ and present our results below. 

We first start with the parallel components, which turn out to be purely real. For the $x\theta$ component and for $m=0$ we obtain:
\begin{subequations}
    \begin{align}
        & \begin{aligned}
        & \GxthpTh{\ell}{0} = \frac{1}{2}\sqrt{(2\ell+1)\pi}\Bigg(\frac{1}{3}\cos{\Theta}\,\delta_{\ell0} - \frac{1}{15}\cos{\Theta}\,\delta_{\ell2} - \frac{1}{2}\cosnhalf[4]{\Theta}\cscnhalf[2]{\Theta}F_{2,1,\ell,0}^+ + \frac{1}{2}\cosnhalf[2]{\Theta}F_{1,0,\ell,0}^+ \\ 
        & \qquad + \frac{1}{2}\sinnhalf[4]{\Theta}\secnhalf[2]{\Theta}F_{2,1,\ell,0}^- - \frac{1}{2}\sinnhalf[2]{\Theta}F_{1,0,\ell,0}^-\Bigg).
        \end{aligned} 
        \intertext{For $m=1$:}
        & \begin{aligned}
        & \GxthpTh{\ell}{1} = \frac{1}{2}\sqrt{(2\ell+1)\pi\frac{(\ell-1)!}{(\ell+1)!}}\Bigg(\frac{1}{5}\sin{\Theta}\,\delta_{\ell2} + \frac{1}{4}\cosnhalf[5]{\Theta}\cscnhalf[3]{\Theta}F_{3,1,\ell,1}^+ - \frac{1}{4}\sinnhalf[]{\Theta}\cosnhalf[]{\Theta}F_{1,-1,\ell,1}^+ \\ 
        &\qquad - \frac{1}{4}\sinnhalf[5]{\Theta}\secnhalf[3]{\Theta}F_{3,1,\ell,1}^- + \frac{1}{4}\sinnhalf[]{\Theta}\cosnhalf[]{\Theta}F_{1,-1,\ell,1}^-\Bigg). 
        \end{aligned} 
        \intertext{For $m=2$:}
        & \begin{aligned}
        & \GxthpTh{\ell}{2} = \frac{1}{2}\sqrt{(2\ell+1)\pi\frac{(\ell-2)!}{(\ell+2)!}}\Bigg(\frac{2}{5}\cos{\Theta}\,\delta_{\ell2} -  \frac{1}{4}\cosnhalf[6]{\Theta}\cscnhalf[4]{\Theta}F_{4,1,\ell,2}^+ + \frac{1}{4}\cosnhalf[4]{\Theta}\cscnhalf[2]{\Theta}F_{3,0,\ell,2}^+ \\ 
        &\qquad + \frac{1}{4}\sinnhalf[6]{\Theta}\secnhalf[4]{\Theta}F_{4,1,\ell,2}^- - \frac{1}{4}\sinnhalf[4]{\Theta}\secnhalf[2]{\Theta}F_{3,0,\ell,2}^-\Bigg). 
        \end{aligned} \\
        \intertext{And for $m=3,4,\cdots$:}
        & \begin{aligned}
        & \GxthpTh{\ell}{m} = (-1)^{m}\frac{1}{2}\sqrt{(2\ell+1)\pi\frac{(\ell-m)!}{(\ell+m)!}}\Bigg(\frac{1}{4}\cosnhalf[m]{\Theta}\cscnhalf[m]{\Theta}\bigg[-\cosnhalf[4]{\Theta}\cscnhalf[2]{\Theta}F_{m+2,1,\ell,m}^+ \\ 
        &\qquad + \cosnhalf[2]{\Theta}F_{m+1,0,\ell,m}^+ + \sinnhalf[2]{\Theta}F_{m,-1,\ell,m}^+ - \sinnhalf[4]{\Theta}\secnhalf[2]{\Theta}F_{m-1,-2,\ell,m}^+\bigg] \\ 
        &\qquad + \frac{1}{4}\sinnhalf[m]{\Theta}\secnhalf[m]{\Theta}\bigg[\sinnhalf[4]{\Theta}\secnhalf[2]{\Theta}F_{m+2,1,\ell,m}^- - \sinnhalf[2]{\Theta}F_{m+1,0,\ell,m}^- \\
        &\qquad - \cosnhalf[2]{\Theta}F_{m,-1,\ell,m}^- + \cosnhalf[4]{\Theta}\cscnhalf[2]{\Theta}F_{m-1,-2,\ell,m}^-\bigg]\Bigg).
        \end{aligned} 
        \intertext{For the $y\phi$ component and for $m=0$ we obtain:}
        & \begin{aligned}
        & \GyphpTh{\ell}{0} = \frac{1}{2}\sqrt{(2\ell+1)\pi}\Bigg(-\frac{2}{3}\delta_{\ell0} + \frac{1}{3}\cos{\Theta}\,\delta_{\ell1} -\frac{1}{15}\delta_{\ell2} + \frac{1}{2}\cosnhalf[4]{\Theta}\cscnhalf[2]{\Theta}F_{2,1,\ell,0}^+ - \cosnhalf[2]{\Theta}F_{1,0,\ell,0}^+ \\ 
        &\qquad + \frac{1}{2}\sinnhalf[2]{\Theta}F_{0,-1,\ell,0}^+ + \frac{1}{2}\sinnhalf[4]{\Theta}\secnhalf[2]{\Theta}F_{2,1,\ell,0}^- - \sinnhalf[2]{\Theta}F_{1,0,\ell,0}^- + \frac{1}{2}\cosnhalf[2]{\Theta}F_{0,-1,\ell,0}^-\Bigg).
        \end{aligned} 
        \intertext{For $m=1$:}
        & \begin{aligned}
        & \GyphpTh{\ell}{1} = \frac{1}{2}\sqrt{(2\ell+1)\pi\frac{(\ell-1)!}{(\ell+1)!}}\Bigg(-\frac{1}{4}\cosnhalf[5]{\Theta}\cscnhalf[3]{\Theta}F_{3,1,\ell,1}^+ + \frac{1}{2}\cosnhalf[3]{\Theta}\cscnhalf[]{\Theta}F_{2,0,\ell,1}^+ \\ 
        &\qquad - \frac{1}{4}\sinnhalf[]{\Theta}\cosnhalf[]{\Theta}F_{1,-1,\ell,1}^+ - \frac{1}{4}\sinnhalf[5]{\Theta}\secnhalf[3]{\Theta}F_{3,1,\ell,1}^- + \frac{1}{2}\sinnhalf[3]{\Theta}\secnhalf[]{\Theta}F_{2,0,\ell,1}^- \\
        &\qquad - \frac{1}{4}\sinnhalf[]{\Theta}\cosnhalf[]{\Theta}F_{1,-1,\ell,1}^-\Bigg).
        \end{aligned}
        \intertext{For $m=2$:}
        & \GyphpTh{\ell}{2} = \frac{1}{2}\sqrt{(2\ell+1)\pi\frac{(\ell-2)!}{(\ell+2)!}}\Bigg(-\frac{2}{5}\delta_{\ell2} + \frac{1}{2}\left(1 + (-1)^{\ell}\right) - \frac{1}{2}\cos{\Theta}\left(1 + (-1)^{\ell+1}\right) + \frac{1}{4}\cosnhalf[6]{\Theta}\cscnhalf[4]{\Theta}F_{4,1,\ell,2}^+ \notag\\ 
        &\qquad - \frac{3}{4}\cosnhalf[4]{\Theta}\cscnhalf[2]{\Theta}F_{3,0,\ell,2}^+ + \frac{3}{4}\cosnhalf[2]{\Theta}F_{2,-1,\ell,2}^+ - \frac{1}{4}\sinnhalf[2]{\Theta}F_{1,-2,\ell,2}^+ \\
        &\qquad + \frac{1}{4}\sinnhalf[6]{\Theta}\secnhalf[4]{\Theta}F_{4,1,\ell,2}^- - \frac{3}{4}\sinnhalf[4]{\Theta}\secnhalf[2]{\Theta}F_{3,0,\ell,2}^- + \frac{3}{4}\sinnhalf[2]{\Theta}F_{2,-1,\ell,2}^- - \frac{1}{4}\cosnhalf[2]{\Theta}F_{1,-2,\ell,2}^-\Bigg).
        \notag\\
        \intertext{And for $m=3,4,\cdots$:}
        & \begin{aligned}
        & \GyphpTh{\ell}{m} = (-1)^{m}\frac{1}{2}\sqrt{(2\ell+1)\pi\frac{(\ell-m)!}{(\ell+m)!}}\Bigg(\frac{1}{4}\cosnhalf[m]{\Theta}\cscnhalf[m]{\Theta}\bigg[\cosnhalf[4]{\Theta}\cscnhalf[2]{\Theta}F_{m+2,1,\ell,m}^+ \\ 
        &\qquad - 3\cosnhalf[2]{\Theta}F_{m+1,0,\ell,m}^++ 3\sinnhalf[2]{\Theta}F_{m,-1,\ell,m}^+ -\sinnhalf[4]{\Theta}\secnhalf[2]{\Theta}F_{m-1,-2,\ell,m}^+\bigg] \\ 
        &\qquad + \frac{1}{4}\sinnhalf[m]{\Theta}\secnhalf[m]{\Theta}\bigg[\sinnhalf[4]{\Theta}\secnhalf[2]{\Theta}F_{m+2,1,\ell,m}^- - 3\sinnhalf[2]{\Theta}F_{m+1,0,\ell,m}^- \\ 
        &\qquad + 3\cosnhalf[2]{\Theta}F_{m,-1,\ell,m}^- - \cosnhalf[4]{\Theta}\cscnhalf[2]{\Theta}F_{m-1,-2,\ell,m}^-\bigg]\Bigg).
        \end{aligned} \\
        \intertext{The perpendicular components turn out to be purely imaginary. For the $x\phi$ component and for $m=0$ the correlations vanish:}
        & \GxphpTh{\ell}{0} = 0. \\
        \intertext{For $m=1$ we obtain:}
        & \begin{aligned}
        & \GxphpTh{\ell}{1} = \frac{\I}{2}\sqrt{(2\ell+1)\pi\frac{(\ell-1)!}{(\ell+1)!}}\Bigg(-\frac{1}{4}\cosnhalf[5]{\Theta}\cscnhalf[3]{\Theta}F_{3,1,\ell,1}^+ + \frac{1}{2}\cosnhalf[3]{\Theta}\cscnhalf[]{\Theta}F_{2,0,\ell,1}^+ \\ 
        &\qquad - \frac{1}{4}\sinnhalf[]{\Theta}\cosnhalf[]{\Theta}F_{1,-1,\ell,1}^+- \frac{1}{4}\sinnhalf[5]{\Theta}\secnhalf[3]{\Theta}F_{3,1,\ell,1}^- + \frac{1}{2}\sinnhalf[3]{\Theta}\secnhalf[]{\Theta}F_{2,0,\ell,1}^-\\
        &\qquad - \frac{1}{4}\sinnhalf[]{\Theta}\cosnhalf[]{\Theta}F_{1,-1,\ell,1}^-\Bigg). 
        \end{aligned} \\
        \intertext{For $m=2:$}
        & \GxphpTh{\ell}{2} = \frac{\I}{2}\sqrt{(2\ell+1)\pi\frac{(\ell-2)!}{(\ell+2)!}}\Bigg(\frac{2}{5}\delta_{\ell2} - \frac{1}{2}\left(1 + (-1)^{\ell}\right) + \frac{1}{2}\cos{\Theta}\left(1 + (-1)^{\ell+1}\right) + \frac{1}{4}\cosnhalf[6]{\Theta}\cscnhalf[4]{\Theta}F_{4,1,\ell,2}^+\notag \\ 
        &\qquad - \frac{1}{4}\cosnhalf[4]{\Theta}\cscnhalf[2]{\Theta}F_{3,0,\ell,2}^+ - \frac{1}{4}\cosnhalf[2]{\Theta}F_{2,-1,\ell,2}^+ + \frac{1}{4}\sinnhalf[2]{\Theta}F_{1,-2,\ell,2}^+ \\ 
        &\qquad + \frac{1}{4}\sinnhalf[6]{\Theta}\secnhalf[4]{\Theta}F_{4,1,\ell,2}^- - \frac{1}{4}\sinnhalf[4]{\Theta}\secnhalf[2]{\Theta}F_{3,0,\ell,2}^- - \frac{1}{4}\sinnhalf[2]{\Theta}F_{2,-1,\ell,2}^- + \frac{1}{4}\cosnhalf[2]{\Theta}F_{1,-2,\ell,2}^-\Bigg).
        \notag\\
        \intertext{And for $m=3,4,\cdots$:}
        & \begin{aligned}
        & \GxphpTh{\ell}{m} = (-1)^{m}\frac{\I}{2}\sqrt{(2\ell+1)\pi\frac{(\ell-m)!}{(\ell+m)!}}\Bigg(\frac{1}{4}\cosnhalf[m]{\Theta}\cscnhalf[m]{\Theta}\bigg[\cosnhalf[4]{\Theta}\cscnhalf[2]{\Theta}F_{m+2,1,\ell,m}^+ \\ 
        &\qquad - \cosnhalf[2]{\Theta}F_{m+1,0,\ell,m}^+ - \sinnhalf[2]{\Theta}F_{m,-1,\ell,m}^+ + \sinnhalf[4]{\Theta}\secnhalf[2]{\Theta}F_{m-1,-2,\ell,m}^+\bigg] \\ 
        &\qquad + \frac{1}{4}\sinnhalf[m]{\Theta}\secnhalf[m]{\Theta}\bigg[\sinnhalf[4]{\Theta}\secnhalf[2]{\Theta}F_{m+2,1,\ell,m}^- - \sinnhalf[2]{\Theta}F_{m+1,0,\ell,m}^- \\ 
        &\qquad -\cosnhalf[2]{\Theta}F_{m,-1,\ell,m}^- + \cosnhalf[4]{\Theta}\cscnhalf[2]{\Theta}F_{m-1,-2,\ell,m}^-\bigg]\Bigg).
        \end{aligned} 
        \intertext{For $y\theta$ and $m=0$ the correlations also vanish:}
        & \GythpTh{\ell}{0} = 0. \\
        \intertext{For $m=1$ we obtain:}
        & \begin{aligned}
        & \GythpTh{\ell}{1} = \frac{\I}{2}\sqrt{(2\ell+1)\pi\frac{(\ell-1)!}{(\ell+1)!}}\Bigg(\frac{1}{5}\sin{\Theta}\,\delta_{\ell2} - \frac{1}{4}\cosnhalf[5]{\Theta}\cscnhalf[3]{\Theta}F_{3,1,\ell,1}^+ + \cosnhalf[3]{\Theta}\cscnhalf[]{\Theta}F_{2,0,\ell,1}^+ \\
        &\qquad - \frac{3}{4}\sinnhalf[]{\Theta}\cosnhalf[]{\Theta}F_{1,-1,\ell,1}^+ +\frac{1}{4}\sinnhalf[5]{\Theta}\secnhalf[3]{\Theta}F_{3,1,\ell,1}^- -\sinnhalf[3]{\Theta}\secnhalf[]{\Theta}F_{2,0,\ell,1}^- \\
        &\qquad + \frac{3}{4}\sinnhalf[]{\Theta}\cosnhalf[]{\Theta}F_{1,-1,\ell,1}^-\Bigg).
        \end{aligned} \\
        \intertext{For $m=2$:}
        & \begin{aligned}
        & \GythpTh{\ell}{2} = \frac{\I}{2}\sqrt{(2\ell+1)\pi\frac{(\ell-2)!}{(\ell+2)!}}\Bigg(\frac{2}{5}\cos{\Theta}\, \delta_{\ell2} + \frac{1}{4}\cosnhalf[6]{\Theta}\cscnhalf[4]{\Theta}F_{4,1,\ell,2}^+ - \frac{3}{4}\cosnhalf[4]{\Theta}\cscnhalf[2]{\Theta}F_{3,0,\ell,2}^+ \\ 
        &\qquad + \frac{1}{2}\cosnhalf[2]{\Theta}F_{2,-1,\ell,2}^+ - \frac{1}{4}\sinnhalf[6]{\Theta}\secnhalf[4]{\Theta}F_{4,1,\ell,2}^- + \frac{3}{4}\sinnhalf[4]{\Theta}\secnhalf[2]{\Theta}F_{3,0,\ell,2}^- \\ 
        &\qquad - \frac{1}{2}\sinnhalf[2]{\Theta}F_{2,-1,\ell,2}^-\Bigg).
        \end{aligned} \\
        \intertext{And for $m=3,4,\cdots$:}
        & \begin{aligned}
        & \GythpTh{\ell}{m} = (-1)^{m}\frac{\I}{2}\sqrt{(2\ell+1)\pi\frac{(\ell-m)!}{(\ell+m)!}}\Bigg(\frac{1}{4}\cosnhalf[m]{\Theta}\cscnhalf[m]{\Theta}\bigg[\cosnhalf[4]{\Theta}\cscnhalf[2]{\Theta}F_{m+2,1,\ell,m}^+ \\ 
        &\qquad - 3\cosnhalf[2]{\Theta}F_{m+1,0,\ell,m}^++3\sinnhalf[2]{\Theta}F_{m,-1,\ell,m}^+ -\sinnhalf[4]{\Theta}\secnhalf[2]{\Theta}F_{m-1,-2,\ell,m}^+\bigg] \\ 
        &\qquad + \frac{1}{4}\sinnhalf[m]{\Theta}\secnhalf[m]{\Theta}\bigg[-\sinnhalf[4]{\Theta}\secnhalf[2]{\Theta}F_{m+2,1,\ell,m}^- + 3\sinnhalf[2]{\Theta}F_{m+1,0,\ell,m}^- \\
        &\qquad - 3\cosnhalf[2]{\Theta}F_{m,-1,\ell,m}^- + \cosnhalf[4]{\Theta}\cscnhalf[2]{\Theta}F_{m-1,-2,\ell,m}^-\bigg]\Bigg).
        \end{aligned}
    \end{align}
\end{subequations}
Since $\GxthpTh{\ell}{m}$ and $\GyphpTh{\ell}{m}$ are real, and $\GxphpTh{\ell}{m}$ and $\GythpTh{\ell}{m}$ are imaginary, using the definition of the generalized overlap reduction functions in Eqs.~\eqref{eq:gorf} and $\left({Y_{\ell}^m}\right)^*=\left(-1\right)^mY_{\ell}^{-m}$, we can express the astrometric correlations for negative $m$ as 

\vspace{0pt}
\begin{subequations}\label{eq:gamma_plus_neg}
\noindent
\begin{minipage}[t]{0.5\linewidth}
    \setlength{\abovedisplayskip}{0pt}
    \setlength{\belowdisplayskip}{0pt}
    \begin{align}
        \GxthpTh{\ell}{-m} &= \left(-1\right)^m\GxthpTh{\ell}{m} , \\
        \GyphpTh{\ell}{-m} & =\left(-1\right)^m\GyphpTh{\ell}{m} ,
    \end{align}
\end{minipage}%
\hfill
\begin{minipage}[t]{0.5\linewidth}
    \setlength{\abovedisplayskip}{0pt}
    \setlength{\belowdisplayskip}{0pt}
    \begin{align}
        \GxphpTh{\ell}{-m} &= \left(-1\right)^{m+1}\GxphpTh{\ell}{m} , \\
        \GythpTh{\ell}{-m} &= \left(-1\right)^{m+1}\GythpTh{\ell}{m} . 
    \end{align}
\end{minipage}
\end{subequations}
\par
\vspace{\belowdisplayskip}
\noindent
We then calculated the astrometric deflection components for an \sgwb\ consisting of \gw s with $\times$ polarization. The calculations are the same, except that we now use the polarization tensor in Eq.~\eqref{subeq:cross_pol}. We find that there is a symmetry between the two polarizations

\vspace{0pt}
\begin{subequations}
\noindent
\begin{minipage}[t]{0.5\linewidth}
    \setlength{\abovedisplayskip}{0pt}
    \setlength{\belowdisplayskip}{0pt}
    \begin{align}
        \delta n_x^{\times}\left(\theta,\phi\right)&=\delta n_y^+\left(\theta, \phi\right),\\
        \delta n_y^{\times}\left(\theta,\phi\right)&=-\delta n_x^+\left(\theta, \phi\right),
    \end{align}
\end{minipage}%
\hfill
\begin{minipage}[t]{0.5\linewidth}
    \setlength{\abovedisplayskip}{0pt}
    \setlength{\belowdisplayskip}{0pt}
    \begin{align}
        \delta m_{\theta}^{\times}\left(\theta,\phi\right)&=\delta m_{\phi}^+\left(\theta, \phi\right)\\
        \delta m_{\phi}^{\times}\left(\theta,\phi\right)&=-\delta m_{\theta}^+\left(\theta, \phi\right).
    \end{align}
\end{minipage}
\end{subequations}
\par
\vspace{\belowdisplayskip}
\noindent
It then follows that the correlation integrals for the $\times$ polarization are related to those of the $+$ polarization by

\vspace{0pt}
\begin{subequations}\label{eq:gamma_plus_cross_sym}
\noindent
\begin{minipage}[t]{0.5\linewidth}
    \setlength{\abovedisplayskip}{0pt}
    \setlength{\belowdisplayskip}{0pt}
    \begin{align}
        \Gorf{x}{\theta}[\times][\Theta][\ell][m] &=\GyphpTh{\ell}{m} , \\
        \Gorf{y}{\phi}[\times][\Theta][\ell][m] &= \GxthpTh{\ell}{m} ,
    \end{align}
\end{minipage}%
\hfill
\begin{minipage}[t]{0.5\linewidth}
    \setlength{\abovedisplayskip}{0pt}
    \setlength{\belowdisplayskip}{0pt}
    \begin{align}
        \Gorf{x}{\phi}[\times][\Theta][\ell][m] &= -\GythpTh{\ell}{m} , \\
        \Gorf{y}{\theta}[\times][\Theta][\ell][m] &= -\GxphpTh{\ell}{m} .
    \end{align}
\end{minipage}
\end{subequations}
\par
\vspace{\belowdisplayskip}
\noindent
Therefore, it follows that $\GxthpcTh{\ell}{m}=\GyphpcTh{\ell}{m}$ and $\GxphpcTh{\ell}{m}=-\GythpcTh{\ell}{m}$. Summing the pairs of correlation functions leads to Eqs.~\eqref{eq:gr_results} in the main text. Using Eqs.~\eqref{eq:gamma_plus_neg} and \eqref{eq:gamma_plus_cross_sym}, we obtain the generalized overlap reduction functions for negative $m$ to be

\vspace{0pt}
\begin{subequations}
\noindent
\begin{minipage}[t]{0.5\linewidth}
    \setlength{\abovedisplayskip}{0pt}
    \setlength{\belowdisplayskip}{0pt}
    \begin{align}
        \GxthpcTh{\ell}{-m} &= \left(-1\right)^m\GxthpcTh{\ell}{m} , \\
        \GyphpcTh{\ell}{-m} & =\left(-1\right)^m\GyphpcTh{\ell}{m} ,
    \end{align}
\end{minipage}%
\hfill
\begin{minipage}[t]{0.5\linewidth}
    \setlength{\abovedisplayskip}{0pt}
    \setlength{\belowdisplayskip}{0pt}
    \begin{align}
        \GxphpcTh{\ell}{-m} &= \left(-1\right)^{m+1}\GxphpcTh{\ell}{m} , \\
        \GythpcTh{\ell}{-m} &= \left(-1\right)^{m+1}\GythpcTh{\ell}{m} .
    \end{align}
\end{minipage}
\end{subequations}
\par
\vspace{\belowdisplayskip}
\noindent

\section{Analytic forms of \texorpdfstring{$F_{q,r,\ell,m}^{\pm}(\cos{\Theta})$}{F functions}}
\label[appendix]{app:Fpm_functions}
\noindent
In this appendix, we quote analytic results for the $F_{q,r,\ell,m}^{\pm}(\cos{\Theta})$ functions used to express the generalized overlap reduction functions. We define the functions as
\begin{subequations}
        \begin{align}
            F_{q,r,\ell,m}^+ \equiv F_{q,r,\ell,m}^+(\cos{\Theta})&\equiv\int_{\cos{\Theta}}^1 \mathrm{d}x\frac{(1-x)^q}{(1+x)^r}\frac{\mathrm{d}^m}{\mathrm{d}x^m}P_\ell(x), \\
            F_{q,r,\ell,m}^- \equiv F_{q,r,\ell,m}^-(\cos{\Theta})&\equiv\int_{-1}^{\cos{\Theta}} \mathrm{d}x\frac{(1+x)^q}{(1-x)^r}\frac{\mathrm{d}^m}{\mathrm{d}x^m}P_\ell(x).
        \end{align}
\end{subequations}
In Appendix E.1 of \rcite{Gair:2014rwa} the authors find explicit forms for $\mathcal{F}_{q,r,\ell,m}^{\pm}(\cos{\Theta})=F_{q,r,\ell,m}^{\pm}(-\cos{\Theta})$ while calculating correlation matrices for timing residuals of \pta s. 
We verified these results and quote them without proof here:
\begin{align}
    F_{q,r,\ell,m}^{-}(\cos{\Theta}) &= \sum_{i=0}^{q} \sum_{j=m}^{\ell} 2^{i-j}\left(-1\right)^{q-i+j+m} \frac{q!\left(\ell+j\right)!}{i!\left(q-i\right)! \, j! \left(\ell-j\right)!\left(j-m\right)!}\int_{-1}^{\cos{\Theta}}\mathrm{d}x\,\left(1-x\right)^{q-i-r+j-m}, \\
    F_{q,r,\ell,m}^{+}(\cos{\Theta})&=\sum_{i=0}^q\sum_{j=m}^{\ell} 2^{i-j}\left(-1\right)^{\ell+q-i+j}\frac{q!\left(\ell+j\right)!}{i!\left(q-i\right)!j!\left(\ell-j\right)!\left(j-m\right)!}\int_{\cos{\Theta}}^{1}\mathrm{d}x\,\left(1+x\right)^{q-i-r+j-m},
\end{align}
where after evaluating the integral the functions become
\begin{subequations}\label{eq:F_results}
    \begin{align}
        F^-_{q,0,\ell,m}\left(\cos{\Theta}\right)&=\sum_{i=0}^q\sum_{j=m}^{\ell}2^{i-j}\left(-1\right)^{q-i+j+m}\frac{q!\left(\ell+j\right)!\left(2^{q-i+j-m+1}-\left(1-\cos{\Theta}\right)^{q-i+j-m+1}\right)}{i!\left(q-i\right)!j!\left(\ell-j\right)!\left(j-m\right)!\left(q-i+j-m+1\right)}, \\[4ex]
        F^-_{q,1,\ell,m}\left(\cos{\Theta}\right)&=\sum_{i=0}^{q-1}\sum_{j=m}^{\ell}2^{i-j}\left(-1\right)^{q-i+j+m}\frac{q!\left(\ell+j\right)!\left(2^{q-i+j-m}-\left(1-\cos{\Theta}\right)^{q-i+j-m}\right)}{i!\left(q-i\right)!j!\left(\ell-j\right)!\left(j-m\right)!\left(q-i+j-m\right)}\notag \\
        &+\sum_{j=m+1}^{\ell}2^{q-j}\left(-1\right)^{j+m}\frac{\left(\ell+j\right)!\left(2^{j-m}-\left(1-\cos{\Theta}\right)^{j-m}\right)}{j!\left(\ell-j\right)!\left(j-m\right)!\left(j-m\right)} \\
        &+\frac{2^{q-m}\left(\ell+m\right)!}{m!\left(\ell-m\right)!}\ln{\left(\frac{2}{1-\cos{\Theta}}\right)}, \notag\\[4ex]
        F^+_{q,0,\ell,m}\left(\cos{\Theta}\right)&=\sum_{i=0}^q\sum_{j=m}^{\ell}2^{i-j}\left(-1\right)^{\ell+q-i+j}\frac{q!\left(\ell+j\right)!\left(2^{q-i+j-m+1}-\left(1+\cos{\Theta}\right)^{q-i+j-m+1}\right)}{i!\left(q-i\right)!j!\left(\ell-j\right)!\left(j-m\right)!\left(q-i+j-m+1\right)}, \\[4ex]
        F^+_{q,1,\ell,m}\left(\cos{\Theta}\right)&=\sum_{i=0}^{q-1}\sum_{j=m}^{\ell}2^{i-j}\left(-1\right)^{\ell+q-i+j}\frac{q!\left(\ell+j\right)!\left(2^{q-i+j-m}-\left(1+\cos{\Theta}\right)^{q-i+j-m}\right)}{i!\left(q-i\right)!j!\left(\ell-j\right)!\left(j-m\right)!\left(q-i+j-m\right)}\notag \\
        &+\sum_{j=m+1}^{\ell}2^{q-j}\left(-1\right)^{\ell+j}\frac{\left(\ell+j\right)!\left(2^{j-m}-\left(1+\cos{\Theta}\right)^{j-m}\right)}{j!\left(\ell-j\right)!\left(j-m\right)!\left(j-m\right)} \\
        &+\frac{\left(-1\right)^{\ell+m}2^{q-m}\left(\ell+m\right)!}{m!\left(\ell-m\right)!}\ln{\left(\frac{2}{1+\cos{\Theta}}\right)}. \notag
    \end{align}
\end{subequations}
We also need expressions for $r=-1$ and $r=-2$. Using $1\pm x=2-\left(1\mp x\right)$, we obtain the recurrence relations
\begin{subequations}\label{eq:F_recursive}
\begin{align}
    F^{\pm}_{q,-1,\ell,m}\left(\cos{\Theta}\right)&=2F^{\pm}_{q,0,\ell,m}\left(\cos{\Theta}\right)-F^{\pm}_{q+1,0,\ell,m}\left(\cos{\Theta}\right),\\
            F^{\pm}_{q,-2,\ell,m}\left(\cos{\Theta}\right)&=2F^{\pm}_{q,-1,\ell,m}\left(\cos{\Theta}\right)-F^{\pm}_{q+1,-1,\ell,m}\left(\cos{\Theta}\right).
\end{align}
\end{subequations}

\section{Calculation of astrometric integrals for transverse-scalar polarization}
\label[appendix]{app:s_calculations}
\noindent
In this appendix, we show how to calculate Eqs.~\eqref{s_results}, the generalized overlap reduction functions for an \sgwb\ that contains \gw s with transverse-scalar polarization. As in the main text, we consider a coordinate system in which one star is at the North pole, while the other is in the xz plane, the position vectors described by Eqs.~\eqref{coordinates_nm}, so that the astrometric correlations are just a function of the angle between the two stars, $\Theta$. The integral we want to calculate is
\begin{align}
    \Gij[S][\Theta][\ell][m] &= \int_{S^2}\mathrm{d}\Omega_{\vq}\,Y_{\ell}^m\left(\vq\right)\delta n_i^S\left(\vq\right)\delta m_j^S\left(\vq\right)=N^m_{\ell}\int_{0}^{\pi}\mathrm{d}\theta\,\sin{\theta}\,P_{\ell}^m\left(\cos{\theta}\right)\int_0^{2\pi}\mathrm{d}\phi\,e^{\I m\phi}\delta n_i^S\left(\vq\right)\delta m_j^S\left(\vq\right),
\end{align}
where $\vq=\left(\sin{\theta}\cos{\phi},\sin{\theta}\sin{\phi},\cos{\theta}\right)$ is the direction from which the \gw s originate and $\delta n_i^S\left(\vq\right)$ and $\delta m_j^S\left(\vq\right)$ are the angular displacements defined in Eq.~\eqref{subeq:delta_ni} for $P=S$, where we henceforth drop the dependence on $\vn$ and $\vm$. $Y_{\ell}^m\left(\vq\right)=N^m_{\ell}P^m_{\ell}\left(\cos{\theta}\right)e^{\I m\phi}$ are the spherical harmonic functions where $N^m_{\ell}$ are normalization constants and $P^m_{\ell}\left(\cos{\theta}\right)$ are the associated Legendre polynomials defined in \cref{app:math}. If we define $x\equiv\cos{\theta}$, we can rewrite the above integral as follows
\begin{align}\label{eq:integral_S}
    \Gij[S][\Theta][\ell][m] &= N^m_{\ell}\int_{-1}^{1}\mathrm{d}x\,P_{\ell}^m\left(x\right)\int_0^{2\pi}\mathrm{d}\phi\,e^{\I m\phi}\delta n_i^S\left(\vq\right)\delta m_j^S\left(\vq\right).
\end{align}

The calculations are similar to those for the \gr\ polarizations in \cref{app:gr_calculations}. We calculate the angular displacements for the two stars, $\delta n_i^S\left(\vq\right)$ and $\delta m_i^S\left(\vq\right)$, using Eq.~\eqref{subeq:delta_ni}, plugging in the definition of $\Delta_i^{jk}\left(\vn,\vq\right)$ in Eq.~\eqref{eq:triangle} and the definition of the polarization tensor for transverse-scalar \gw s in Eq.~\eqref{subeq:s_pol} defined in terms of the spherical coordinate basis vectors in Eqs.~\eqref{eq:spherical_basis}. We consider the basis vectors on the sphere defined in Eqs.~\eqref{basis_vectors_def}. 
Dotting the angular displacement vectors with the basis vectors on the sphere using Eqs.~\eqref{eq:basis_vectors_dotting}, we obtain the components of the astrometric deflections to be

\vspace{0pt}
\begin{subequations}
\noindent
\begin{minipage}[t]{0.4\linewidth}
    \setlength{\abovedisplayskip}{0pt}
    \setlength{\belowdisplayskip}{0pt}
    \begin{align}
        \delta n_x^S \left(\theta, \phi\right) &= -\frac{1}{2}\sin{\theta}\cos{\phi}, \\
         \delta n_y^S  \left(\theta, \phi\right) &= -\frac{1}{2}\sin{\theta}\sin{\phi},
    \end{align}
\end{minipage}%
\hfill
\begin{minipage}[t]{0.56\linewidth}
    \setlength{\abovedisplayskip}{0pt}
    \setlength{\belowdisplayskip}{0pt}
    \begin{align}
        \delta m_{\theta}^S  \left(\theta, \phi\right) &= \frac{1}{2}\sin{\Theta}\cos{\theta}-\frac{1}{2}\cos{\Theta}\sin{\theta}\cos{\phi}, \\
         \delta m_{\phi}^S \left(\theta, \phi\right) &= -\frac{1}{2}\sin{\theta}\sin{\phi} .
    \end{align}
\end{minipage}
\end{subequations}
\par
\vspace{\belowdisplayskip}
\noindent
The products of the angular displacements are then
\begin{subequations}
    \begin{align}
         \delta n_x^S\delta m_{\theta}^S&=-\frac{1}{4}\sin{\Theta}\sin{\theta}\cos{\theta}\cos{\phi}+\frac{1}{8}\cos{\Theta}\sin^2{\theta}\left(1+\cos{\left(2\phi\right)}\right),
         \\[2ex]
         \delta n_y^S\delta m_{\phi}^S&=\frac{1}{8}\sin^2{\theta}\left(1-\cos{\left(2\phi\right)}\right),
         \\[2ex]
         \delta n_x^S\delta m_{\phi}^S&=\frac{1}{8}\sin^2{\theta}\sin{\left(2\phi\right)},
         \\[2ex]
          \delta n_y^S\delta m_{\theta}^S&=-\frac{1}{4}\sin{\Theta}\sin{\theta}\cos{\theta}\sin{\phi}+\frac{1}{8}\cos{\Theta}\sin^2{\theta}\sin{\left(2\phi\right)}.
    \end{align}
\end{subequations}
There is no need for contour integration, and hence the azimuthal integrals are trivial (again focusing on $m\geq0$):
\begin{subequations}
    \begin{align}
        \int_0^{2\pi}\mathrm{d}\phi\,e^{\I m\phi}\delta n_x^S\left(\vq\right)\delta m_{\theta}^S\left(\vq\right)&=-\frac{\pi}{4}\sin{\Theta}\sin{\theta}\cos{\theta}\,\delta_{m1}+\frac{\pi}{8}\cos{\Theta}\sin^2{\theta}\left(2\delta_{m0}+\delta_{m2}\right),
        \\[2ex]
        \int_0^{2\pi}\mathrm{d}\phi\,e^{\I m\phi}\delta n_y^S\left(\vq\right)\delta m_{\phi}^S\left(\vq\right)&=\frac{\pi}{8}\sin^2{\theta}\left(2\delta_{m0}-\delta_{m2}\right),
        \\[2ex]
        \int_0^{2\pi}\mathrm{d}\phi\,e^{\I m\phi}\delta n_x^S\left(\vq\right)\delta m_{\phi}^S\left(\vq\right)&=\frac{\I\pi}{8}\sin^2{\theta}\,\delta_{m2},
        \\[2ex]
         \int_0^{2\pi}\mathrm{d}\phi\,e^{\I m\phi}\delta n_y^S\left(\vq\right)\delta m_{\theta}^S\left(\vq\right)&=-\frac{\I\pi}{4}\sin{\Theta}\sin{\theta}\cos{\theta}\,\delta_{m1}+\frac{\I\pi}{8}\cos{\Theta}\sin^2{\theta}\,\delta_{m2}.
    \end{align}
\end{subequations}
After the azimuthal integral, to do the integral over $x$, we rewrite the trigonometric functions of $\theta$ in terms of polynomials of $x\equiv\cos{\theta}$, which we then rewrite in terms of linear combinations of associated Legendre polynomials, $P_{\ell}^m\left(x\right)$, and use the orthogonality relations in Eq.~\eqref{eq:as_leg_norm} (or, if not possible, use the definition of $P_{\ell}^m$ in Eq.~\eqref{eq:associated_legendre} and integrate by parts to get ordinary Legendre polynomials for which we can use the orthogonality relations Eq.~\eqref{eq:leg_norm}). Doing these calculations we obtain Eqs.~\eqref{s_results} in the main text. For negative $m$ we can express our results using indicator functions, using the symmetry of the generalized overlap reduction functions:

\vspace{0pt}
\begin{subequations}
\noindent
\begin{minipage}[t]{0.5\linewidth}
    \setlength{\abovedisplayskip}{0pt}
    \setlength{\belowdisplayskip}{0pt}
    \begin{align}
        \GxthsTh{\ell}{-m} &= \left(-1\right)^m\GxthsTh{\ell}{m},\\
        \GyphsTh{\ell}{-m} &= \left(-1\right)^m\GyphsTh{\ell}{m},
    \end{align}
\end{minipage}%
\hfill
\begin{minipage}[t]{0.5\linewidth}
    \setlength{\abovedisplayskip}{0pt}
    \setlength{\belowdisplayskip}{0pt}
    \begin{align}
        \GxphsTh{\ell}{-m} &= \left(-1\right)^{m+1}\GxphsTh{\ell}{m},\\
        \GythsTh{\ell}{-m} &= \left(-1\right)^{m+1}\GythsTh{\ell}{m}.
    \end{align}
\end{minipage}
\end{subequations}
\par
\vspace{\belowdisplayskip}
\noindent

\section{Calculation of astrometric integrals for vectorial polarizations}
\label[appendix]{app:xy_calculations}
\noindent
In this appendix, we show how to calculate the generalized overlap reduction functions for an \sgwb\ that contains \gw s with vectorial polarizations, $X$ and $Y$. 
As in the main text, we consider a coordinate system in which one star is on the North pole, while the other is in the xz plane, the position vectors described by Eqs.~\eqref{coordinates_nm}, so that the astrometric correlations are just a function of the angle between the two stars, $\Theta$. 
The integrals we want to calculate are
\begin{subequations}
    \begin{align}
        \Gij[X][\Theta][\ell][m] &= \int_{S^2}\mathrm{d}\Omega_{\vq}\,Y_{\ell}^m\left(\vq\right)\delta n_i^X\left(\vq\right)\delta m_j^X\left(\vq\right)=N^m_{\ell}\int_{0}^{\pi}\mathrm{d}\theta\,\sin{\theta}\,P_{\ell}^m\left(\cos{\theta}\right)\int_0^{2\pi}\mathrm{d}\phi\,e^{\I m\phi}\delta n_i^X\left(\vq\right)\delta m_j^X\left(\vq\right),\\
        \Gij[Y][\Theta][\ell][m] &= \int_{S^2}\mathrm{d}\Omega_{\vq}\,Y_{\ell}^m\left(\vq\right)\delta n_i^Y\left(\vq\right)\delta m_j^Y\left(\vq\right)=N^m_{\ell}\int_{0}^{\pi}\mathrm{d}\theta\,\sin{\theta}\,P_{\ell}^m\left(\cos{\theta}\right)\int_0^{2\pi}\mathrm{d}\phi\,e^{\I m\phi}\delta n_i^Y\left(\vq\right)\delta m_j^Y\left(\vq\right),
    \end{align}
\end{subequations}
where $\vq=\left(\sin{\theta}\cos{\phi},\sin{\theta}\sin{\phi},\cos{\theta}\right)$ is the direction from which the \gw s originate and $\delta n_i^X\left(\vq\right)$, $\delta m_j^X\left(\vq\right)$, $\delta n_i^Y\left(\vq\right)$ and $\delta m_j^Y\left(\vq\right)$ are the angular displacements defined in Eq.~\eqref{subeq:delta_ni} for $P=\left\{X,Y\right\}$. $Y_{\ell}^m\left(\vq\right)=N^m_{\ell}P^m_{\ell}\left(\cos{\theta}\right)e^{\I m\phi}$ are the spherical harmonic functions, where $N^m_{\ell}$ are normalization constants and $P^m_{\ell}\left(\cos{\theta}\right)$ are the associated Legendre polynomials defined in \cref{app:math}. If we define $x\equiv\cos{\theta}$, we can rewrite the above integrals as follows
\begin{subequations}
    \begin{align}\label{eq:integral_XY}
        \Gij[X][\Theta][\ell][m] &= N^m_{\ell}\int_{-1}^{1}\mathrm{d}x\,P_{\ell}^m\left(x\right)\int_0^{2\pi}\mathrm{d}\phi\,e^{\I m\phi}\delta n_i^X\left(\vq\right)\delta m_j^X\left(\vq\right),\\
         \Gij[Y][\Theta][\ell][m] &= N^m_{\ell}\int_{-1}^{1}\mathrm{d}x\,P_{\ell}^m\left(x\right)\int_0^{2\pi}\mathrm{d}\phi\,e^{\I m\phi}\delta n_i^Y\left(\vq\right)\delta m_j^Y\left(\vq\right).
    \end{align}
\end{subequations}
We assume equal power in each of the polarizations and define the correlation functions:
\begin{subequations}
    \begin{align}
        \Gij[X,Y][\Theta][\ell][m] &= \Gij[X][\Theta][\ell][m] +  \Gij[Y][\Theta][\ell][m].
    \end{align}
\end{subequations}
We first calculate the astrometric deflections due to $X$ and $Y$ polarized \gw s. Using calculations identical to those for the \gr\ polarizations in \cref{app:gr_calculations} except that we now use the polarization tensors in Eqs.~\eqref{subeq:x_pol} and \eqref{subeq:y_pol}, we obtain the angular deflection components to be, first for the $X$ polarization (Eqs.~(E2c) and (E2d) in \rcite{Mihaylov:2018uqm} have typos):
\begin{subequations}
    \begin{align}
        & \delta n_x^X \left(\theta, \phi\right) = \frac{1}{2}(1+2\cos{\theta})\cos{\phi},
        \\[3ex]
        & \delta n_y^X \left(\theta, \phi\right) = \frac{1}{2}(1+2\cos{\theta})\sin{\phi}, 
        \\[3ex]
        & \begin{aligned}
        & \delta m_{\theta}^X \left(\theta, \phi\right) = \frac{1}{1 - \cos\Theta \cos\theta - \sin\Theta \sin\theta \cos\phi}\Bigg(\frac{1}{8}\left(2\sin{\Theta}\sin{\theta} - 3\sin{(2\Theta)}\sin{(2\theta)}\right) \\ 
        &\qquad + \frac{1}{2}\left(\cos{\Theta}\cos{\theta}-\cos{(2\Theta)}\cos{(2\theta)}\right)\cos{\phi} + \frac{1}{8}\left(2\sin{\Theta}\sin{\theta}-\sin{(2\Theta)}\sin{(2\theta)}\right)\cos{(2\phi)}\Bigg), 
        \end{aligned} \\[3ex]
        & \begin{aligned}
        & \delta m_{\phi}^X \left(\theta, \phi\right) = \frac{1}{1 - \cos\Theta \cos\theta - \sin\Theta \sin\theta \cos\phi}\Bigg(\frac{1}{2}\cos{\Theta}\left(\cos{\Theta}\cos{\theta} - \cos{\left(2\theta\right)}\right)\sin{\phi} \\
        &\qquad + \frac{1}{8}\left(\sin{\left(2\Theta\right)}\sin{\theta} - 2\sin{\Theta}\sin{\left(2\theta\right)}\right)\sin{\left(2\phi\right)}\Bigg).
        \end{aligned} \\
        \intertext{And then for the $Y$ polarization:}
        & \delta n_x^Y \left(\theta, \phi\right) = \frac{1}{2}\cos{\theta}\sin{\phi}, 
        \\[3ex]
        & \delta n_y^Y \left(\theta, \phi\right) = -\frac{1}{2}\cos{\theta}\cos{\phi}, 
        \\[3ex]
        & \begin{aligned}
        & \delta m_{\theta}^Y \left(\theta, \phi\right) = \frac{1}{1 - \cos\Theta \cos\theta - \sin\Theta \sin\theta \cos\phi}\Bigg(\frac{1}{2}\left(\cos{(2\Theta)} - \cos{\Theta}\cos{\theta}\right)\cos{\theta}\sin{\phi} \\ 
        &\qquad + \frac{1}{4}\sin{\Theta}\left(2\cos{\Theta} - \cos{\theta}\right)\sin{\theta}\sin{(2\phi)}\Bigg),  
        \end{aligned} \\[3ex]
        & \begin{aligned}
        & \delta m_{\phi}^Y \left(\theta, \phi\right) = \frac{1}{1 - \cos\Theta \cos\theta - \sin\Theta \sin\theta \cos\phi}\Bigg(\frac{3}{16}\sin{\left(2\Theta\right)}\sin{\left(2\theta\right)} + \frac{1}{4}\left(1 - 2\cos{\Theta}\cos{\theta} + \cos{\left(2\Theta\right)}\cos{\left(2\theta\right)}\right)\cos{\phi} \\
        &\qquad +\frac{1}{16}\left(\sin{\left(2\Theta\right)}\sin{\left(2\theta\right)} - 8\sin{\Theta}\sin{\theta}\right)\cos{\left(2\phi\right)}\Bigg). 
        \end{aligned}
    \end{align}
\end{subequations}
The products of the angular displacements can be rewritten as, first for the $X$ polarization:
\begin{subequations}\label{ang_prod_xy}
    \begin{align}
        & \begin{aligned}
        & \delta n_{x}^X \delta m_{\theta}^X = \frac{1}{8}\left(1+2\cos{\theta}\right)(1+2\cos{\theta}\cos{\Theta})\left(1 + \cos{(2\phi)}\right) + \frac{1}{2}\left(\sin{\theta} + \sin{(2\theta)}\right)\sin{\Theta}\cos{\phi} \\
        &\qquad - \frac{\left(\sin{\theta} + \sin{(2\theta)}\right)\sin{\Theta}}{8(1 - \cos{\Theta}\cos{\theta} - \sin{\Theta}\sin{\theta}\cos{\phi})}(\cos{\phi} - \cos{(3\phi)}), 
        \end{aligned} \\[3ex]
        & \delta n_{y}^X \delta m_{\phi}^X = \frac{1}{8}\left(1+2\cos{\theta}\right)\left(2\cos{\theta} - \cos{\Theta}\right)\left(1-\cos{(2\phi)}\right) - \frac{\left(1 + 2\cos{\theta}\right)\left(\cos{\theta} - \cos{\Theta}\right)}{4(1 - \cos{\Theta}\cos{\theta} - \sin{\Theta}\sin{\theta}\cos{\phi})}(1 - \cos{(2\phi)}),
        \\[3ex]
        & \delta n_{x}^X \delta m_{\phi}^X = \frac{1}{8}\left(1 + 2\cos{\theta}\right)\left(2\cos{\theta} - \cos{\Theta}\right)\sin{(2\phi)} - \frac{\left(1 + 2\cos{\theta}\right)\left(\cos{\theta} - \cos{\Theta}\right)}{4(1 - \cos{\Theta}\cos{\theta } -\sin{\Theta}\sin{\theta}\cos{\phi})}\sin{(2\phi)}, 
        \\[3ex]
        & \begin{aligned}
        & \delta n_{y}^X \delta m_{\theta}^X = \frac{1}{8}\left(1 + 2\cos{\theta}\right)(1 + 2\cos{\theta}\cos{\Theta})\sin{(2\phi)} + \frac{1}{2}\left(\sin{\theta} + \sin{(2\theta)}\right)\sin{\Theta}\sin{\phi}\\
        &\qquad - \frac{\left(\sin{\theta} + \sin{(2\theta)}\right)\sin{\Theta}}{8(1 - \cos{\Theta}\cos{\theta} - \sin{\Theta}\sin{\theta}\cos{\phi})}(3\sin{\phi} - \sin{(3\phi)}).
        \end{aligned} \\
        \intertext{And then for the $Y$ polarization:}
        & \delta n_{x}^Y \delta m_{\theta}^Y = \frac{1}{8}\cos{\theta}\left(\cos{\theta} - 2\cos{\Theta}\right)\left(1 - \cos{\left(2\phi\right)}\right) + \frac{\cos{\theta}\left(\cos{\Theta} - \cos{\theta}\right)}{4(1-\cos{\Theta}\cos{\theta} - \sin{\Theta}\sin{\theta}\cos{\phi})}\left(1 - \cos{\left(2\phi\right)}\right),
        \\[3ex]
        & \delta n_{y}^Y \delta m_{\phi}^Y = \frac{1}{8}\cos{\Theta}\cos^2{\theta}\left(1 + \cos{\left(2\phi\right)}\right) + \frac{1}{8}\sin{\Theta}\sin{\left(2\theta\right)}\cos{\phi} - \frac{\sin{\Theta}\sin{\theta}\cos{\theta}}{8(1 - \cos{\Theta}\cos{\theta} - \sin{\Theta}\sin{\theta}\cos{\phi})}\left(\cos{\phi} - \cos{\left(3\phi\right)}\right),
        \\[3ex]
        & \delta n_{x}^Y \delta m_{\phi}^Y = -\frac{1}{8}\cos{\Theta}\cos^2{\theta}\sin{\left(2\phi\right)} - \frac{1}{8}\sin{\Theta}\sin{\left(2\theta\right)}\sin{\phi} + \frac{\sin{\Theta}\sin{\theta}\cos{\theta}}{8(1 - \cos{\Theta}\cos{\theta} - \sin{\Theta}\sin{\theta}\cos{\phi})}\left(3\sin{\phi} - \sin{\left(3\phi\right)}\right),
        \\[3ex]
        & \delta n_{y}^Y \delta m_{\theta}^Y = -\frac{1}{8}\cos{\theta}\left(\cos{\theta} - 2\cos{\Theta}\right)\sin{\left(2\phi\right)} - \frac{\cos{\theta}\left(\cos{\Theta} - \cos{\theta}\right)}{4(1 - \cos{\Theta}\cos{\theta}-\sin{\Theta}\sin{\theta}\cos{\phi})}\sin{\left(2\phi\right)}.
    \end{align}
\end{subequations}
Using the rewritten forms of the angular displacement products in Eqs.~\eqref{ang_prod_xy} and following similar calculations as those in \cref{app:gr_calculations}, we present the generalized overlap reduction functions for an \sgwb\ with vectorial polarizations.  
\begin{subequations}\label{eq:vectorial_results}
    \begin{align}
        \intertext{We start with the parallel correlations, which, similar to the $+$ and $\times$ polarizations, also turn out to be purely real. For the $x\theta$ component and for $m=0$ we obtain:}
        & \begin{aligned}
        & \GxthxyTh{\ell}{0} = \frac{1}{2}\sqrt{(2\ell+1)\pi}\Bigg(\frac{2}{3}\left(1 + \cosn{\Theta}\right)\delta_{\ell0} + \frac{1}{3}\delta_{\ell1} + \frac{1}{15}\left(1 + 4\cosn{\Theta}\right)\delta_{\ell2} + \frac{1}{4}\left(1-2\cosnhalf[2]{\Theta}\cscnhalf[2]{\Theta}\right)F_{1,0,\ell,0}^+ \\
        &\qquad - \frac{1}{4}F_{0,-1,\ell,0}^+ - \sinnhalf[2]{\Theta}\secnhalf[2]{\Theta}F_{2,1,\ell,0}^- + \frac{1}{4}\left(1 + 2\sinnhalf[2]{\Theta}\secnhalf[2]{\Theta}\right)F_{1,0,\ell,0}^- - \frac{1}{4}F_{0,-1,\ell,0}^-\Bigg). 
        \end{aligned} \\
        \intertext{For $m=1$:}
        & \begin{aligned}
        & \GxthxyTh{\ell}{1} = \frac{1}{2}\sqrt{(2\ell+1)\pi\frac{(\ell-1)!}{(\ell+1)!}}\Bigg(-\frac{2}{3}\sinn{\Theta}\,\delta_{\ell1} -\frac{4}{5}\sinn{\Theta}\,\delta_{\ell2} \\ 
        &\qquad + \left(-\frac{1}{4}\cosnhalf[]{\Theta}\cscnhalf[]{\Theta} + \frac{1}{4}\cosnhalf[3]{\Theta}\cscnhalf[3]{\Theta}\right)F_{2,0,\ell,1}^+ + \frac{1}{2}\cosnhalf[]{\Theta}\cscnhalf[]{\Theta}F_{1,-1,\ell,1}^+ \\
        &\qquad + \frac{1}{2}\sinnhalf[3]{\Theta}\secnhalf[3]{\Theta}F_{3,1,\ell,1}^- + \left(\frac{1}{4}\sinnhalf[]{\Theta}\secnhalf[]{\Theta} - \frac{1}{4}\sinnhalf[3]{\Theta}\secnhalf[3]{\Theta}\right)F_{2,0,\ell,1}^-\Bigg).
        \end{aligned} \\
        \intertext{For $m=2$:}
        & \begin{aligned}
        & \GxthxyTh{\ell}{2} = \frac{1}{2}\sqrt{(2\ell+1)\pi\frac{(\ell-2)!}{(\ell+2)!}}\Bigg(\frac{1}{5}\left(2 - 8\cosn{\Theta}\right)\delta_{\ell2} + \frac{1}{2}\left(1+4\cosn{\Theta}\right) + \frac{1}{2}\left(-1\right)^{\ell+1} \\ 
        &\qquad + \frac{1}{4}\cosnhalf[4]{\Theta}\cscnhalf[4]{\Theta}\left(-2 + \secnhalf[2]{\Theta}\right)F_{3,0,\ell,2}^+ + \frac{1}{8}\left(1 - 2\cscnhalf[2]{\Theta}\right)F_{2,-1,\ell,2}^+ \\
        &\qquad + \frac{1}{8}F_{1,-2,\ell,2}^+ - \frac{1}{2}\sinnhalf[4]{\Theta}\secnhalf[4]{\Theta}F_{4,1,\ell,2}^- + \frac{1}{4}\sinnhalf[2]{\Theta}\secnhalf[4]{\Theta}F_{3,0,\ell,2}^- \\
        &\qquad + \frac{1}{8}\left(1-2\secnhalf[2]{\Theta}\right)F_{2,-1,\ell,2}^- + \frac{1}{8}F_{1,-2,\ell,2}^-\Bigg).
        \end{aligned} \\
        \intertext{And for $m=3,4,\cdots$:}
        & \begin{aligned}
        & \GxthxyTh{\ell}{m} = \left(-1\right)^{m}\frac{1}{2}\sqrt{(2\ell+1)\pi\frac{(\ell-m)!}{(\ell+m)!}}\Bigg(\cosnhalf[m]{\Theta}\cscnhalf[m]{\Theta}\bigg[\left(\frac{1}{4}\cscnhalf[2]{\Theta} - \frac{1}{2}\cosnhalf[2]{\Theta}\cscnhalf[2]{\Theta}\right)F_{m+1,0,\ell,m}^+ \\
        &\qquad - \frac{1}{4}\secnhalf[2]{\Theta} F_{m,-1,\ell,m}^+ + \frac{1}{2}\sinnhalf[2]{\Theta}\secnhalf[2]{\Theta}F_{m-1,-2,\ell,m}^+\bigg] + \sinnhalf[m]{\Theta}\secnhalf[m]{\Theta} \\
        &\qquad\bigg[-\frac{1}{2}\sinnhalf[2]{\Theta}\secnhalf[2]{\Theta}F_{m+2,1,\ell,m}^- + \frac{1}{4}\secnhalf[2]{\Theta}F_{m+1,0,\ell,m}^- \\
        &\qquad + \left(-\frac{1}{4}\cscnhalf[2]{\Theta}+\frac{1}{2}\cosnhalf[2]{\Theta}\cscnhalf[2]{\Theta}\right)F_{m,-1,\ell,m}^- \bigg] \Bigg). 
        \end{aligned}\raisetag{4\baselineskip} \\
        \intertext{For the $y\phi$ component and for $m=0$ we obtain:}
        & \begin{aligned}
        & \GyphxyTh{\ell}{0} = \frac{1}{2}\sqrt{(2\ell+1)\pi}\Bigg(\left(\frac{2}{3}-\frac{1}{3}\cosn{\Theta}\right)\delta_{\ell0} + \frac{1}{3}\left(1 - \cosn{\Theta}\right)\delta_{\ell1}+\left(\frac{4}{15} + \frac{1}{15}\cosn{\Theta}\right)\delta_{\ell2} \\
        &\qquad + \left(\frac{1}{4} + \frac{1}{2}\cosnhalf[2]{\Theta}\cscnhalf[2]{\Theta}\right)F_{1,0,\ell,0}^+ - \frac{3}{4}F_{0,-1,\ell,0}^+ - \sinnhalf[2]{\Theta}\secnhalf[2]{\Theta}F_{2,1,\ell,0}^- \\
        &\qquad + \left(\frac{3}{4} + \frac{1}{2}\sinnhalf[2]{\Theta}\secnhalf[2]{\Theta}\right)F_{1,0,\ell,0}^- - \frac{1}{4}F_{0,-1,\ell,0}^-\Bigg). 
        \end{aligned} \\
        \intertext{For $m=1$:}
        & \begin{aligned}
        & \GyphxyTh{\ell}{1} = \frac{1}{2}\sqrt{(2\ell+1)\pi\frac{(\ell-1)!}{(\ell+1)!}}\Bigg(-\frac{1}{5}\sinn{\Theta}\,\delta_{\ell2} - \frac{1}{4}\cosnhalf[]{\Theta}\cscnhalf[3]{\Theta}F_{2,0,\ell,1}^+ + \frac{1}{2}\cosnhalf[]{\Theta}\cscnhalf[]{\Theta}F_{1,-1,\ell,1}^+ \\
        &\qquad + \frac{1}{2}\sinnhalf[3]{\Theta}\secnhalf[3]{\Theta}F_{3,1,\ell,1}^- - \frac{1}{4}\sinnhalf[]{\Theta}\secnhalf[3]{\Theta}F_{2,0,\ell,1}^-\Bigg).
        \end{aligned} \\
        \intertext{For $m=2$:}
        & \begin{aligned}
        & \GyphxyTh{\ell}{2} = \frac{1}{2}\sqrt{(2\ell+1)\pi\frac{(\ell-2)!}{(\ell+2)!}}\Bigg(\left(\frac{8}{5}-\frac{2}{5}\cosn{\Theta}\right)\delta_{\ell2} + \left(-\frac{3}{2}+\cosn{\Theta}\right) + \frac{1}{2}\left(-1\right)^{\ell+1} \\
        &\qquad + \frac{1}{4}\cosnhalf[2]{\Theta}\cscnhalf[4]{\Theta}F_{3,0,\ell,2}^++\left(-\frac{1}{8} - \frac{3}{4}\cosnhalf[2]{\Theta}\cscnhalf[2]{\Theta}\right)F_{2,-1,\ell,2}^+ + \frac{3}{8}F_{1,-2,\ell,2}^+ \\
        &\qquad - \frac{1}{2}\sinnhalf[4]{\Theta}\secnhalf[4]{\Theta}F_{4,1,\ell,2}^- + \left(\frac{3}{4}\sinnhalf[2]{\Theta}\secnhalf[2]{\Theta} + \frac{1}{4}\sinnhalf[4]{\Theta}\secnhalf[4]{\Theta}\right)F_{3,0,\ell,2}^- \\
        &\qquad + \left(-\frac{3}{8} - \frac{1}{4}\sinnhalf[2]{\Theta}\secnhalf[2]{\Theta}\right)F_{2,-1,\ell,2}^- + \frac{1}{8}F_{1,-2,\ell,2}^-\Bigg).
        \end{aligned} \\
        \intertext{And for $m=3,4,\cdots$:}
        & \begin{aligned}
        & \GyphxyTh{\ell}{m} = (-1)^{m}\frac{1}{2}\sqrt{(2\ell+1)\pi\frac{(\ell-m)!}{(\ell+m)!}}\Bigg(\cosnhalf[m]{\Theta}\cscnhalf[m]{\Theta}\bigg[\frac{1}{4}\cscnhalf[2]{\Theta}F_{m+1,0,\ell,m}^+ \\
        &\qquad + \left(-\frac{3}{4}-\frac{1}{4}\sinnhalf[2]{\Theta}\secnhalf[2]{\Theta}\right)F_{m,-1,\ell,m}^+ + \frac{1}{2}\sinnhalf[2]{\Theta}\secnhalf[2]{\Theta}F_{m-1,-2,\ell,m}^+\bigg] \\
        &\qquad + \sinnhalf[m]{\Theta}\secnhalf[m]{\Theta}\bigg[-\frac{1}{2}\sinnhalf[2]{\Theta}\secnhalf[2]{\Theta}F_{m+2,1,\ell,m}^- \\
        &\qquad + \left(\frac{3}{4} + \frac{1}{4}\sinnhalf[2]{\Theta}\secnhalf[2]{\Theta}\right)F_{m+1,0,\ell,m}^- - \frac{1}{4}\cscnhalf[2]{\Theta}F_{m,-1,\ell,m}^-\bigg]\Bigg).
        \end{aligned} \\
        \intertext{The perpendicular correlations are purely imaginary. For the $x\phi$ component and for $m=0$ the correlations vanish:}
        & \GxphxyTh{\ell}{0} = 0. \\
        \intertext{For $m=1$ we obtain:}
        & \begin{aligned}
        & \GxphxyTh{\ell}{1} = \frac{\I}{2}\sqrt{(2\ell+1)\pi\frac{(\ell-1)!}{(\ell+1)!}}\Bigg(\frac{1}{5}\sinn{\Theta}\,\delta_{\ell2} + \left(\frac{1}{4}\cosnhalf[]{\Theta}\cscnhalf[]{\Theta} - \frac{1}{4}\cosnhalf[3]{\Theta}\cscnhalf[3]{\Theta}\right)F_{2,0,\ell,1}^+ \\
        &\qquad + \frac{1}{2}\sinnhalf[3]{\Theta}\secnhalf[3]{\Theta}F_{3,1,\ell,1}^- + \left(-\frac{3}{4}\sinnhalf[]{\Theta}\secnhalf[]{\Theta} - \frac{1}{4}\sinnhalf[3]{\Theta}\secnhalf[3]{\Theta}\right)F_{2,0,\ell,1}^- \\
        &\qquad + \frac{1}{2}\sinnhalf[]{\Theta}\secnhalf[]{\Theta}F_{1,-1,\ell,1}^-\Bigg).
        \end{aligned} \\
        \intertext{For $m=2$:}
        & \begin{aligned}
        & \GxphxyTh{\ell}{2} = \frac{\I}{2}\sqrt{(2\ell+1)\pi\frac{(\ell-2)!}{(\ell+2)!}}\Bigg(\left(-\frac{8}{5} + \frac{2}{5}\cosn{\Theta}\right)\delta_{\ell2} + \left(\frac{3}{2}-\cosn{\Theta}\right) + \frac{1}{2}\left(-1\right)^{\ell} \\ 
        &\qquad + \left(-\frac{1}{4}\cosnhalf[2]{\Theta}\cscnhalf[2]{\Theta} + \frac{1}{4}\cosnhalf[4]{\Theta}\cscnhalf[4]{\Theta}\right)F_{3,0,\ell,2}^+ \\
        &\qquad + \left(\frac{1}{8}+\frac{1}{4}\cosnhalf[2]{\Theta}\cscnhalf[2]{\Theta}\right)F_{2,-1,\ell,2}^+ - \frac{3}{8}F_{1,-2,\ell,2}^+ - \frac{1}{2}\sinnhalf[4]{\Theta}\secnhalf[4]{\Theta}F_{4,1,\ell,2}^- \\
        &\qquad + \left(\frac{1}{4}\sinnhalf[2]{\Theta}\secnhalf[2]{\Theta} + \frac{1}{4}\sinnhalf[4]{\Theta}\secnhalf[4]{\Theta}\right)F_{3,0,\ell,2}^- \\
        &\qquad + \left(\frac{3}{8} - \frac{1}{4}\sinnhalf[2]{\Theta}\secnhalf[2]{\Theta}\right)F_{2,-1,\ell,2}^- - \frac{1}{8}F_{1,-2,\ell,2}^-\Bigg).
        \end{aligned} \\
        \intertext{And for $m=3,4,\cdots$:}
        & \begin{aligned}
        & \GxphxyTh{\ell}{m} = \left(-1\right)^{m}\frac{\I}{2}\sqrt{(2\ell+1)\pi\frac{(\ell-m)!}{(\ell+m)!}}\Bigg(\cosnhalf[m]{\Theta}\cscnhalf[m]{\Theta}\bigg[\left(-\frac{1}{4}\cscnhalf[2]{\Theta} + \frac{1}{2}\cosnhalf[2]{\Theta}\cscnhalf[2]{\Theta}\right)F_{m+1,0,\ell,m}^+ \\ 
        &\qquad + \frac{1}{4}\secnhalf[2]{\Theta}F_{m,-1,\ell,m}^+ - \frac{1}{2}\sinnhalf[2]{\Theta}\secnhalf[2]{\Theta}F_{m-1,-2,\ell,m}^+\bigg] + \sinnhalf[m]{\Theta}\secnhalf[m]{\Theta} \\
        &\qquad \bigg[-\frac{1}{2}\sinnhalf[2]{\Theta}\secnhalf[2]{\Theta}F_{m+2,1,\ell,m}^- + \frac{1}{4}\secnhalf[2]{\Theta}F_{m+1,0,\ell,m}^- \\
        &\qquad + \left(-\frac{1}{4}\cscnhalf[2]{\Theta} + \frac{1}{2}\cosnhalf[2]{\Theta}\cscnhalf[2]{\Theta}\right)F_{m,-1,\ell,m}^- \bigg] \Bigg).
        \end{aligned}\raisetag{4\baselineskip} \\
        \intertext{For the $y\theta$ component and for $m=0$ the correlations also vanish:}
        & \GythxyTh{\ell}{0} = 0. \\
        \intertext{For $m=1$ we obtain:}
        & \begin{aligned}
        & \GythxyTh{\ell}{1} = \frac{\I}{2}\sqrt{(2\ell+1)\pi\frac{(\ell-1)!}{(\ell+1)!}}\Bigg(-\frac{2}{3}\sinn{\Theta}\,\delta_{\ell1} - \frac{4}{5}\sinn{\Theta}\,\delta_{\ell2} - \frac{1}{4}\cosnhalf[]{\Theta}\cscnhalf[3]{\Theta}F_{2,0,\ell,1}^+ \\ 
        &\qquad + \cosnhalf[]{\Theta}\cscnhalf[]{\Theta}F_{1,-1,\ell,1}^+ - \frac{1}{2}\sinnhalf[3]{\Theta}\secnhalf[3]{\Theta}F_{3,1,\ell,1}^- \\
        &\qquad + \left(\sinnhalf[]{\Theta}\secnhalf[]{\Theta} + \frac{1}{4}\sinnhalf[]{\Theta}\secnhalf[3]{\Theta}\right)F_{2,0,\ell,1}^- - \frac{1}{2}\sinnhalf[]{\Theta}\secnhalf[]{\Theta}F_{1,-1,\ell,1}^-\Bigg). 
        \end{aligned} \\
        \intertext{For $m=2$:}
        & \begin{aligned}
        & \GythxyTh{\ell}{2} = \frac{\I}{2}\sqrt{(2\ell+1)\pi\frac{(\ell-2)!}{(\ell+2)!}}\Bigg(\frac{1}{5}\left(2 - 8\cosn{\Theta}\right)\delta_{\ell2} + \frac{1}{2}\left(1 + 4\cosn{\Theta}\right) + \frac{1}{2}\left(-1\right)^{\ell+1} \\
        &\qquad + \frac{1}{4}\cosnhalf[2]{\Theta}\cscnhalf[4]{\Theta}F_{3,0,\ell,2}^+ + \left(-\frac{1}{8} - \frac{3}{4}\cosnhalf[2]{\Theta}\cscnhalf[2]{\Theta}\right)F_{2,-1,\ell,2}^+ + \frac{1}{8}F_{1,-2,\ell,2}^+ \\
        &\qquad + \frac{1}{2}\sinnhalf[4]{\Theta}\secnhalf[4]{\Theta}F_{4,1,\ell,2}^- + \left(-\frac{3}{4}\sinnhalf[2]{\Theta}\secnhalf[2]{\Theta} - \frac{1}{4}\sinnhalf[4]{\Theta}\secnhalf[4]{\Theta}\right)F_{3,0,\ell,2}^- \\
        &\qquad + \left(-\frac{1}{8} + \frac{1}{4}\sinnhalf[2]{\Theta}\secnhalf[2]{\Theta}\right)F_{2,-1,\ell,2}^- + \frac{1}{8}F_{1,-2,\ell,2}^- \Bigg).
        \end{aligned} \\
        \intertext{And for $m=3,4,\cdots$:}
        & \begin{aligned}
        & \GythxyTh{\ell}{m} = (-1)^{m}\frac{\I}{2}\sqrt{(2\ell+1)\pi\frac{(\ell-m)!}{(\ell+m)!}}\Bigg(\cosnhalf[m]{\Theta}\cscnhalf[m]{\Theta}\bigg[\frac{1}{4}\cscnhalf[2]{\Theta}F_{m+1,0,\ell,m}^+ \\ 
        &\qquad + \left(-\frac{3}{4}-\frac{1}{4}\sinnhalf[2]{\Theta}\secnhalf[2]{\Theta}\right)F_{m,-1,\ell,m}^+ + \frac{1}{2}\sinnhalf[2]{\Theta}\secnhalf[2]{\Theta}F_{m-1,-2,\ell,m}^+\bigg] \\
        &\qquad + \sinnhalf[m]{\Theta}\secnhalf[m]{\Theta}\bigg[\frac{1}{2}\sinnhalf[2]{\Theta}\secnhalf[2]{\Theta}F_{m+2,1,\ell,m}^- + \left(-\frac{3}{4} - \frac{1}{4}\sinnhalf[2]{\Theta}\secnhalf[2]{\Theta}\right)F_{m+1,0,\ell,m}^- \\
        &\qquad + \frac{1}{4}\cscnhalf[2]{\Theta}F_{m,-1,\ell,m}^-\bigg]\Bigg).
        \end{aligned}
    \end{align}
\end{subequations}
For negative values of \(m\), the astrometric correlations are related to those with positive $m$ via

\vspace{0pt}
\begin{subequations}
\noindent
\begin{minipage}[t]{0.5\linewidth}
    \setlength{\abovedisplayskip}{0pt}
    \setlength{\belowdisplayskip}{0pt}
    \begin{align}
        \GxthxyTh{\ell}{-m}  &= \left(-1\right)^{m}\GxthxyTh{\ell}{m}, \\
        \GyphxyTh{\ell}{-m}  &= \left(-1\right)^{m}\GyphxyTh{\ell}{m},
    \end{align}
\end{minipage}%
\hfill
\begin{minipage}[t]{0.5\linewidth}
    \setlength{\abovedisplayskip}{0pt}
    \setlength{\belowdisplayskip}{0pt}
    \begin{align}
        \GxphxyTh{\ell}{-m}  &= \left(-1\right)^{m+1}\GxphxyTh{\ell}{m}, \\
        \GythxyTh{\ell}{-m}  &= \left(-1\right)^{m+1}\GythxyTh{\ell}{m}.
    \end{align}
\end{minipage}
\end{subequations}
\par
\vspace{\belowdisplayskip}
\noindent
The parallel and perpendicular generalized overlap reduction functions up to $\ell = 3$ and $m = 3$ are plotted in \autoref{fig:XY_parallel_results} and \autoref{fig:XY_perpendicular_results} in the main text, respectively.

\section{Calculation of astrometric integrals for scalar longitudinal polarization}
\label[appendix]{app:l_calculations}
\noindent
In this appendix, we show how to calculate the generalized overlap reduction functions for an \sgwb\ that contains \gw s with scalar longitudinal polarization. The relevant integral is
\begin{align}
    \Gij[L][\Theta][\ell][m] &= \int_{S^2}\mathrm{d}\Omega_{\vq}\,Y_{\ell}^m\left(\vq\right)\delta n_i^L\left(\vq\right)\delta m_j^L\left(\vq\right)=N^m_{\ell}\int_{0}^{\pi}\mathrm{d}\theta\,\sin{\theta}\,P_{\ell}^m\left(\cos{\theta}\right)\int_0^{2\pi}\mathrm{d}\phi\,e^{\I m\phi}\delta n_i^L\left(\vq\right)\delta m_j^L\left(\vq\right),
\end{align}
where $\vq=\left(\sin{\theta}\cos{\phi},\sin{\theta}\sin{\phi},\cos{\theta}\right)$ is the direction from which the \gw s originate. $Y_{\ell}^m\left(\vq\right)=N^m_{\ell}P^m_{\ell}\left(\cos{\theta}\right)e^{\I m\phi}$ are the spherical harmonic functions where $N^m_{\ell}$ are normalization constants and $P^m_{\ell}\left(\cos{\theta}\right)$ are the associated Legendre polynomials defined in \cref{app:math}. If we define $x\equiv\cos{\theta}$, we can rewrite the above integral as follows
\begin{align}
    \Gij[L][\Theta][\ell][m] &= N^m_{\ell}\int_{-1}^{1}\mathrm{d}x\,P_{\ell}^m\left(x\right)\int_0^{2\pi}\mathrm{d}\phi\,e^{\I m\phi}\delta n_i^L\left(\vq\right)\delta m_j^L\left(\vq\right).
\end{align}
Using calculations identical to those for the \gr\ polarizations in \cref{app:gr_calculations}, except that we now use the polarization tensor in Eq.~\eqref{subeq:l_pol}, we obtain the angular deflection components to be (Eqs.~(F2c) and (F2d) in \rcite{Mihaylov:2018uqm} have typos):
\begin{subequations}
    \begin{align}
         & \delta n^L_x\left(\theta, \phi\right) = -\frac{1}{\sqrt{2}}\frac{\sin{\theta}\cos{\theta}\cos{\phi}}{1-\cos{\theta}}, 
         \\[3ex]
         & \delta n_y^L\left(\theta,\phi\right) = -\frac{1}{\sqrt{2}}\frac{\sin{\theta}\cos{\theta}\sin{\phi}}{1-\cos{\theta}}, 
         \\[3ex]
         & \begin{aligned}
         & \delta m_{\theta}^L\left(\theta,\phi\right) = \frac{1}{\sqrt{2}\left(1-\cos{\Theta}\cos{\theta}-\sin{\Theta}\sin{\theta}\cos{\phi}\right)}\Bigg(\frac{1}{8}\sin{\left(2\Theta\right)}\left(1+3\cos{\left(2\theta\right)}\right)-\frac{1}{2}\cos{\left(2\Theta\right)}\sin{\left(2\theta\right)}\cos{\phi} \\
         &\qquad + \frac{1}{8}\sin{\left(2\Theta\right)}\left(\cos{\left(2\theta\right)} - 1\right)\cos{\left(2\phi\right)}\Bigg), 
         \end{aligned} \\[3ex]
         & \delta m_{\phi}^L\left(\theta,\phi\right) = - \frac{1}{\sqrt{2}\left(1-\cos{\Theta}\cos{\theta}-\sin{\Theta}\sin{\theta}\cos{\phi}\right)}\Bigg(\frac{1}{2}\cos{\Theta}\sin{\left(2\theta\right)}\sin{\phi}+\frac{1}{4}\sin{\Theta}\left(1-\cos{\left(2\theta\right)}\right)\sin{\left(2\phi\right)}\Bigg).
    \end{align}
\end{subequations}
The products of the angular displacements can be rewritten as:
\begin{subequations}\label{eq:ang_prod_l}
    \begin{align}
        & \begin{aligned}
        & \delta n_x^L \delta m_{\theta}^L = -\frac{1}{2}\cosnhalf[2]{\theta}\cos{\theta}\cos{\Theta}+\frac{\left(\sin{\Theta}\cos{\theta} - \cot{\Theta}\right)\sin{\theta}\cos{\theta}}{2\left(1 - \cos{\theta}\right)}\cos{\phi} - \frac{1}{2}\cosnhalf[2]{\theta}\cos{\theta}\cos{\Theta}\cos{\left(2\phi\right)} \\
        &\qquad - \frac{\csc{\Theta}\left(\cos{\theta} - \cos{\Theta}\right)\sin{\theta}\cos{\theta}}{2\left(1 - \cos{\theta}\right)\left(1 - \cos\Theta \cos\theta - \sin\Theta \sin\theta \cos\phi\right)}\cos{\phi},
        \end{aligned} \\[6ex]
        & \delta n_y^L \delta m_{\phi}^L = -\frac{1}{2}\cos^2{\left(\theta/2\right)}\cos{\theta} + \frac{1}{2}\cos^2{\left(\theta/2\right)}\cos{\theta}\cos{\left(2\phi\right)} + \frac{\cos^2{\left(\theta/2\right)}\cos{\theta}}{2\left(1 - \cos\Theta \cos\theta - \sin\Theta \sin\theta \cos\phi\right)}\left(1 - \cos{\left(2\phi\right)}\right),
        \\[6ex]
        & \delta n_x^L \delta m_{\phi}^L = -\frac{1}{2}\cos^2{\left(\theta/2\right)}\cos{\theta}\sin{\left(2\phi\right)} + \frac{\cos^2{\left(\theta/2\right)}\cos{\theta}}{2\left(1 - \cos\Theta \cos\theta - \sin\Theta \sin\theta \cos\phi\right)}\sin{\left(2\phi\right)},
        \\[6ex]
        & \begin{aligned}
        & \delta n_y^L \delta m_{\theta}^L = \frac{\left(\sin{\Theta}\cos{\theta} - \cot{\Theta}\right)\sin{\theta}\cos{\theta}}{2\left(1 - \cos{\theta}\right)}\sin{\phi} - \frac{1}{2}\cosnhalf[2]{\theta}\cos{\theta}\cos{\Theta}\sin{\left(2\phi\right)} \\
        &\qquad - \frac{\csc{\Theta}\left(\cos{\theta} - \cos{\Theta}\right)\sin{\theta}\cos{\theta}}{2\left(1 - \cos{\theta}\right)\left(1 - \cos\Theta \cos\theta - \sin\Theta \sin\theta \cos\phi\right)}\sin{\phi}.
        \end{aligned}
    \end{align}
\end{subequations}
Using the rewritten forms of the angular displacement products in Eqs.~\eqref{eq:ang_prod_l} and similar calculations as those in \cref{app:gr_calculations}, we present the generalized overlap reduction functions for an \sgwb\ with \gw s with scalar longitudinal polarization.
\begin{subequations}\label{eq:scalar_longitudinal_results}
    \begin{align}
        \intertext{For the $x\theta$ component and for $m=0$:}
        & \begin{aligned}
        & \GxthlTh{\ell}{0} = \frac{1}{2}\sqrt{\left(2\ell+1\right)\pi}\Bigg(-\frac{1}{3}\cos{\Theta}\,\delta_{\ell0} - \frac{1}{3}\cos{\Theta}\,\delta_{\ell1} - \frac{2}{15}\cos{\Theta}\,\delta_{\ell2} + \frac{1}{4}\cscnhalf[2]{\Theta}F_{1,0,\ell,0}^+ \\
        &\qquad -\frac{1}{4}\cscnhalf[2]{\Theta}F_{0,-1,\ell,0}^+ + \frac{1}{4}\secnhalf[2]{\Theta}F_{2,1,\ell,0}^- - \frac{1}{4}\secnhalf[2]{\Theta}F_{1,0,\ell,0}^-\Bigg).
        \end{aligned} \\
        \intertext{For $m=1,2,\cdots$:}
        & \begin{aligned}
        & \GxthlTh{\ell}{m} = \frac{1}{2}\sqrt{\left(2\ell+1\right)\pi\frac{\left(\ell-m\right)!}{\left(\ell+m\right)!}}\Bigg(\delta_{m1}\bigg[\left(\cot{\Theta} - \sin{\Theta}\right) - \frac{2}{3}\left(\cot{\Theta} -\sin{\Theta}\right)\delta_{\ell1} + \frac{2}{5}\sin{\Theta}\,\delta_{\ell2}\bigg] \\
        &\qquad + \delta_{m2}\bigg[-\cos{\Theta} + \frac{4}{5}\cos{\Theta}\,\delta_{\ell2}\bigg] + \left(-1\right)^m\cosnhalf[m]{\Theta}\cscnhalf[m]{\Theta}\Bigg[\frac{1}{8}\cscnhalf[2]{\Theta}F_{m+1,0,\ell,m}^+ \\
        &\qquad + \left(\frac{1}{8}\secnhalf[2]{\Theta} - \frac{1}{8}\cscnhalf[2]{\Theta}\right)F_{m,-1,\ell,m}^+ - \frac{1}{8}\secnhalf[2]{\Theta}F_{m-1,-2,\ell,m}^+\Bigg] + \left(-1\right)^m\sinnhalf[m]{\Theta}\secnhalf[m]{\Theta} \\
        &\qquad \Bigg[\frac{1}{8}\secnhalf[2]{\Theta}F_{m+2,1,\ell,m}^- + \left(\frac{1}{8}\cscnhalf[2]{\Theta} - \frac{1}{8}\secnhalf[2]{\Theta}\right)F_{m+1,0,\ell,m}^- -\frac{1}{8}\cscnhalf[2]{\Theta}F_{m,-1,\ell,m}^-\Bigg]\Bigg).
        \end{aligned}\raisetag{4.75\baselineskip} \\
        \intertext{For the $y\phi$ component and for $m=0$:}
        & \begin{aligned}
        & \GyphlTh{\ell}{0} = \frac{1}{2}\sqrt{\left(2\ell+1\right)\pi}\Bigg(-\frac{1}{3}\delta_{\ell0} - \frac{1}{3}\delta_{\ell1} - \frac{2}{15}\delta_{\ell2} - \frac{1}{4}\cscnhalf[2]{\Theta}F_{1,0,\ell,0}^+ + \frac{1}{4}\cscnhalf[2]{\Theta}F_{0,-1,\ell,0}^+ \\
        &\qquad + \frac{1}{4}\secnhalf[2]{\Theta}F_{2,1,\ell,0}^- - \frac{1}{4}\secnhalf[2]{\Theta}F_{1,0,\ell,0}^-\Bigg). 
        \end{aligned} \\
        \intertext{For $m=1$:}
        & \begin{aligned}
        & \GyphlTh{\ell}{1} = \frac{1}{2}\sqrt{\left(2\ell+1\right)\pi\frac{\left(\ell-1\right)!}{\left(\ell+1\right)!}}\Bigg(\frac{1}{8}\cosnhalf[]{\Theta}\cscnhalf[3]{\Theta}F_{2,0,\ell,1}^+ - \frac{1}{8}\cosnhalf[]{\Theta}\cscnhalf[3]{\Theta}F_{1,-1,\ell,1}^+ \\
        &\qquad - \frac{1}{8}\sinnhalf[]{\Theta}\secnhalf[3]{\Theta}F_{3,1,\ell,1}^- + \frac{1}{8}\sinnhalf[]{\Theta}\secnhalf[3]{\Theta}F_{2,0,\ell,1}^-\Bigg).
        \end{aligned} \\
        \intertext{And for $m=2,3,\cdots$:}
        & \begin{aligned}
        & \GyphlTh{\ell}{m} = \frac{1}{2}\sqrt{\left(2\ell+1\right)\pi\frac{\left(\ell-m\right)!}{\left(\ell+m\right)!}}\Bigg(\delta_{m 2}\left(1 - \frac{4}{5}\delta_{\ell2}\right) + \left(-1\right)^m\cosnhalf[m]{\Theta}\cscnhalf[m]{\Theta}\Bigg[-\frac{1}{8}\cscnhalf[2]{\Theta}F_{m+1,0,\ell,m}^+ \\
        &\qquad + \frac{1}{8}\cscnhalf[2]{\Theta}\secnhalf[2]{\Theta}F_{m,-1,\ell,m}^+ - \frac{1}{8}\secnhalf[2]{\Theta}F_{m-1,-2,\ell,m}^+\Bigg] + \left(-1\right)^m\sinnhalf[m]{\Theta}\secnhalf[m]{\Theta} \\
        &\qquad \Bigg[\frac{1}{8}\secnhalf[2]{\Theta}F_{m+2,1,\ell,m}^- - \frac{1}{8}\cscnhalf[2]{\Theta}\secnhalf[2]{\Theta}F_{m+1,0,\ell,m}^- + \frac{1}{8}\cscnhalf[2]{\Theta}F_{m,-1,\ell,m}^-\Bigg]\Bigg).
        \end{aligned} \\
        \intertext{For the $x\phi$ component and for $m=0$:}
        & \GxphlTh{\ell}{0} = 0. \\
        \intertext{For $m=1$:}
        & \begin{aligned}
        & \GxphlTh{\ell}{1} = \frac{\I}{2}\sqrt{\left(2\ell+1\right)\pi\frac{\left(\ell-1\right)!}{\left(\ell+1\right)!}}\Bigg(\frac{1}{8}\cosnhalf[]{\Theta}\cscnhalf[3]{\Theta}F_{2,0,\ell,1}^+ - \frac{1}{8}\cosnhalf[]{\Theta}\cscnhalf[3]{\Theta}F_{1,-1,\ell,1}^+ \\
        &\qquad - \frac{1}{8}\sinnhalf[]{\Theta}\secnhalf[3]{\Theta}F_{3,1,\ell,1}^- + \frac{1}{8}\sinnhalf[]{\Theta}\secnhalf[3]{\Theta}F_{2,0,\ell,1}^-\Bigg).
        \end{aligned} \\
        \intertext{And for $m=2,3,\cdots$:}
        & \GxphlTh{\ell}{m} = \frac{\I}{2}\sqrt{\left(2\ell+1\right)\pi\frac{\left(\ell-m\right)!}{\left(\ell+m\right)!}}\Bigg(\delta_{m 2}\left(-1 + \frac{4}{5}\delta_{\ell2}\right) + \left(-1\right)^m\cosnhalf[m]{\Theta}\cscnhalf[m]{\Theta}\Bigg[-\frac{1}{8}\cscnhalf[2]{\Theta}F_{m+1,0,\ell,m}^+\notag\\
        &\qquad + \left(\frac{1}{8}\cscnhalf[2]{\Theta}-\frac{1}{8}\secnhalf[2]{\Theta}\right)F_{m,-1,\ell,m}^+ + \frac{1}{8}\secnhalf[2]{\Theta}F_{m-1,-2,\ell,m}^+\Bigg] + \left(-1\right)^m\sinnhalf[m]{\Theta}\secnhalf[m]{\Theta}\\
        &\qquad\Bigg[\frac{1}{8}\secnhalf[2]{\Theta}F_{m+2,1,\ell,m}^- + \left(\frac{1}{8}\cscnhalf[2]{\Theta} - \frac{1}{8}\secnhalf[2]{\Theta}\right)F_{m+1,0,\ell,m}^-
        -\frac{1}{8}\cscnhalf[2]{\Theta}F_{m,-1,\ell,m}^-\Bigg]\Bigg).
        \notag
        \intertext{For the $y\theta$ component and for $m=0$:}
        & \GythlTh{\ell}{0} = 0.
        \\
        \intertext{{For $m=1, 2, \cdots$:}}
        & \begin{aligned}
        & \GythlTh{\ell}{m} = \frac{\I}{2}\sqrt{\left(2\ell+1\right)\pi\frac{\left(\ell-m\right)!}{\left(\ell+m\right)!}}\Bigg(\delta_{m1}\bigg[\left(\cot{\Theta} - \sin{\Theta}\right) - \frac{2}{3}\left(\cot{\Theta} -\sin{\Theta}\right)\delta_{\ell1} + \frac{2}{5}\sin{\Theta}\,\delta_{\ell2}\bigg] \\
        &\qquad + \delta_{m2}\bigg[-\cos{\Theta} + \frac{4}{5}\cos{\Theta}\,\delta_{\ell2}\bigg] + \left(-1\right)^m\cosnhalf[m]{\Theta}\cscnhalf[m]{\Theta}\Bigg[-\frac{1}{8}\cscnhalf[2]{\Theta}F_{m+1,0,\ell,m}^+ \\
        &\qquad + \frac{1}{8}\cscnhalf[2]{\Theta}\secnhalf[2]{\Theta}F_{m,-1,\ell,m}^+ - \frac{1}{8}\secnhalf[2]{\Theta}F_{m-1,-2,\ell,m}^+\Bigg]+\left(-1\right)^m\sinnhalf[m]{\Theta}\secnhalf[m]{\Theta} \\
        &\qquad \Bigg[-\frac{1}{8}\secnhalf[2]{\Theta}F_{m+2,1,\ell,m}^- + \frac{1}{8}\cscnhalf[2]{\Theta}\secnhalf[2]{\Theta}F_{m+1,0,\ell,m}^- - \frac{1}{8}\cscnhalf[2]{\Theta}F_{m,-1,\ell,m}^-\Bigg]\Bigg).
        \end{aligned}\raisetag{4.75\baselineskip}
    \end{align}
\end{subequations}
As mentioned in the main text, $\GxthlTh{\ell}{0}$ and $\GyphlTh{\ell}{0}$ diverge as $\Theta\rightarrow0$. The results for negative $m$ are related to those with positive $m$ via:

\vspace{0pt}
\begin{subequations}
\noindent
\begin{minipage}[t]{0.5\linewidth}
    \setlength{\abovedisplayskip}{0pt}
    \setlength{\belowdisplayskip}{0pt}
    \begin{align}
        \GxthlTh{\ell}{-m} &= \left(-1\right)^m\GxthlTh{\ell}{m},\\
        \GyphlTh{\ell}{-m} &= \left(-1\right)^m\GyphlTh{\ell}{m},
    \end{align}
\end{minipage}%
\hfill
\begin{minipage}[t]{0.5\linewidth}
    \setlength{\abovedisplayskip}{0pt}
    \setlength{\belowdisplayskip}{0pt}
    \begin{align}
        \GxphlTh{\ell}{-m} &= \left(-1\right)^{m+1}\GxphlTh{\ell}{m},\\
        \GythlTh{\ell}{-m} &= \left(-1\right)^{m+1}\GythlTh{\ell}{m}.
    \end{align}
\end{minipage}
\end{subequations}
\par
\vspace{\belowdisplayskip}
\noindent
All astrometric correlations for $m\geq0$ up to $\ell=3$ and $m=3$ are plotted in \autoref{fig:gLxtheta_results} and \autoref{fig:gLxphi_results} in the main text.

\subsection{Codirectional case}
\label[appendix]{app:scalar_long_y0}
\noindent
The generalized overlap reduction functions obtained by using the exact astrometric response in Eq.~\eqref{eq:triangle_exact} are difficult to solve analytically. 
However, the calculations are much easier when we consider the codirectional case where $\Theta=0$. 
In this subsection, we calculate the exact generalized overlap reduction functions in Eq.~\eqref{eq:gorf_complex} in the codirectional case for the scalar longitudinal polarization for the $x\theta$ and $y\phi$ components when $m=0$, for which we get divergent expressions in the distant-light source approximation as shown in \autoref{fig:gLxtheta_results}. 
Our calculations are similar to those of Appendices H and J.1 in \rcite{Gair:2015hra} where the authors calculate the generalized overlap reduction functions in the context of \pta s for the codirectional case for the vector polarizations for $m=0$ and for the scalar longitudinal polarization for all $m$ (for \pta s these correlations diverge in the distant-light source approximation). The relevant integrals are 
\begin{subequations}\label{eq:scalar_l_exact_integrals}
    \begin{align}
        \mathcal{L}^L_{x\theta, \ell 0}\left(\dn,\dm,\Theta=0\right)&=N^0_{\ell}\int_{-1}^{1}\mathrm{d}x\,P_{\ell}\left(x\right)\int_0^{2\pi}\mathrm{d}\phi\,\delta n_x^L\left(\vq,\dn\right)\delta m_{\theta}^{L*}\left(\vq,\dm\right),\\
        \mathcal{L}^L_{y\phi,\ell 0}\left(\dn,\dm,\Theta=0\right)&=N^0_{\ell}\int_{-1}^{1}\mathrm{d}x\,P_{\ell}\left(x\right)\int_0^{2\pi}\mathrm{d}\phi\,\delta n_y^L\left(\vq,\dn\right)\delta m_{\phi}^{L*}\left(\vq,\dm\right).
    \end{align}
\end{subequations}
where as in the previous cases we have done the substitution $x\equiv\cos{\theta}$. Substituting Eq.~\eqref{eq:triangle_exact} and Eq.~\eqref{subeq:l_pol} into Eq.~\eqref{subeq:exact_response_definition}, we obtain the exact angular response functions to be:
\begin{subequations}\label{eq:scalar_l_exact_responses}
    \begin{align}
        \delta n_x^L\left(\vq,\dn\right) &= -\sqrt{2}\left(\frac{1}{2}-\frac{ \I \left(1-e^{ \I \dn\left(1-x\right)}\right)}{\dn\left(1-x\right)}\right)\sqrt{1-x^2}\,x\,\cos\phi+\left(-1+\frac{ \I \left(1-e^{ \I \,\dn\left(1-x\right)}\right)}{\dn\left(1-x\right)}\right)\frac{\sqrt{1-x^2}\,x^{2}\,\cos\phi}{\sqrt{2}\left(1-x\right)},\\
        \delta m_{\theta}^{L*}\left(\vq,\dm\right)&=-\sqrt{2}\left(\frac{1}{2}+\frac{ \I \left(1-e^{- \I \,\dm\left(1-x\right)}\right)}{\dm\left(1-x\right)}\right)\sqrt{1-x^2}\,x\,\cos\phi+\left(-1-\frac{ \I \left(1-e^{- \I \,\dm\left(1-x\right)}\right)}{\dm\left(1-x\right)}\right)\frac{\sqrt{1-x^2}\,x^2\,\cos\phi}{\sqrt{2}\left(1-x\right)},\\
        \delta n_y^L\left(\vq,\dn\right)&=-\sqrt{2}\left(\frac{1}{2}-\frac{ \I \left(1-e^{ \I \,\dn\left(1-x\right)}\right)}{\dn\left(1-x\right)}\right)\sqrt{1-x^2}\,x\,\sin\phi+\left(-1+\frac{ \I \left(1-e^{ \I \,\dn\left(1-x\right)}\right)}{\dn\left(1-x\right)}\right)\frac{\sqrt{1-x^2}\,x^2\,\sin\phi}{\sqrt{2}\left(1-x\right)},\\
        \delta m_{\phi}^{L*}\left(\vq,\dm\right)&=-\sqrt{2}\left(\frac{1}{2}+\frac{ \I \left(1-e^{- \I \,\dm\left(1-x\right)}\right)}{\dm\left(1-x\right)}\right)\sqrt{1-x^2}\,x\,\sin\phi+\left(-1-\frac{ \I \left(1-e^{- \I \,\dm\left(1-x\right)}\right)}{\dm\left(1-x\right)}\right)\frac{\sqrt{1-x^2}\,x^2\,\sin\phi}{\sqrt{2}\left(1-x\right)}.
    \end{align}
\end{subequations}

The integral over $\phi$ is trivial and it can easily be seen that $\mathcal{L}^L_{x\theta, \ell 0}\left(\dn,\dm,\Theta=0\right)=\mathcal{L}^L_{y\phi, \ell 0}\left(\dn,\dm,\Theta=0\right)$. 
This is expected geometrically, since these are the parallel correlations of two stars located on the North pole, the two components describing perpendicular configurations (cf. \autoref{fig:unit_vectors}). 
But these configurations are equivalent since for $m=0$ the background has no azimuthal dependence, and hence the two correlations are equal.
\begin{figure*}
\centering
\includegraphics{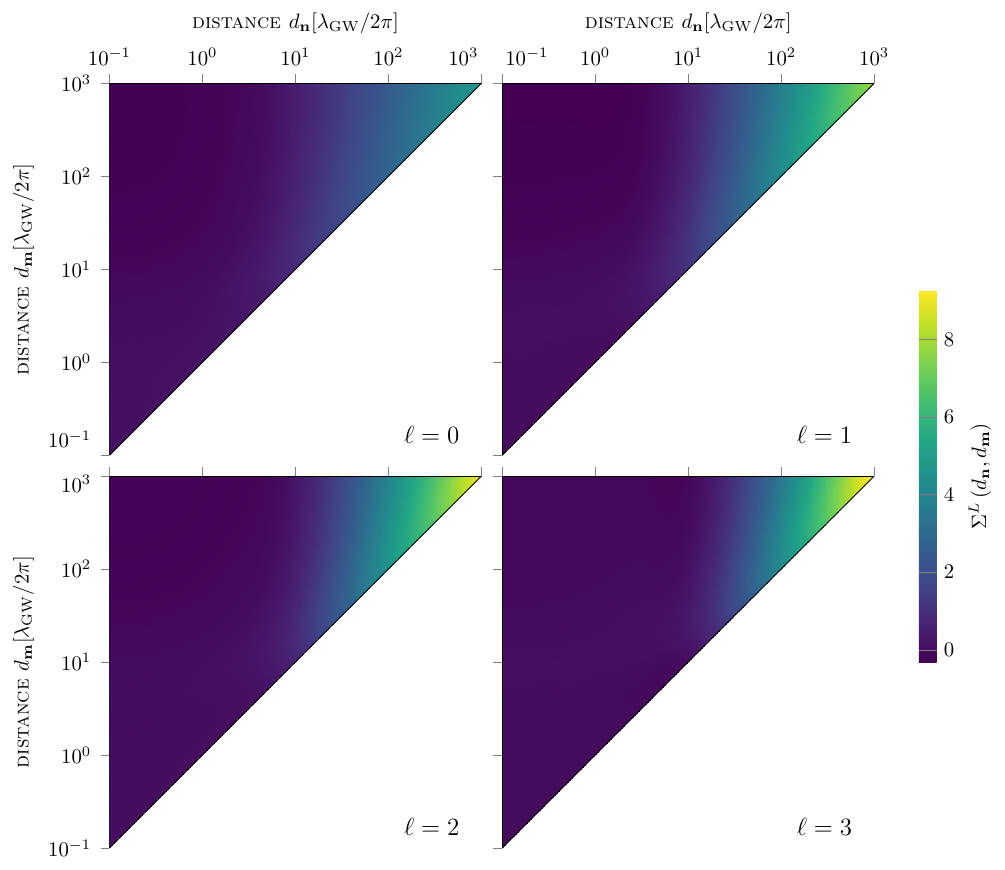}
\caption{Color plot of the real part of the codirectional astrometric longitudinal correlations, $\Sigma^L\left(\dn,\dm\right)\equiv\Re\Big[\mathcal{L}^L_{x\theta, \ell 0}\left(\dn, \dm, 0\right)\Big]$, from monopole to octopole anisotropy in the \sgwb. The two distances are normalized as $d\equiv2\pi|\bm{\mathrm{x}}_S| / \lambda_{\mathrm{GW}}$, where $\lambda_{\mathrm{GW}}$ is the wavelength of the GWs.} 
\label{fig:l_y0_exact}
\end{figure*}
We can then plug Eqs.~\eqref{eq:scalar_l_exact_responses} into the integrals in Eq.~\eqref{eq:scalar_l_exact_integrals}. To do the integral over $x$ we use the orthogonality relations of the Legendre polynomials in Eq.~\eqref{eq:leg_norm} and the following results:
\begin{align}
    &\int_{-1}^1\mathrm{d}x\,P_{\ell}\left(x\right)e^{\I yx} = 2 \I^{\ell} j_{\ell}\left(y\right),\label{eq:bessel1} \\
    &\int_{-1}^1\mathrm{d}x\,x^n P_{\ell}\left(x\right)e^{\I yx}=2 \I^{\ell-n} j^{\left(n\right)}_{\ell}\left(y\right), \label{eq:bessel2} \\
    &\int_{-1}^1\mathrm{d}x\frac{P_{\ell}\left(x\right)-1}{1-x}=-2 \Hnum{\ell}, \label{eq:harmonic_integral}
\end{align}
where \(j_{\ell}\!\left(y\right)\) are the spherical Bessel functions of the first kind, $j^{\left(n\right)}_{\ell}\left(y\right)$ are their n-th derivative w.r.t. y and $\Hnum{\ell} = \sum_{k = 1}^{\ell} \sfrac{1}{k}$ is the harmonic number (the sum of the first $\ell$ terms of the harmonic series). 
For Eqs.~\eqref{eq:bessel1} and \eqref{eq:harmonic_integral} see e.g. Refs. \cite[\href{https://dlmf.nist.gov/10.54.E2}{Eq.~(10.54.2)}]{NIST:DLMF} and \cite[\href{https://dlmf.nist.gov/18.17.E44}{Eq.~(18.17.44)}]{NIST:DLMF}, respectively. We obtain Eq.~\eqref{eq:bessel2} by differentiating Eq.~\eqref{eq:bessel1} w.r.t. $y$ $n$ times. Using these results, we obtain the codirectional generalized overlap reduction functions to be
\begin{align}\label{eq:gorf_y0}
        &\mathcal{L}^L_{x\theta, \ell 0}\left(\dn, \dm, 0\right) = \mathcal{L}^L_{y\phi, \ell 0}\left(\dn, \dm, 0\right) = \frac{1}{2}\sqrt{\left(2\ell + 1\right)\pi}\Bigg(\bigg[-\frac{7}{3}\delta_{\ell0}-\frac{2}{3}\delta_{\ell1}-\frac{2}{15}\delta_{\ell2}-2\Hnum{\ell} + \int_{-1}^1\frac{\mathrm{d}x}{1-x}\bigg]\notag\\
        &~
        -\frac{\I}{\dn}\bigg[\frac{2}{3}\delta_{\ell0}-\frac{1}{3}\delta_{\ell1} - \frac{2}{15} \delta_{\ell 2} + \frac{\I^{\ell}}{\dn^2}e^{\I \dn}\Big(\ell\left(1-\ell + \I \dn\right)j_{\ell}\left(-\dn\right)+\dn\left(2 + \I \dn\right) j_{\ell+1}\left(-\dn\right)\Big)-\frac{3}{2}\widetilde{H}_{\ell}\left(-\dn\right) + \widetilde{K}_{\ell}\left(-\dn\right)\!\bigg]\notag\\
        &~
        +\frac{\I}{\dm}\bigg[\frac{2}{3}\delta_{\ell0}-\frac{1}{3}\delta_{\ell 1} - \frac{2}{15} \delta_{\ell 2} + \frac{\I^{\ell}}{\dm^2} e^{- \I \dm}\Big(\ell\left(1 - \ell - \I \dm \right) j_{\ell}\left(\dm\right)+\dm\left(-2 + \I \dm\right)j_{\ell+1}\left(\dm\right)\Big)-\frac{3}{2}\widetilde{H}_{\ell}\left(\dm\right)+\widetilde{K}_{\ell}\left(\dm\right)\!\bigg]\notag\\
        &~
        +\frac{1}{\dn \dm}\bigg[A_{\ell}\left(-\dn\right)+A_{\ell}\left(\dm\right)-A_{\ell}\left(\dm-\dn\right)\!\bigg]\Bigg),
\end{align}
where we define the functions:
\begin{subequations}
    \begin{align}
        A_{\ell}\left(y\right)&\equiv\frac{8}{3}\delta_{\ell0}-\frac{2}{15}\delta_{\ell2}+\frac{\I^{\ell}}{y^2}e^{- \I y}\Big(\left(\ell\left(1-\ell\right)-2y^2\right)j_{\ell}\left(y\right)-2yj_{\ell+1}\left(y\right)\Big)-2\widetilde{H}_{\ell}\left(y\right)-\frac{1}{2}\widetilde{K}_{\ell}\left(y\right)+\widetilde{U}_{\ell}\left(y\right),
        \\[2ex]
        \widetilde{H}_{\ell}\left(y\right)&\equiv\int_{-1}^{1}\mathrm{d}x\frac{1}{\left(1-x\right)}P_{\ell}\left(x\right)\left(1-e^{- \I y(1-x)}\right),
        \\[2ex]
        \widetilde{K}_{\ell}\left(y\right)&\equiv\int_{-1}^{1}\mathrm{d}x\frac{1}{\left(1-x\right)^2}P_{\ell}\left(x\right)\left(1-e^{- \I y(1-x)}\right),
        \\[2ex]
        \widetilde{U}_{\ell}\left(y\right)&\equiv\int_{-1}^{1}\mathrm{d}x\frac{1}{\left(1-x\right)^3}P_{\ell}\left(x\right)\left(1-e^{- \I y(1-x)}\right).
    \end{align}
\end{subequations}
We can express the last three functions recursively as
\begin{subequations}
    \begin{align}
        & \widetilde{H}_0\left(y\right) = \Cin\left(2y\right) + \I \, \Si\left(2y\right), \\
        & \widetilde{H}_1\left(y\right) = \widetilde{H}_0\left(y\right)-2-\frac{\I}{y}\left(1-e^{-2 \I y}\right), \\
        & \widetilde{H}_{\ell}\left(y\right) = \frac{\left(2\ell-1\right)}{\ell}\widetilde{H}_{\ell-1}\left(y\right)-\frac{\left(\ell-1\right)}{\ell}\widetilde{H}_{\ell-2}\left(y\right)+2 \I^{\ell-1} \frac{2\ell - 1}{\ell}e^{- \I y}j_{\ell-1}\left(y\right), \,\, \mathrm{for} \, \ell\geq2, 
        \\[4ex]
        & \widetilde{K}_0\left(y\right) = \left(\frac{\cos{\left(2y\right)}-1}{2}+y\mathrm{Si}\left(2y\right)\right)+\I\left(-\frac{1}{2}\sin{\left(2y\right)}-y\mathrm{Cin}\left(2y\right)+y\Bigg[1+\int_{-1}^1\frac{\mathrm{d}x}{1-x}\Bigg]\right),\\
        & \widetilde{K}_1\left(y\right) = -\widetilde{H}_0\left(y\right)+\widetilde{K}_0\left(y\right), \\
        & \widetilde{K}_{\ell}\left(y\right ) = \frac{\left(2\ell-1\right)}{\ell}\widetilde{K}_{\ell-1}\left(y\right)-\frac{\left(\ell-1\right)}{\ell}\widetilde{K}_{\ell-2}\left(y\right)-\frac{\left(2\ell-1\right)}{\ell}\widetilde{H}_{\ell-1}\left(y\right), \,\, \mathrm{for} \, \ell\geq2, 
        \\[4ex]
        & \begin{aligned}
        & \widetilde{U}_0\left(y\right) = -\frac{1}{8}\left(1-\cos{\left(2y\right)}\right)-\frac{y}{4}\sin{\left(2y\right)}+\frac{3}{4}y^2-\frac{1}{2}y^2\mathrm{Cin}\left(2y\right)+\I\left(-\frac{1}{8}\sin{\left(2y\right)}-\frac{y}{4}\cos{\left(2y\right)}-\frac{1}{2}y^2\mathrm{Si}\left(2y\right)\right)\\
        &\quad
        +y\times\mathrm{b.t.}+\frac{y^2}{2}\int_{-1}^1\frac{\mathrm{d}x}{1-x}, 
        \end{aligned} \\
        & \widetilde{U}_1\left(y\right) = -\widetilde{K}_0\left(y\right)+\widetilde{U}_0\left(y\right), \\
        & \widetilde{U}_{\ell}\left(y\right) = \frac{\left(2\ell-1\right)}{\ell}\widetilde{U}_{\ell-1}\left(y\right)-\frac{\left(\ell-1\right)}{\ell}\widetilde{U}_{\ell-2}\left(y\right)-\frac{\left(2\ell-1\right)}{\ell}\widetilde{K}_{\ell-1}\left(y\right), \,\, \mathrm{for} \, \ell\geq2,
    \end{align}
\end{subequations}
where $\mathrm{Si}\left(x\right)$ and $\mathrm{Cin}\left(x\right)$ denote the sine and cosine integrals respectively, defined by

\vspace{0pt}
\begin{subequations}
\noindent
\begin{minipage}[t]{0.45\linewidth}
    \setlength{\abovedisplayskip}{0pt}
    \setlength{\belowdisplayskip}{0pt}
    \begin{equation}
        \Si\left(x\right) = \int_{0}^{x} \dd t \, \frac{\sin{t}}{t},
    \end{equation}
\end{minipage}%
\hfill
\begin{minipage}[t]{0.55\linewidth}
    \setlength{\abovedisplayskip}{0pt}
    \setlength{\belowdisplayskip}{0pt}
    \begin{equation}
        \Cin\left(x\right) = \int_{0}^{x} \dd t \, \frac{1-\cos{t}}{t},
    \end{equation}
\end{minipage}
\end{subequations}
\par
\vspace{\belowdisplayskip}
\noindent
and where to define the recursive relations above we have used the standard recursive relation for Legendre polynomials in Eq.~\eqref{eq:legendre_recursive}. The boundary terms labeled by b.t. and the divergent integrals $\int_{-1}^1\frac{\mathrm{d}x}{1-x}$ cancel out when combining all the terms and can thus be safely ignored when evaluating the astrometric correlations. A surface plot of $\Sigma^L\left(\dn,\dm\right)\equiv\Re\Big[\mathcal{L}^L_{x\theta, \ell 0}\left(\dn, \dm, 0\right)\Big]$ is shown in \autoref{fig:l_y0_exact}.
\end{appendices}
\end{document}